\documentclass{aa}  
\usepackage{siunitx}
\usepackage{rotating}

\usepackage{graphicx}
\usepackage{txfonts}
\usepackage[title]{appendix}

\titlerunning{First Dust Grains in Galaxies}
\authorrunning{Burgarella, Nanni et al.}

\begin{document}

   \title{Observational and theoretical constraints on the formation and early evolution of the first dust grains in galaxies at 5 $<$ z $<$ 10}


   \author{
          D. Burgarella\inst{1},
          A. Nanni\inst{1},
          H. Hirashita\inst{2},
          P. Theul\'e\inst{1},
          A. K. Inoue\inst{3, 4},
          \and
          T. T. Takeuchi\inst{5}
          }

   \institute{
         Aix Marseille Univ, CNRS, CNES, LAM, Marseille, France, \email{denis.burgarella@lam.fr}
         \and
         Institute of Astronomy and Astrophysics, Academia Sinica, Astronomy-Mathematics Building, AS/NTU, No.\ 1, Sec.\ 4, Roosevelt Road, Taipei 10617, Taiwan
         \and
         Department of Physics, School of Advanced Science and Engineering, Faculty of Science and Engineering, Waseda University, 3-4-1 Okubo, Shinjuku, Tokyo 169-8555, Japan
         \and 
         Waseda Research Institute for Science and Engineering, Faculty of Science and Engineering, Waseda University, 3-4-1 Okubo, Shinjuku, Tokyo 169-8555, Japan
         \and
         Division of Particle and Astrophysical Science, Nagoya University, Furo-cho, Chikusa-ku, Nagoya 464-8602, Japan
              }

   \date{Received November , 2019; accepted February 4, 2020}

 
  \abstract
   {The first generation of stars were born a few hundred million years after the big bang. These stars synthesized elements heavier than H and He, that are later expelled into the interstellar medium, initiating the rise of metals. Within this enriched medium, the first dust grains formed. This event is cosmological crucial for molecule formation as dust plays a major role by cooling low-metallicity star-forming clouds which can fragment to create lower mass stars. Collecting information on these first dust grains is difficult because of the negative alliance of large distances and low dust masses.}
   {We combine the observational information from galaxies at redshifts 5 $\lesssim$ z $\lesssim$ 10 to constrain their dust emission and theoretically understand the first evolutionary phases of the dust cycle.}
   {Spectral energy distributions (SEDs) are fitted with CIGALE and the physical parameters and their evolution are modelled. From this SED fitting, we build a dust emission template for this population of galaxies in the epoch of reionization.}
   {Our new models explain why some early galaxies are observed and others are not. We follow in time the formation of the first grains by supernovae later destroyed by other supernova blasts and expelled in the circumgalactic and intergalactic media.}
   {We have found evidence for the first dust grains formed in the universe. But, above all, this paper underlines the need to collect more data and to develop new facilities to further constrain the dust cycle in galaxies in the epoch of reionization.}


   \maketitle
%
 
\section{Introduction}
   \label{Introduction}

Understanding the characteristics of the dust cycle is still an issue. This statement is even truer for Lyman break galaxies at all redshifts (LBGs, e.g. \citealt{Burgarella2007}) and at redshifts z $>$ 5 (\citealt{Bouwens2011}, \citealt{Finkelstein2010}) at the epoch of reionization (EoR). This is due to the severe limits due to sub-millimetre (submm) and millimetre (mm) observations but also because we have a poor idea on the chemical conditions, the dust grain characteristics and even the stellar populations.

Fundamentally, as in the local universe, dust grains in high-redshift (Hi-z) galaxies absorb far ultraviolet (FUV) photons from young and massive stars. By doing so, these grains warm up to dust temperatures of a few tens of degrees and emit far-infrared (FIR) photons. 

This paper re-evaluates the roles of the relevant physical processes in the formation and early evolution of dust grains in these LBGs: formation in supernovae (SNe) explosion and galactic outflows from lower-mass stars  (e.g., \citealt{Todini2001}, \citealt{Matsuura2015}, \citealt{Ventura2012}, \citealt{Nanni2013}, \citealt{Nanni2014}, \citealt{Marassi2019}), destruction by SNe shocks, growth by accretion (e.g., \citealt{Hirashita2009}, \citealt{Dwek2011}, \citealt{Asano2013}) and galactic outflows (e.g., \citealt{Jones2018}, \citealt{Ohyama2019}). 


\vspace*{-0.2cm}
\subsection{What is the dust cycle in high redshift objects?}

Two types of stellar sources form and eject dust grains  in the interstellar medium (ISM) of galaxies: stars with an initial mass in the range 1 - 8 M$_\odot$ during the asymptotic giant branch (AGB) phase and stars with an initial mass in the range 8 - 40 M$_\odot$ during the core-collapse SNe phase. Dust grains also form seeds that grow in the ISM (e.g., \citealt{Dwek1998}, \citealt{Draine2009}, \citealt{Jones2011}, \citealt{Asano2013}). AGB dominate dust production in galaxies old enough to allow low-mass stars to evolve to the AGB phase (\citealt{Valiante2009}). The AGB contribution to the global dust production in metal-poor environments reach at most 30\% after about 2 Gyr (\citealt{DellAgli2019}). This is confirmed in other works (e.g. \citealt{Dwek2011}, \citealt{Dweketal2011}). However, SNe blasts produce a reverse shock wave able to destroy dust grains in the ISM (e.g., \citealt{Slavin2015}, \citealt{Matsuura2019}, \citealt{Nozawa2003}, \citealt{Dwek2011}).

There are numerous works (e.g. review by \citealt{Rupke2018}) that show that galactic winds are ubiquitous at low redshifts. These multiphase winds carries neutral and ionised gas and dust grains outside the galaxies. For instance, In M82, dust is found well beyond the radius where gas is thought to be outflowing from the galaxy. This suggests that dust escapes from the gravitational potential of galaxies into the intergalactic space (e.g., \citealt{Engelbracht2006}, \citealt{Yoshida2019}). Dust is also observed in the outflows from other local star-forming galaxies and strong evidence for dust in the outflowing neutral clouds is also found at z = 3 (\citealt{Shapley2003}). Very recently, in a very detailed analysis, \citet{Jones2018} find that their measurements favour a picture where the majority of heavy elements are ejected in a predominantly low ionization outflow, which regulates galactic chemical evolution.

The origin and the evolution of dust at high redshift has become a prominent issue since the early works of \citet{Watson2015}. At z $>$ 5, large amounts of dust ($\sim$10$^8$ M$_\odot$) are found in galaxies (e.g., \citealt{Riechers2013}, \citealt{Mortlock2011},  \citealt{Venemans2012}, \citealt{Michalowski2015}) or quasars (e.g., \citealt{Bertoldi2003}, \citealt{Priddey2003}, \citealt{Robson2004}, \citealt{Beelen2006}, \citealt{Michalowski2010}, \citealt{Cullen2017}, \citealt{Lesniewska2019}). Recently, we started to collect new and interesting data on the galaxy host of $\gamma$-ray bursts that will be very useful to study the properties of dust and the comparison of dust extinction and dust attenuation laws in these objects (e.g., \citealt{Stratta2011}, \citealt{Hjorth2013}, \citealt{Bolmer2018}).

\subsection{What explains the rest-frame UV colour properties and detectability of Hi-z LBGs?}

Most Hi-z LBGs observed in FIR are not detected (e.g. \citealt{Capak2015}, \citealt{Bouwens2016}, \citealt{Hirashita2017}). Explanations are proposed to understand the non-detections. \citet{Ferrara2017} propose that these objects can contain a large molecular gas mass. While dust in the diffuse ISM attains relatively high temperatures (T $\approx$ 70  - 80K), dust embedded in dense gas remains cold (T $\approx$ 30  - 40K). \citet{Faisst2017} suggest that radiation pressure causes a spatial offset between dust clouds and young star-forming regions within the lifetime of O/B stars. These offsets modify the radiation balance and create viewing-angle effects that can change UV colours at fixed IRX (IRX = log (L$_{dust}$ / L$_{FUV}$)). 

We use a 7-year WMAP7 cosmology (\citealt{Komatsu2011}) as adopted by Astropy, of $\Omega_M$ = 0.273, $\Omega_\Lambda$ = 0.727 and H$_0$ = 70.4. We assume a Chabrier initial mass function (IMF, \citealt{Chabrier2003}) for the SED fitting because top-heavy IMF (m$^{\alpha}$ with $\alpha$ = -1.5, -1.35, -1.0 , \citealt{Dabringhausen2010}, \citealt{Gall2011}) single stellar populations (SSPs) models are not available in CIGALE. The Chabrier IMF for the SED fitting is fixed in the one assumed by \citep{Bruzual2003}, i.e., the  lower and upper mass cut-offs m$_{Low} = 0.1 M_{\odot}$ and m$_{Up} = 100 M_{\odot}$. A Chabrier and a top-heavy IMFs are assumed for the dust modelling lower and upper mass cut-offs m$_{Low} = 1.3 M_{\odot}$ and m$_{Up} = 30 M_{\odot}$.
\section{Building spectral energy distributions}
  \label{Building_SEDs}
The Hi-z LBG sample is inhomogeneous and collected from several papers listed in Tab.~\ref{Table_Data}. It corresponds to a FUV selection. The criteria are the following ones: we selected all LBGs to have spectroscopic redshifts in the range 5 $\lesssim$ z $\lesssim$ 10 for which a mm detection was attempted (successful or not) in the rest-frame far-infrared (Tab.~\ref{Table_Data}). Practically, this means that we focused on LBGs observed in ALMA's Band 6 and 7. This redshift range (at EoR) provides constraints in rest-frame FIR, i.e., observed in the mm range. As an additional criteria, we only use LBGs with at least 5 data points in the rest-frame UV and optical. We collected photometric data to build SEDs and derive stellar masses (M$_{star}$), dust masses (M$_{dust}$), star formation rates (SFR), FUV and dust luminosities (L$_{FUV}$ and L$_{dust}$), the age of the dominant stellar population (Age$_{main}$) and the UV slope $\beta$ (\citealt{Calzetti1994}). The set of data is consistently fit with the same code CIGALE (\citealt{Burgarella2005}, \citealt{Noll2009}, \citealt{Boquien2019}), using the same dust models (\citealt{Draine2007}, \citealt{Draine2014}). The CIGALE modules and input parameters used for all the fits are described in Tab.~\ref{Table_CIGALE}. To model the dust emission, we also tried a modified, single dust temperature grey body plus a mid-IR power law which approximates the hot-dust emission from AGN (Active Galactic Nuclei) heating or clumpy, hot star bursting regions \citep{Casey2012}. The impact is negligible and we base this paper on the former to constrain our models. The dust models are calibrated with a sub sample of dwarf (compact  with sizes ranging from 0.08' to 1.2', \citealt{Madden2013}) low-metallicity galaxies (Low-zZ) with $12+\log(O/H) \le 8.4$ in the Dwarf Galaxy Survey (DGS,  \citealt{Madden2013}, \citealt{Remy-Ruyer2015}). The DGS has been originally selected from the Herschel Dwarf Galaxy Survey. It is a compilation of a SPIRE SAG2 guaranteed time (GT) key program plus SPIRE GT2 observations, using the PACS and SPIRE instruments inboard the Herschel Space Observatory to obtain 55–500$\mu$m photometry and spectroscopy of 48 dwarf galaxies in our local Universe, chosen to cover a broad range of physical conditions. We add data from the NASA Extragalactic Database (NED) in the UV (GALEX) and in the optical (B and R bands). When selecting the FUV, NUV, B and R data in NED, we have kept the data that provide total fluxes for our galaxies and when several total fluxes were available we preferred new or newly reprocessed ones. If these criteria are not respected, no data were added for the corresponding objects, which finally meant that we did not keep them in the Low-zZ sample.

\begin{table*}[htp]
\begin{center}
%
\begin{tabular}{|>{\centering\arraybackslash}p{2.7cm}|>{\centering\arraybackslash}p{2.6cm}|>{\centering\arraybackslash}p{1.5cm}|>{\centering\arraybackslash}p{1.6cm}|p{8.0cm}|}
 
  \hline\hline
  {\bf Source} & {\bf Number of objects} & {\bf Selection} & {\bf redshift} &{\bf  Note} \\
  \hline\hline
  \citet{Bouwens2016})  & 3 & UV & $5 \lesssim z \lesssim 10$ & * ALMA~6 upper limits \newline * SNR$_{HST}$ $>$ 1.5.\newline * LBGs with $>\ 5$ data points in UV-optical only\\ 
\hline
  \citet{Capak2015} \& \newline \citet{Faisst2017} & 9 (HZ1 - HZ4 \& \newline HZ6 - HZ10) & UV & z $\sim$ 5.6 & * [CII]158$\mu$ m for all Hi-z LBGs \newline * ALMA~7 detections: HZ4, HZ6 (3) HZ9 \& HZ10 (5) \newline * ALMA~7 upper limits for the others \newline * HZ5 detected in Chandra and not included in the sample \newline * Additional data from \citet{Pavesi2016}\\
\hline
  \citet{Scoville2016} & 1 (566428) & UV & z = 5.89 & * ALMA~6 detection \newline * [CII]158$\mu$ m measurement \\
\hline
  \citet{Willott2015} & 2 (CLM1 \& WMH5) & UV & z $\sim$ 6.1	 &  * ALMA~6 detections \newline * [CII]158$\mu$ m measurement \\
\hline
  \citet{Aravena2016} & 2 (ID27, ID31) & UV & z $\sim$ 7.5  &  * ALMA~6 detections \newline * [CII]158$\mu$ m measurement \\
\hline
  \citet{Hashimoto2018} & 1 (MACSJD1149) & UV & z = 9.1 & * ALMA~7 upper limit \newline * [OIII]88$\mu$ m measurement \\
\hline
  \citet{Remy-Ruyer2015} & 31 & DGS \&\newline Z < 0.1Z$_\odot$ & z $\sim$ 0  & * GALEX FUV and NUV data from NED \newline * B and R data from NED \newline * IR data (including Herschel) from \citet{Remy-Ruyer2013} and \citet{Remy-Ruyer2015}\\
\hline\hline
\end{tabular}
%
\end{center}
  \label{Table_Data}
  \caption{Origins of the data used in this paper. The Hi-z LBG sample contains 18 objects while the Low-zZ sample contains 31 objects for which the final fits are good enough, i.e. $\chi^2_{reduced}\ \le \ 5$. 9 Hi-z LBGs are detected in ALMA in Band 6 or Band 7, they provide 15 detected measurements but only 11 measurements with SNR$_{submm}\ >$ 3 are used to build the IR template.}
\end{table*}

The IR information on these Hi-z LBGs is limited. Our approach relies on the assumption that the physical conditions for dust grains are shared by these LBGs, which means that these LBGs share the same FIR SED shape within uncertainties. Here, we have used luminosity-weighted SEDs normalized at $\lambda = 200 \mu$m, where the dust is optically thin (e.g. \citealt{Casey2012}). 
Our LBG sample presents different FIR-to-FUV ratios and the dust attenuations are different. We leverage the redshift range to combine and build the IR SEDs of our ALMA-detected LBGs and thus gain access to a better spectral information for these objects (Fig. 1). However, we have no observational information on the dust emission, below the peak, i.e. near- and mid-IR. From the assumption that these Hi-z LBGs are metal-poor (\citealt{Castellano2014}, \citealt{Yuan2019}), we use the minimum value for q$_{PAH}$ (fraction of the dust mass in the PAH) because q$_{PAH}$ is correlated with the metallicity (e.g., \citealt{Ciesla2014}). From our SEDs, $\gamma$, the fraction of the dust heated by starlight above the lower cutoff U$_{min}$ (the minimum of the distribution of starlight intensity relative to the local interstellar radiation field), in Draine and Li's models cannot be constrained. We assume that the best value for the DGS sample is valid for the Hi-z LBGs (but the impact of this parameter on our result is limited to $\pm$8\%, estimated using the range found by \citep{Alvarez-Marquez2019} on a sample of 22000 LBGs at z $\sim$ 3.0).  In addition to the dust continuum information, we use the emission in the rest-frame FUV and optical range to model the stellar populations. Whenever available, [CII]158$\mu$m and [OIII]88.3$\mu$m are used to constrain the SED fitting. The nebular module in CIGALE uses nebular templates based on \cite{Inoue2011}, which have been generated using CLOUDY 08.00 (\citealt{Ferland1998}). Each line has a Gaussian shape with a user–defined line width to take rotation into account, which can be especially important for narrowband filters and high–redshift objects due to the apparent line broadening with redshift in the observed frame. This normalised nebular emission line spectrum is rescaled to the appropriate level by multiplying by the ionizing photon luminosity which was computed along with the composite stellar population.

\begin{table*}[htp]
\begin{center}
%
\begin{tabular}{|>{\centering\arraybackslash}p{3.0cm}|>{\centering\arraybackslash}p{3.0cm}|>{\centering\arraybackslash}p{2.1cm}|>{\centering\arraybackslash}p{2.1cm}|>{\centering\arraybackslash}p{2.1cm}|>{\centering\arraybackslash}p{2.1cm}|}
 
  \hline\hline
  {\bf Parameters} & {\bf Symbol} & {\bf Range (Ph.~1, Hi-z LBGs)} & {\bf Range (Ph.~2, Low-zZ)} &  {\bf Range (Ph.~3, Hi-z LBGs)} &  {\bf Range (Ph.~4, Hi-z LBGs)} \\
  \hline
 Target  sample    &                      &  Individual Hi-z LBGs & Individual Low-zZ & Combined Hi-z LBGs & Individual Hi-z LBGs  \\ 
  \hline\hline
    \multicolumn{6}{c}{}\\
    \multicolumn{6}{c}{\bf Delayed SFH and recent burst}\\
  \hline
 e-folding time-scale of the delayed SFH & $\tau_{main}$ [Myr] & 25, 50, 100, 250, 500, 1000  &  25, 50, 100, 250, 500, 1000 &  25, 50, 100, 250, 500, 1000 & 25, 50, 100, 250  \\ 
  \hline
 Age of the main population & Age$_{main}$[Myr]  & 61 log values in [2 - 1000] & 61 log values in [2 - 1500] & 17 log values in  [1, 800] & 101 log values in [50, 1260] \\ 
  \hline
 Burst & f$_{burst}$  &  No burst  &  No burst  & No burst  & No burst \\ 
  \hline
    \multicolumn{6}{c}{}\\
    \multicolumn{6}{c}{\bf SSP}\\
  \hline
  SSP &   & BC03 & BC03 & BC03 & BC03 \\ 
  \hline
  Initial Mass Function &  IMF & Chabrier & Chabrier & Chabrier & Chabrier \\ 
  \hline
  Metallicity     & Z &  0.004 &  0.004 & 0.004 & 0.004 \\ 
  \hline
    \multicolumn{6}{c}{}\\
    \multicolumn{6}{c}{\bf Nebular emission}\\
  \hline
  Ionisation parameter &  logU    &  -1.5  &  -1.5  & -1.5  & -1.5 \\
  Line Width [km/s]    &     ---    &  100 &  100 & 100 & 100 \\
  \hline
    \multicolumn{6}{c}{}\\
    \multicolumn{6}{c}{\bf Dust attenuation law}\\
  \hline
  Colour excess for both the old and young stellar populations &  E\_BV\_lines &  61 log values in [0.001, 0.3] &  61 log values in [0.001, 1.0] & 11 log values in [1e-3, 1e0] & 101 log values in [1e-3, 1e0]\\ 
  \hline
  Bump amplitude &  uv\_bump\_amplitude &  0.0 & 0.0 & 0.0 & 0.0 \\ 
  \hline
  Power law slope & power law\_slope &  -0.7, -0.35, 0.0, 0.35, 0.7 &  0.0 &  0.0 & 0.0 \\ 
  \hline
    \multicolumn{6}{c}{}\\
    \multicolumn{6}{c}{\bf Dust emission}\\
  \hline
  Mass fraction of PAH & $q_{PAH}$ &  0.47,1.12, 1.77, 2.50, 3.19 & 0.47,1.12, 1.77, 2.50, 3.19  &  0.47 & 0.47 \\ 
  \hline
  Minimum radiation field &  U$_{min}$ &  0.1, 0.5, 1.0, 5.0, 10.0, 50.0 &  0.1, 0.5, 1.0, 5.0, 10.0, 50.0 &  4.0 & 4.0 \\ 
  \hline
  Power law slope dU/dM $\approx$ U$^\alpha$ & $\alpha$ & 2.0, 2.2, 2.4, 2.6, 2.8, 3.0 &  2.0, 2.2, 2.4, 2.6, 2.8, 3.0 &  2.2 & 2.2 \\ 
  \hline
   Dust fraction in PDRs  & $\gamma$ &  0.00, 0.01, 0.1, 0.2, 0.3, 0.4, 0.5, 0.6, 0.7, 0.8, 0.9, 1.0 &  0.00, 0.01, 0.1, 0.2, 0.3, 0.4, 0.5, 0.6, 0.7, 0.8, 0.9, 1.0 & 0.75 & 0.75 \\ 
  \hline
    \multicolumn{6}{c}{}\\
    \multicolumn{6}{c}{\bf No AGN emission}\\
\hline\hline
\end{tabular}
%
\end{center}
  \label{Table_CIGALE}
  \caption{CIGALE modules and input parameters used for all the fits. BC03 means \cite{Bruzual2003} and the Chabrier IMF refers to \cite{Chabrier2003}.}
\end{table*}

\section{Methodology to derive the physical parameters}

The SED analysis is performed in five phases. 

\begin{figure*}
   \centering
   \includegraphics[width=2\columnwidth]{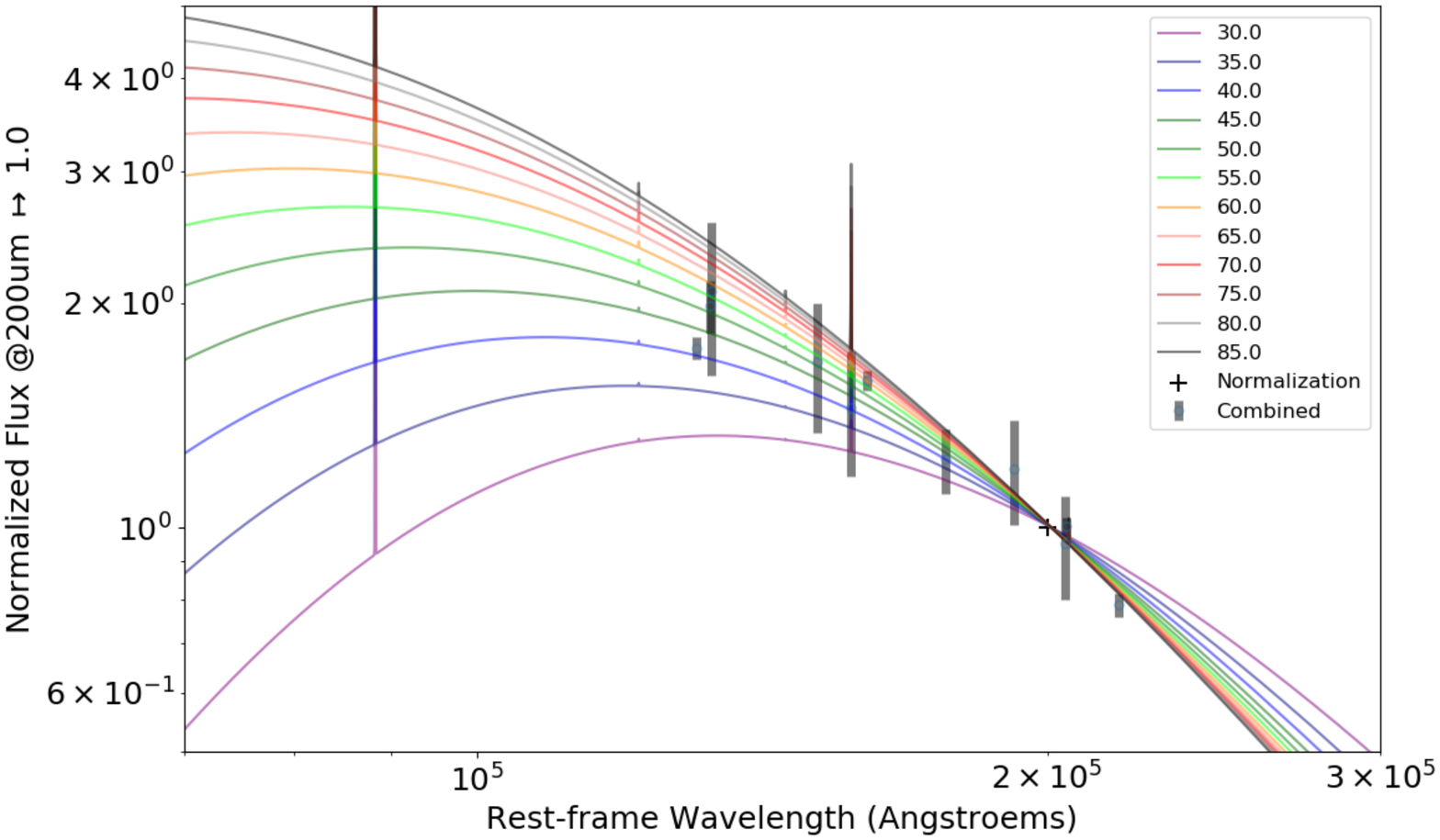}
    \caption{The dust emission of the detected LBGs with SNR${submm}\ >$ 3 (11 data points) are normalized at $\lambda_{rest-frame}$ = 200$\mu$m and combined to form a single Hi-z LBG template. For each detected LBG in our sample, we use the number of SNR$_{submm}\ >$ 3 ALMA detections that are validated in agreement with our criteria, as described in Tab.~\ref{Table_Data}. The redshift range (5 $\lesssim$ z $\lesssim$ 10) allows to cover a wide wavelength range. We over-plot colour-coded curves for SEDs based on modified blackbodies in the range 30 $\le$ T$_{dust}$ [K] $\le$ 85. }
   \label{Fig_Template_Tdust}
    \end{figure*}

\begin{figure}
   \centering
   \includegraphics[width=0.45\columnwidth]{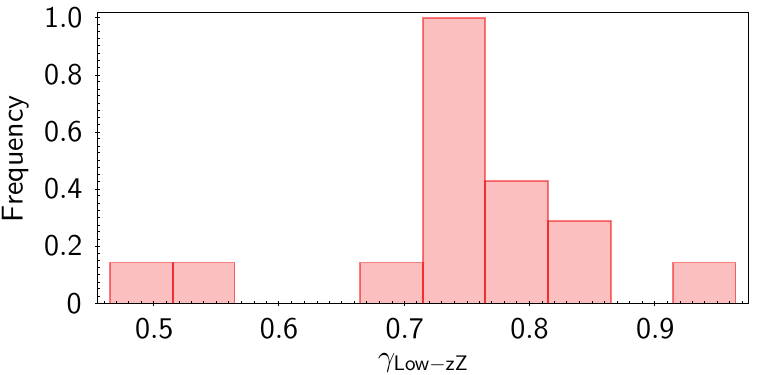}
   \includegraphics[width=0.45\columnwidth]{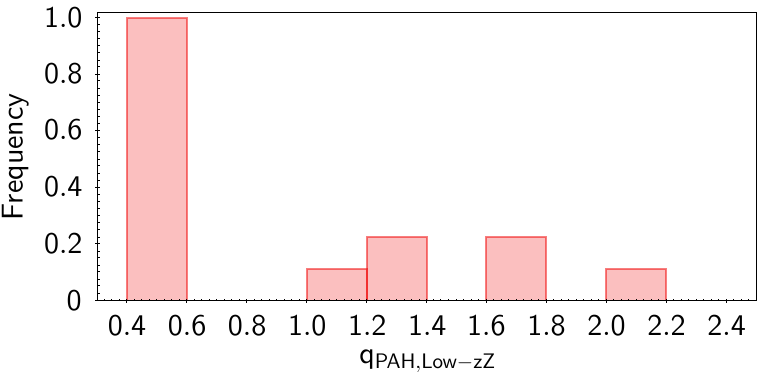}
   \caption{Distribution of the $\gamma_{Low-zZ}$ (left panel) and q$_{PAH\_Low-zZ}$ (right panel) dust parameters from Draine \& Li's models for the Low-zZ galaxy sample. The $\gamma_{Low-zZ}$ distribution peaks at 0.75 $\pm$ 0.07 and most of the q$_{PAH}$ = 0.47 which is the lowest value in Draine \& Li's models. $\gamma_{Low-zZ}$ is assumed to be valid for the Hi-z LBG sample and q$_{PAH\_Low-zZ}$ is fully consistent with our assumed q$_{PAH\_Hi-z}$.}%
    \label{Fig_gamma_PAH_Low-zZ}
   \end{figure}

\begin{figure}
   \centering
   \includegraphics[width=\columnwidth]{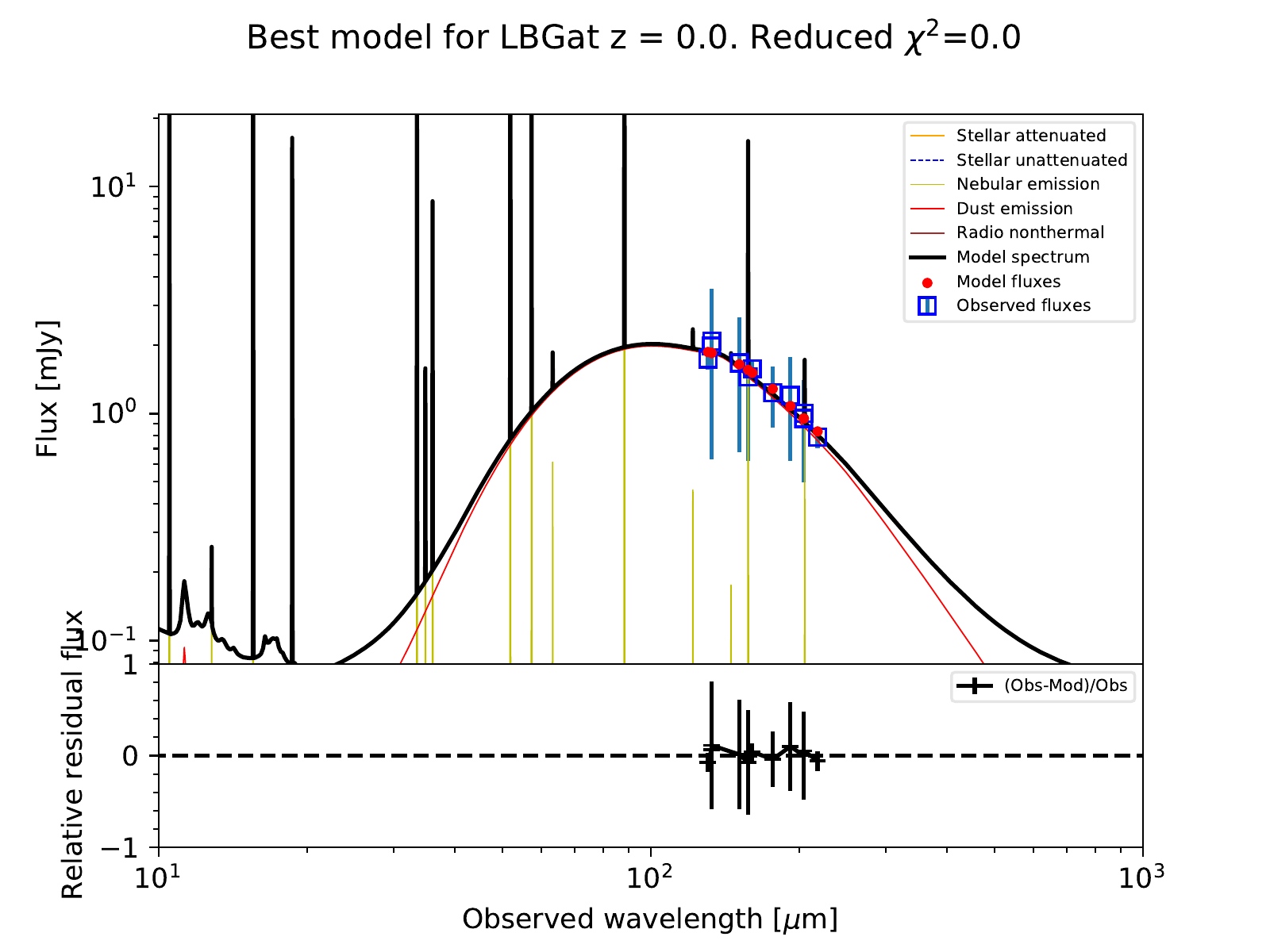}
   \includegraphics[width=\columnwidth]{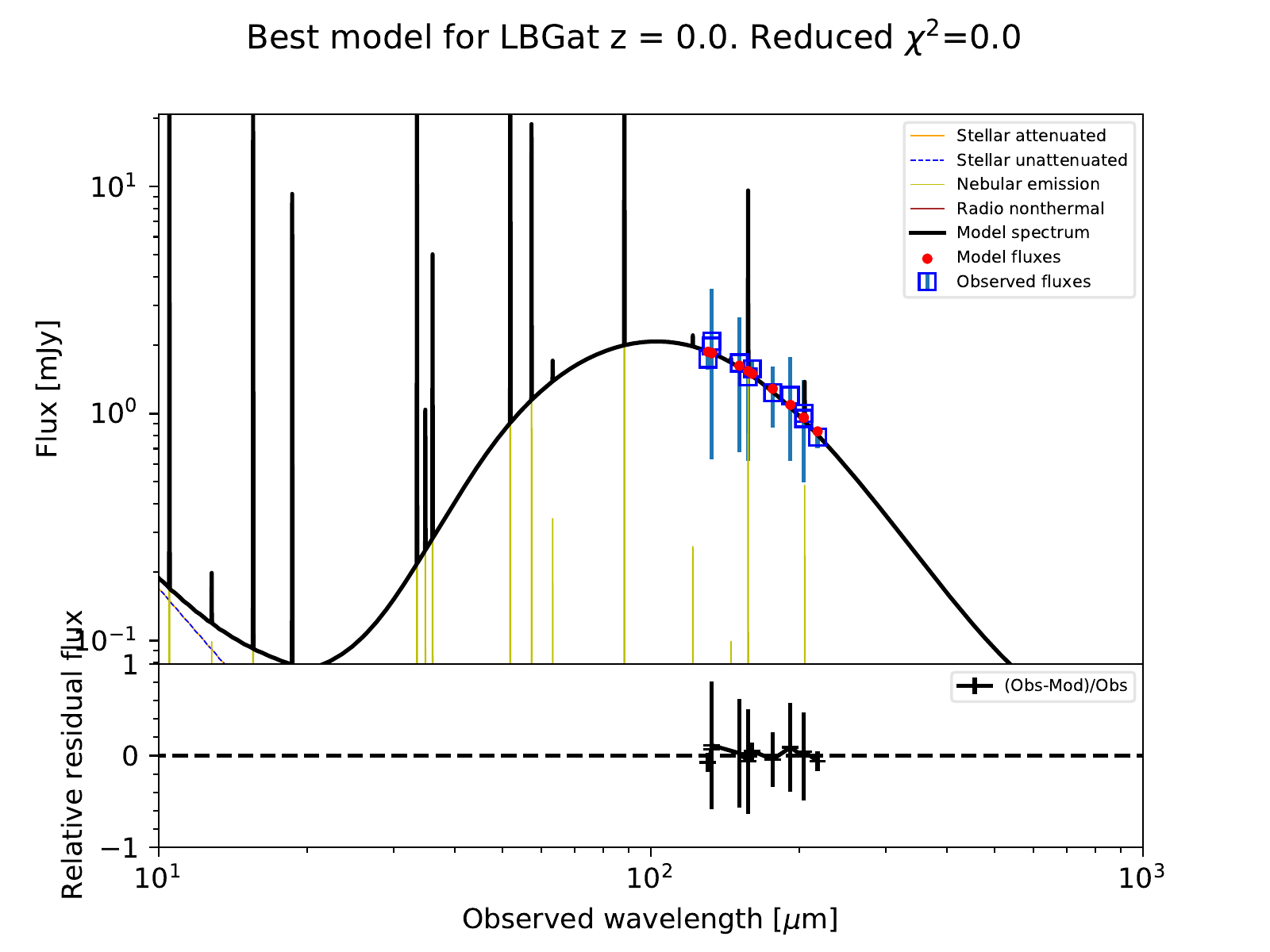}
   \caption{Fit of the IR template SED built from individual ones normalised at 200 $\mu$m and combined. We used Draine and Li's and a modified black body a-la-Casey (2012).}%
    \label{Fig_Template-fit}
   \end{figure}

\begin{enumerate}

\item {Phase 0:} A necessary preliminary phase was to perform a lot of initial fits on individual Low-zZ and Hi-z LBGs to understand the limits of the physical parameters to be explored (star formation history (SFH), dust attenuation, AGN fraction, etc.) in the CIGALE fits. Because of the exploration nature of these fits, these initial fits are not reported in Tab.~\ref{Table_CIGALE}.

\item {Phase 1:} CIGALE fits individual Hi-z LBGs with a large distribution of priors for the dust parameters and SFH (delayed\footnote{$SFR\ = \frac{t}{\tau^2}e^{-\frac{t}{\tau}}$}, delayed + burst, and constant SFHs are assumed). Constant SFHs provide very close results. Fits with a delayed + burst SFH are less good. Because none of the stellar models available in CIGALE feature a top-heavy IMF, we use a Chabrier IMF\footnote{Using a top-heavy IMF in the SED fitting instead of a Chabrier one could change the results of the SED fitting in the UV-optical range but less on the IR range. Future development of CIGALE will address this issue. In a similar way, stellar models that include binaries (e.g., BPASS \citealt{Stanway2018}) would change the UV-optical emission of these galaxies. We will address this possibility in a future paper.} These initial fits provide an estimated rest-frame flux$_{200 \mu m}$ to normalize the SEDs (with FIR detections) at $\lambda = 200 \mu$m where the dust is optically thin (e.g. \citealt{Casey2012}). Detected LBGs at 5 $\lesssim$ z $\lesssim$ 10 allow to sample the spectrum of dust emission (Fig.~\ref{Fig_Template_Tdust}). Although only the Rayleigh-Jeans range is covered, dust temperatures in the range 40K to 70K fit the data better.

\item  {Phase 2:} Local Low-zZ galaxies are often said to be analogues to early galaxies (\citealt{Madden2013}, \citealt{Capak2015}, \citealt{Hou2019}), We compile a sample of Low-zZ galaxies from \citet{Remy-Ruyer2013} and \citet{Remy-Ruyer2015}. We only keep DGS galaxies with 12 + log(O/H) $<$ 8.0. With the same assumptions on the priors than for the LBGs, these galaxies are fit with CIGALE. We check that the main dust physical parameters found for the Low-zZ sample are in agreement with the ones from the Hi-z LBG sample (which confirms the partial analogy) and we leverage the former fits to set the value of the $\gamma$ parameter in the models from \citealt{Draine2007} and \citealt{Draine2014}. We also check that the value estimated for q$_{PAH}$ is consistent with our assumption of a low metallicity (Fig.~\ref{Fig_gamma_PAH_Low-zZ}). 

\item {Phase 3:} q$_{PAH}$ and $\gamma$ are fixed and we fit the IR-combined SEDs of the Hi-z LBGs (Fig.~\ref{Fig_Template-fit}). This Phase 3 provides physical parameters for the average combined SEDs of the Hi-z LBGs used to build the Hi-z IR template. The fit with a modified black body gives: $\beta_{submm}$ = 1.51 (slope on rest-frame submm side of the dust emission) and T$_{dust}$ = 56.6K.

\item {Phase 4:} We re-fit all the Hi-z LBGs over the FUV-to-FIR range with this Hi-z IR template  (Tabs.~\ref{Table1_Hi-z} and~\ref{Table2_Hi-z} ). As a safety check, we also fit the low-zZ galaxies with the same IR template. The quality of the fit is shown in the Annex.

\end{enumerate}

%

%

   \begin{figure}
   \centering
   \includegraphics[width=0.45\columnwidth]{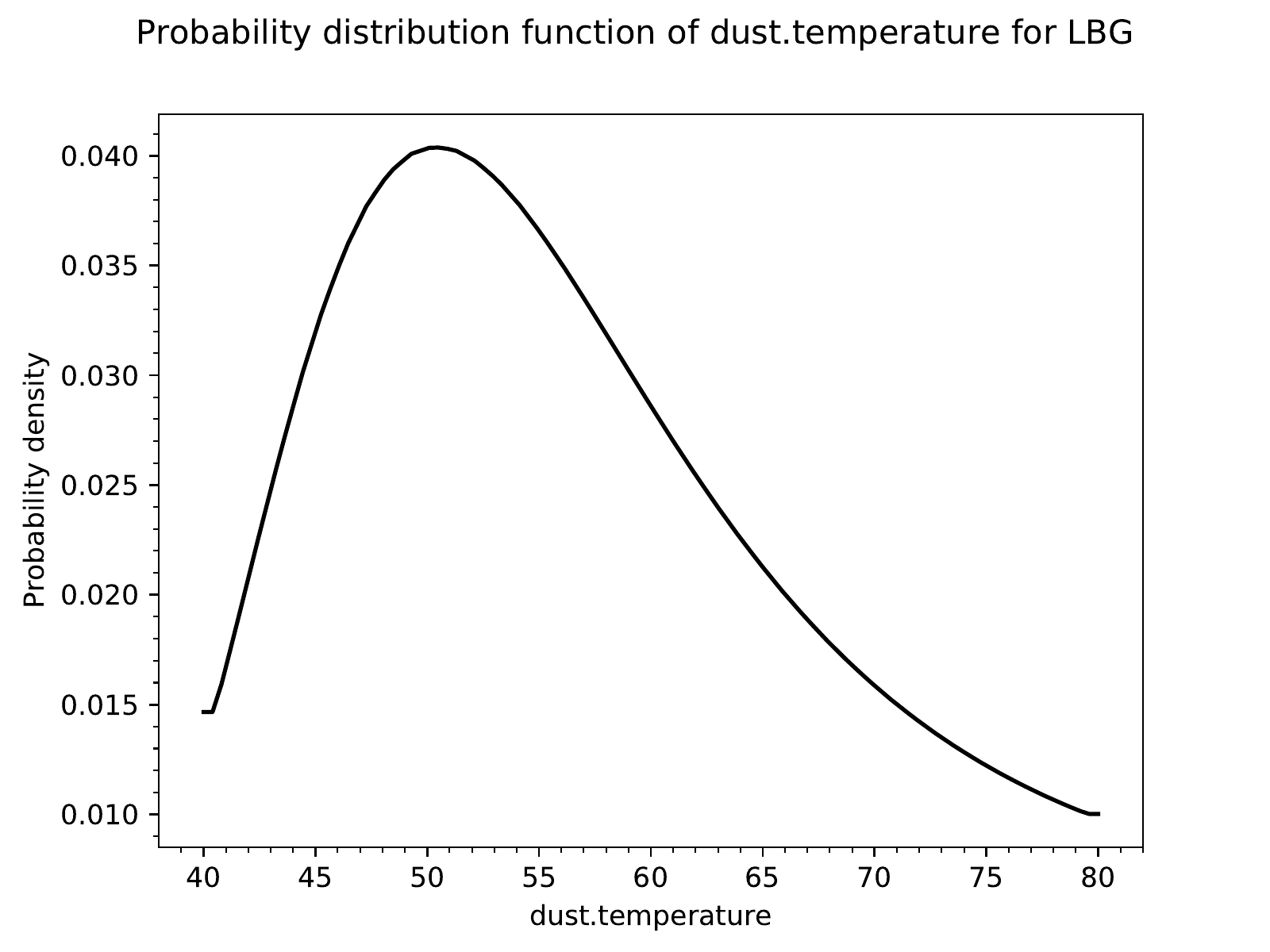}
   \includegraphics[width=0.45\columnwidth]{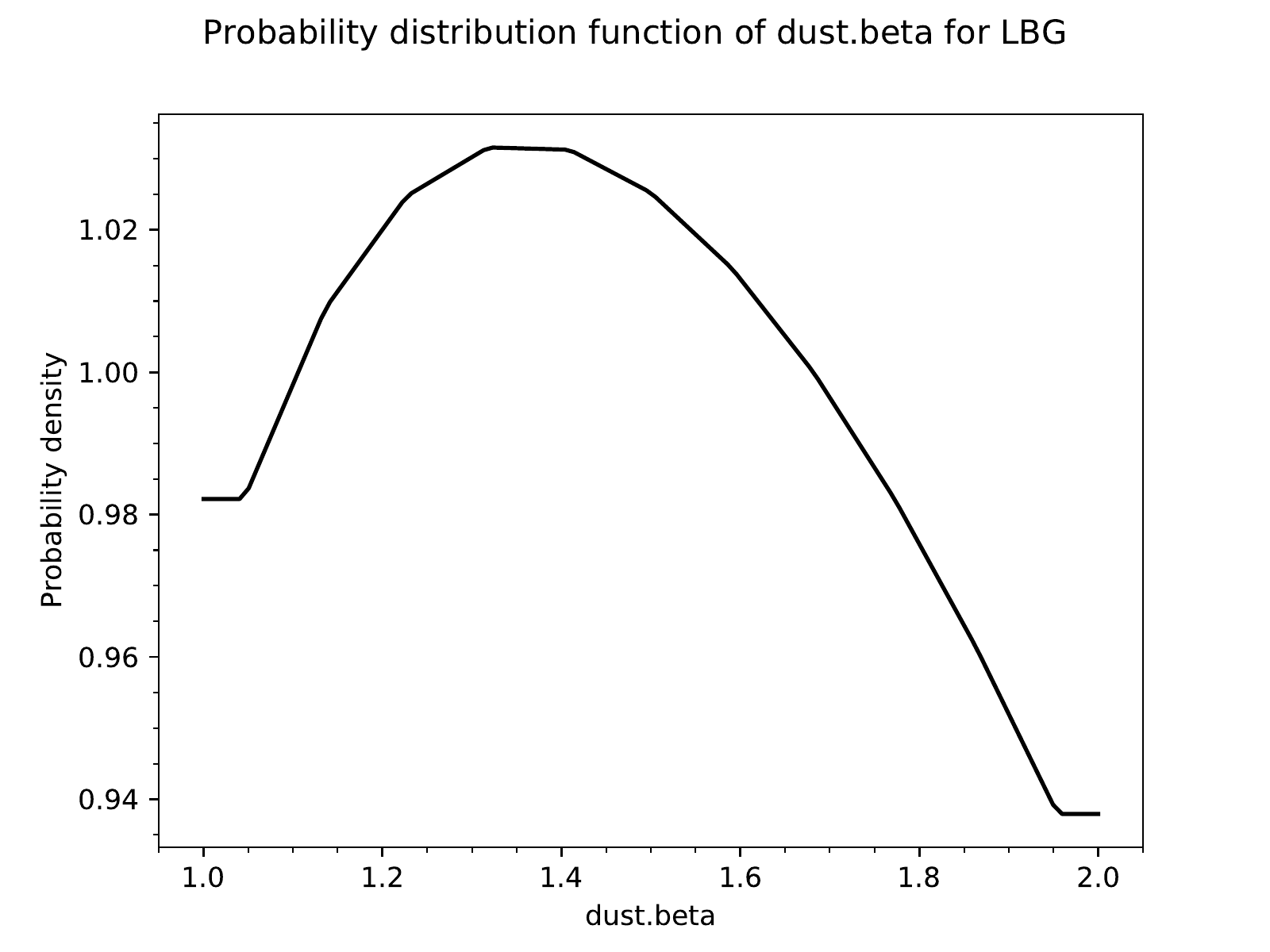}
   \caption{Probability distribution functions for the dust temperature and the emissivity derived from the fit of the IR dust template built from the Hi-z LBGs with CIGALE and a modified black body a-la-Casey (2012).}%
    \label{Fig_dust_PDF}
   \end{figure}
\section{Results}
  \label{Results}

\begin{figure*}
   \centering
   \includegraphics[width=2\columnwidth]{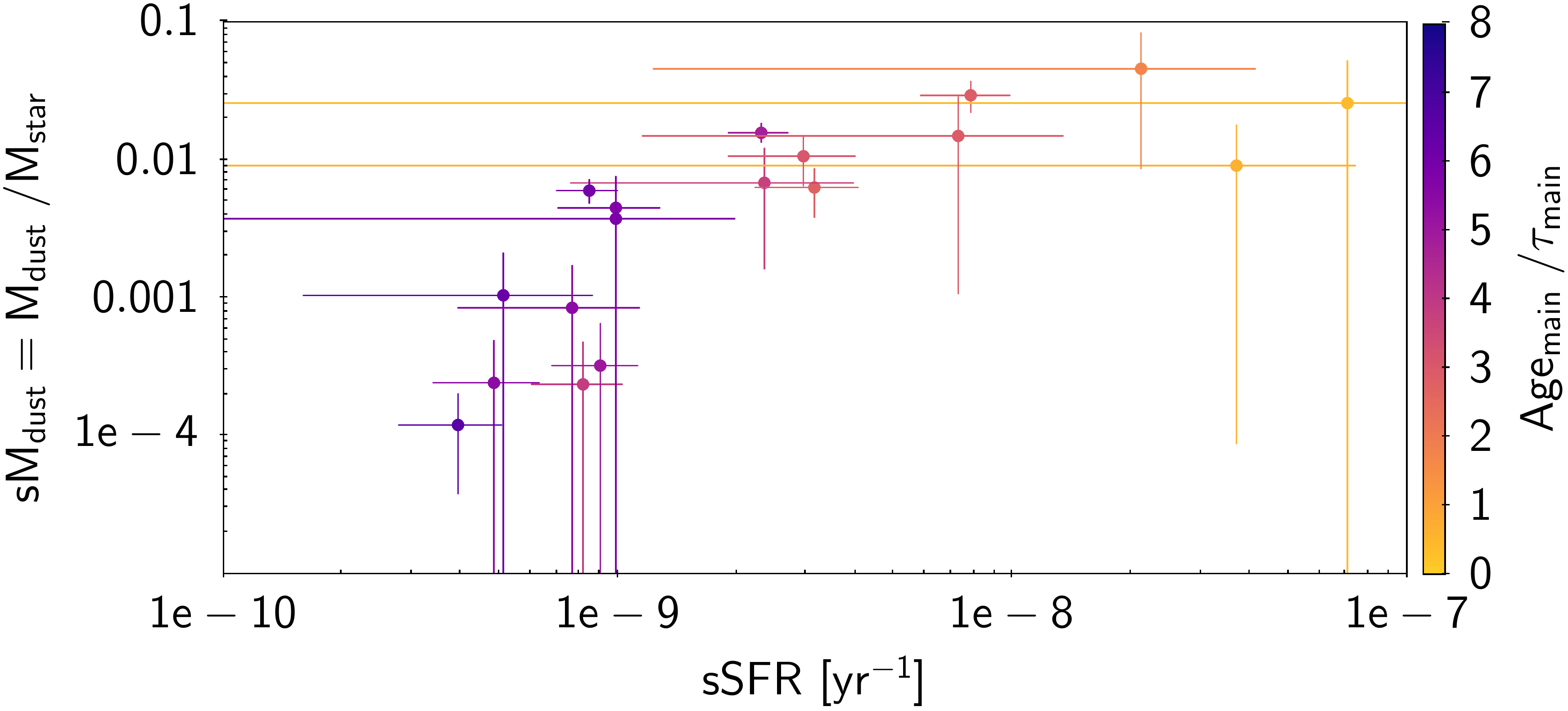}
   \includegraphics[width=2\columnwidth]{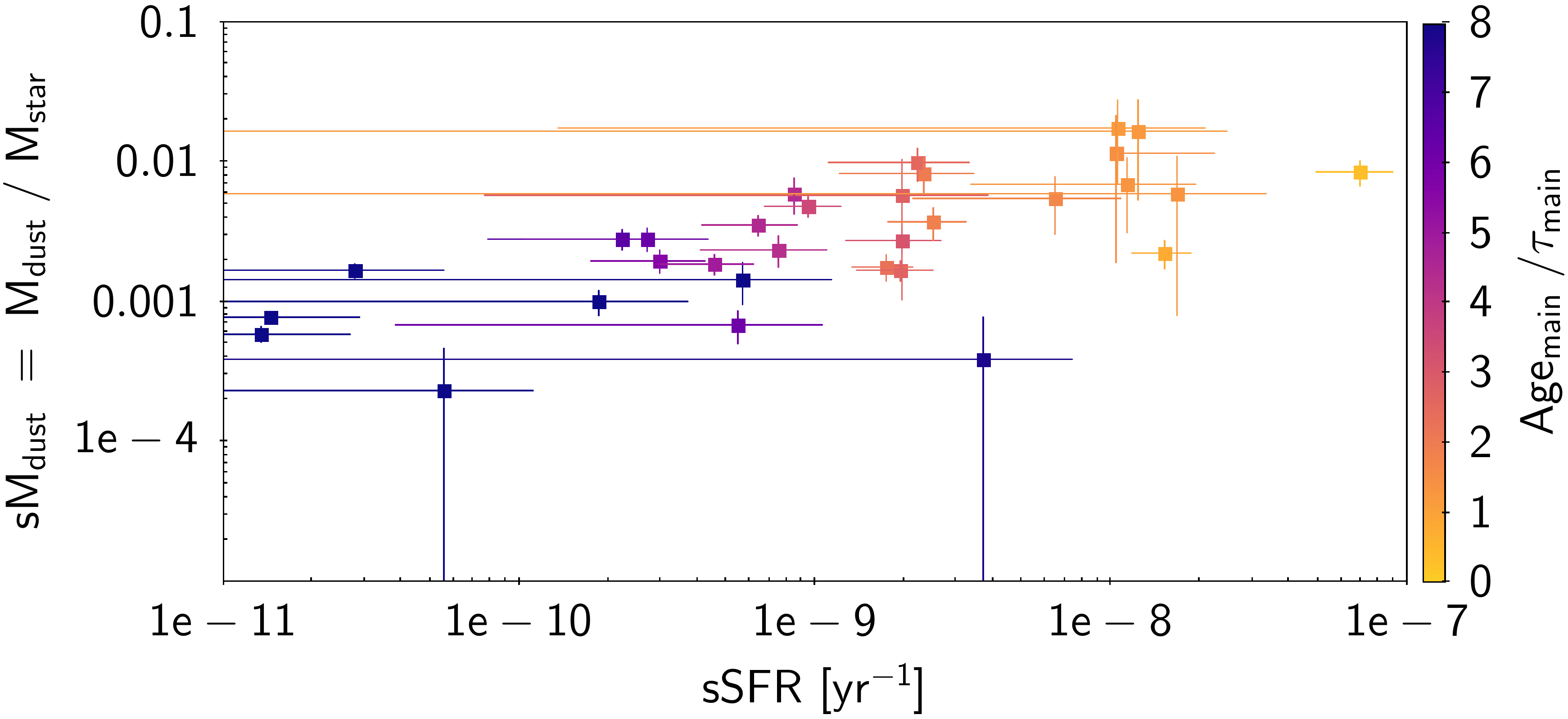}
   \caption{Hi-z LBGs (filled dots, a) and Low-zZ galaxies (open boxes, b) DFRD, sM$_{dust}$ vs. sSFR, colour-coded in age. We see a sequence from top-right to bottom-left. Using a colour-coding based on the age of the dominant stellar population, we see that this sequence corresponds to galaxies getting older and older from top-right to bottom-left. }%
    \label{Fig_DFRD_Age}
    \end{figure*}

\begin{figure*}
   \centering
   \includegraphics[width=2\columnwidth]{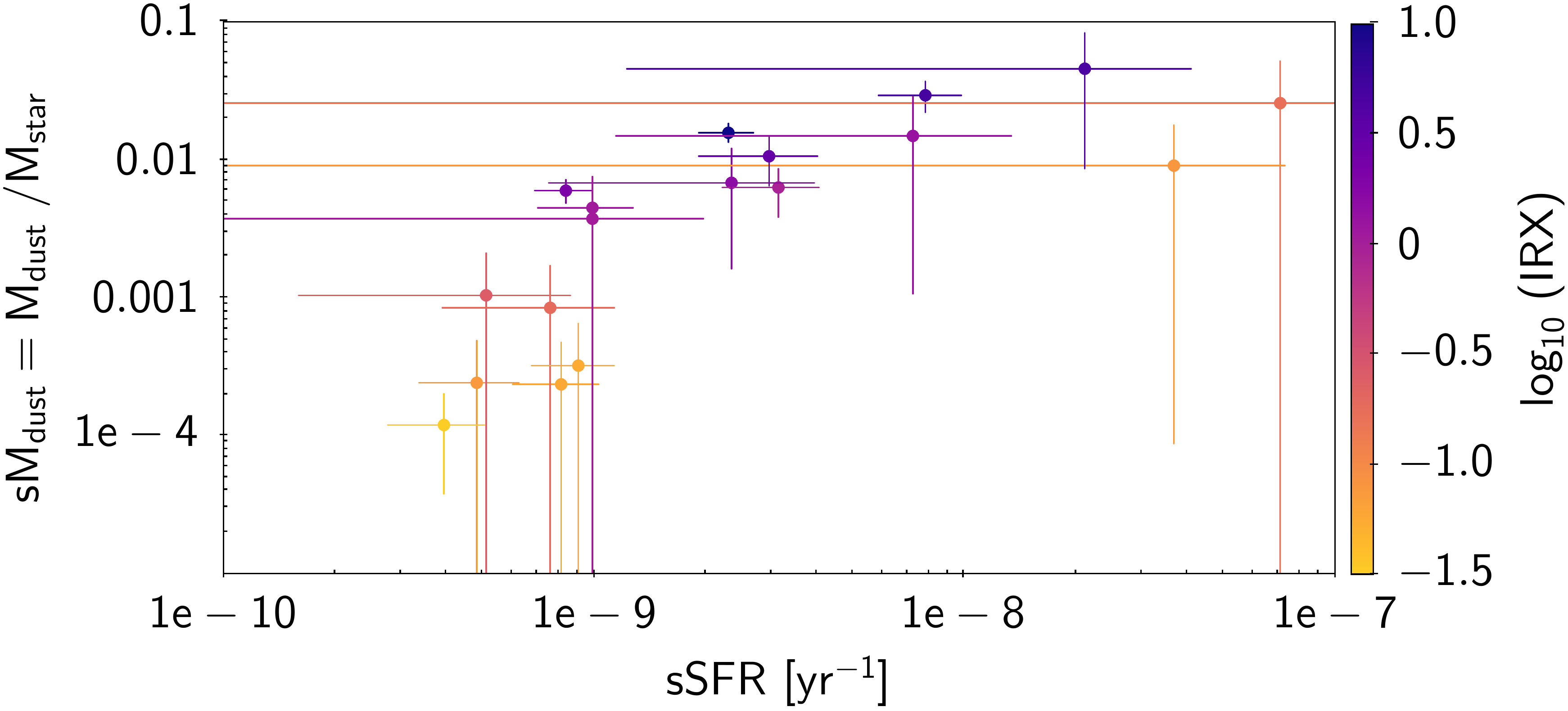}
   \includegraphics[width=2\columnwidth]{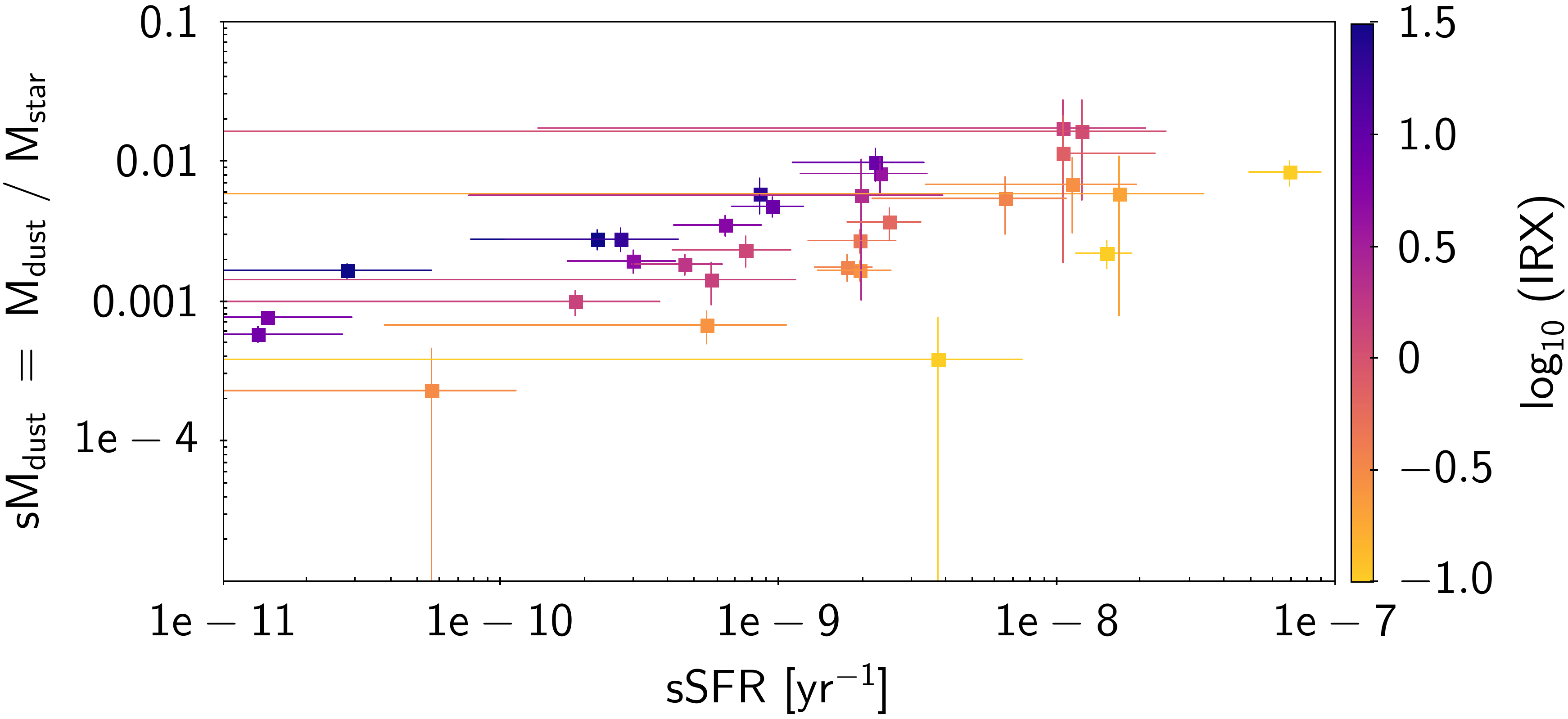}
   \includegraphics[width=1.9\columnwidth]{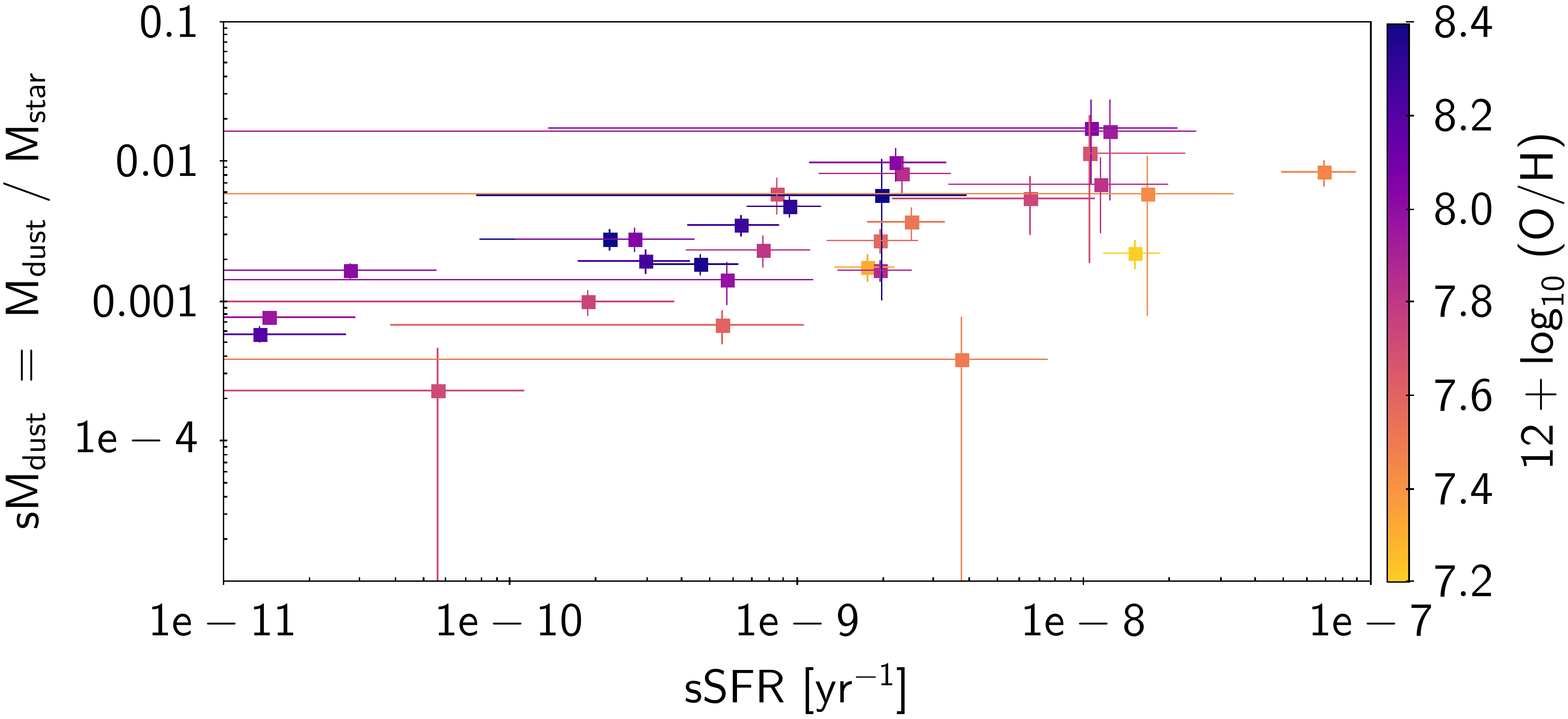}
   \caption{Hi-z LBGs (filled dots, a) and Low-zZ galaxies (open boxes, b) dust formation rate diagram, sM$_{dust}$ vs. sSFR, colour-coded in IRX. Like in Fig.~\ref{Fig_DFRD_Age}, colour-coding the data points helps interpreting the sequence as an evolutionary sequence with low-IRX objects at high sSFR and again at low sSFR. In the middle range, IRX is higher. }%
    \label{Fig_DFRD_IRX}
    \end{figure*}

From Tab.~\ref{Table1_Hi-z} and Tab.~\ref{Table2_Hi-z} we compute the specific SFR (sSFR = SFR/M$_{star}$), the specific dust mass (sM$_{dust}$ = M$_{dust}$/M$_{star}$) and IRX =  L$_{dust}$ / L$_{FUV}$. In Figs.~\ref{Fig_DFRD_Age} and \ref{Fig_DFRD_IRX}, we define the dust formation rate diagram (DFRD): sM$_{dust}$ vs. sSFR. The x-axis provides an information on the stellar population and the y-axis on dust grains. For some of the galaxies, the fit is not good, e.g., SBS 0335-052 (see Sect.~\ref{Results}). Two possible origins for these bad fits are the quality of the data and/or that our template do not match the dust properties of these objects. They are not used in our analysis.

The physical parameters derived for the Hi-z LBG and the Low-zZ DGS samples by CIGALE are listed in Tab.~\ref{Table1_Hi-z}, Tab.~\ref{Table2_Hi-z} and Tab.~\ref{Table_Low-zZ}.

\begin{table*}[t]
\sisetup{table-column-width=12ex,    
         round-mode=places,
         round-precision=3,
         tight-spacing,
         table-format = 2.3e-2,
         table-number-alignment = center
         }
  \resizebox{\linewidth}{0.8\height}{

\begin{tabular}{l|S|S|S|S|S|S|S|S|S|S|S|S|S|S|S|S|}

\hline
  \multicolumn{1}{|c|}{id} &
  \multicolumn{1}{c|}{$\chi_\nu^2$} &
  \multicolumn{1}{c|}{sSFR[yr$^{-1}$]} &
  \multicolumn{1}{c|}{sSFR\_err[yr$^{-1}$]} &
  \multicolumn{1}{c|}{sM$_{dust}$} &
  \multicolumn{1}{c|}{sM$_{dust}$\_err} &
  \multicolumn{1}{c|}{IRX} &
  \multicolumn{1}{c|}{IRX\_err} &
  \multicolumn{1}{c|}{A$_{FUV}$} &
  \multicolumn{1}{c|}{A$_{FUV}$\_err} &
  \multicolumn{1}{c|}{$\beta_{calz94}$} &
  \multicolumn{1}{c|}{$\beta_{calz94}$\_err} &
  \multicolumn{1}{c|}{Age[Myr]} &
  \multicolumn{1}{c|}{Age\_err[Myr]} &
  \multicolumn{1}{c|}{Age/$\tau$} &
  \multicolumn{1}{c|}{Age/$\tau$]\_err} &
  \multicolumn{1}{c|}{12+$\log_{10}\ (O/H)$ $^a$}\\
\hline
\hline
  ID27 & 0.44246197945927773 & 7.125515251980087E-8 & 7.1255E-8 & 0.025557150815114473 & 0.02555 & 0.27675044251906006 & 0.27675 & 0.10527315293081725 & 0.12289464067922648 & -2.615958976496767 & 0.05973876293357417 & 34.498418358563036 & 43.312889023673165 & 0.4431823793803463 & 0.4281615522161736 & 7.627357772186866\\
  ID31 & 0.1545754935798121 & 3.7007320675263604E-8 & 3.698926680202969E-8 & 0.008884478160426822 & 0.0088 & 0.14487342011342091 & 0.1448 & 0.052967156077890064 & 0.06771722267731886 & -2.6125766228830254 & 0.052328358110762765 & 53.980564668143714 & 56.52114438553019 & 0.6509865390693879 & 0.5344585985440479 & 7.540666134203301\\
  lbg10 & 0.27152885848462854 & 2.120385490331588E-8 & 1.9975724971330087E-8 & 0.045693598242776474 & 0.03712743495782498 & 4.596922617547997 & 2.8230732039394533 & 1.4229271726823716 & 0.6216240011252855 & -2.0067823326514316 & 0.10691649261504832 & 106.72082999496436 & 96.8551003120613 & 1.724296469269668 & 1.821796314808553 & 8.00291196892828\\
  HZ4 & 0.5451264778885693 & 3.1581389324720702E-9 & 9.227780854408122E-10 & 0.006176829294231244 & 0.0023784537613786008 & 0.9053446992430267 & 0.29506199796706384 & 0.53778201483052 & 0.14929914809773595 & -2.1998255283994554 & 0.05831544640411405 & 302.9395366717754 & 137.272828176785 & 2.763477849581292 & 1.0180641412130826 & 7.803282670177345\\
  HZ9 & 3.2598827758439892 & 7.866524159775136E-9 & 2.008178964956289E-9 & 0.02907131115232921 & 0.007315912038371634 & 4.9356552670328195 & 0.6375887287967048 & 1.6618038687396985 & 0.1381975741485445 & -1.8475276954738835 & 0.07336338375684429 & 101.5093195054336 & 42.764774808442006 & 2.873534329583004 & 0.7507572230257696 & 8.011647249181767\\
  lbg53 & 1.1339474081519647 & 7.293142310539077E-9 & 6.1412397085773465E-9 & 0.014533618530120267 & 0.013500680904272639 & 1.2852120245140555 & 1.0575030163863064 & 0.6271362220316036 & 0.4292924871592738 & -2.2258860413683776 & 0.08080708490637879 & 134.69961663532607 & 83.36892151865352 & 2.907596281083765 & 1.8798556312440213 & 7.846328965313329\\
  MACSJD1149 & 0.469825840388361 & 2.968935455652832E-9 & 1.058930611211668E-9 & 0.010496933990061937 & 0.004132687011094936 & 3.0023667761917334 & 0.9251496311249913 & 1.2288909700908484 & 0.2508902674842452 & -1.8997404961491322 & 0.09045155242797906 & 278.09534072907655 & 126.28221081029206 & 3.0667555119109156 & 1.0467811042808386 & 7.950574493625772\\
  566428 & 0.3092354200123932 & 2.3576508294722528E-9 & 1.5984675705036265E-9 & 0.006696477110323473 & 0.005116143927744587 & 1.5434557662313764 & 1.1265153180488454 & 0.7341555853129639 & 0.4444533027226875 & -2.038284932509998 & 0.10646762111844059 & 299.7340310423622 & 181.60046833752057 & 3.750587784173394 & 1.754006910560193 & 7.868825005668514\\
  HZ2 & 2.2824328789722004 & 8.207676809536868E-10 & 2.1170477267989363E-10 & 2.3259451894130763E-4 & 2.32E-4 & 0.09377944197555088 & 0.0937 & 0.04136923607265847 & 0.05050282704358323 & -2.219511708661225 & 0.05607149911179164 & 691.9701324492388 & 217.7654509706874 & 3.835889814996654 & 0.7728374117699667 & 7.495986141146543\\
  HZ10 & 3.6516867890006592 & 2.3089894899636916E-9 & 3.9454010166487344E-10 & 0.015535652510792886 & 0.0024217843221919365 & 10.39721210186397 & 1.1199902056188795 & 2.347406234496019 & 0.14044709465250735 & -1.3734244647080234 & 0.08067719463866513 & 119.49344024610943 & 6.843100982066057 & 4.779411361340476 & 0.2725744867929171 & 8.103185792033283\\
  HZ8 & 1.7345709663718178 & 9.079746925772317E-10 & 2.267537465838922E-10 & 3.1801940503233423E-4 & 3.18E-4 & 0.06496651284892277 & 0.06496 & 0.04220098260101025 & 0.04705246729772277 & -2.1961207232388413 & 0.03685222034651838 & 290.78206099056047 & 100.363118631707 & 5.027729295295286 & 0.6041046737445351 & 7.461785976395262\\
  HZ7 & 0.5926646041430074 & 7.686477480654121E-10 & 3.7398258237564214E-10 & 8.429308873186199E-4 & 8.4293E-4 & 0.19998614825173844 & 0.1999 & 0.12973093879740094 & 0.12980363284003027 & -2.1290441829609885 & 0.05414248529051527 & 303.33053394680377 & 159.92168212258133 & 5.414521795442939 & 1.1570495929098128 & 7.609101244615039\\
  HZ1 & 0.48355787354353647 & 4.870603616360092E-10 & 1.4706430801594405E-10 & 2.3669E-4 & 2.366E-4 & 0.0888722234861341 & 0.0888 & 0.05267094432496958 & 0.0599312466687202 & -2.0723529031300387 & 0.049106836304849424 & 440.3248993766254 & 152.34526428626265 & 5.420628506638703 & 0.6669019611849706 & 7.496715194101922\\
  HZ6 & 2.195613763698042 & 9.937119237290451E-10 & 2.8571481232109177E-10 & 0.004371748309383913 & 5.169342907368144E-4 & 1.1043542985562351 & 0.07808964133416271 & 0.6577146088712251 & 0.03447373890132453 & -1.9633450542675015 & 0.04266174613544832 & 174.07893375856287 & 51.716861528002106 & 5.639069478035462 & 0.6668497048161754 & 7.827695373701333\\
  lbg48 & 0.21569035945446857 & 9.946987993139398E-10 & 9.946E-10 & 0.003654914497081512 & 0.00365 & 1.499476132935065 & 1.4994 & 0.4767531462766207 & 0.5205277194948809 & -1.90403440999077 & 0.08445455254944874 & 335.4917471223412 & 150.4528672771617 & 5.824602686612852 & 1.978044442794066 & 7.8249548070261135\\
  WMH5 & 1.951086320784788 & 8.477433978930948E-10 & 1.4934620169416633E-10 & 0.0058824515854745485 & 0.0011476393238122656 & 2.0839539366492392 & 0.30283763568832534 & 1.023542023350634 & 0.12106730558793474 & -1.7870324336983747 & 0.07045947910361673 & 154.0886301779611 & 22.361434286195788 & 5.987588745994803 & 0.2945709991327516 & 7.905713447780134\\
  HZ3 & 0.8374044285798142 & 5.13137688033645E-10 & 3.538898616833057E-10 & 0.0010305321087809957 & 0.00103 & 0.30501242150922997 & 0.305 & 0.16773103902576944 & 0.1841012268018148 & -2.0301853695813863 & 0.07840066695444284 & 289.82378405191486 & 151.5706467109258 & 6.2512800516475515 & 1.4239330144842364 & 7.646651773246171\\
  CLM1 & 0.9896535929623044 & 3.9380478813513456E-10 & 1.1496484891987051E-10 & 1.1660894748063509E-4 & 7.914695835779139E-5 & 0.028880902100740263 & 0.018396494912763565 & 0.023117480826614963 & 0.014540553636068813 & -2.078746469280929 & 0.06245897343510011 & 224.5481821526005 & 70.54636386400891 & 6.627968506773475 & 0.35811465678664595 & 7.380006782415309\\
  \hline\end{tabular}}
   \caption{Parameters derived from fitting the SEDs of the Hi-z LBG sample. Note that the parameters for which the uncertainty (e.g., param\_err) is $\ge$ parameter (i.e., param) should be considered as upper limits. $^a$estimated from the relation described in the text. However, this method being very indirect (see text), these values should only be considered as indicative.}%
    \label{Table1_Hi-z}
\end{table*}

\begin{table*}[t]
\sisetup{table-column-width=10ex,    
         round-mode=places,
         round-precision=3,
         tight-spacing,
         table-format = 2.3e-2,
         table-number-alignment = center
         }
  \resizebox{\linewidth}{0.8\height}{

\begin{tabular}{lS|S|S|S|S|S|S|S|S|S|S|S||}

\hline
  \multicolumn{1}{|c|}{id} &
  \multicolumn{1}{c|}{$\chi_\nu^2$} &
  \multicolumn{1}{c|}{L$_{dust}$[L$_{\odot}$]} &
  \multicolumn{1}{c|}{L$_{dust}$\_err[L$_{\odot}$]} &
  \multicolumn{1}{c|}{L$_{FUV}$[L$_\odot$]} &
  \multicolumn{1}{c|}{L$_{FUV}$\_err[L$_\odot$]} &
  \multicolumn{1}{c|}{M$_{dust}$[M$_{\odot}$]} &
  \multicolumn{1}{c|}{M$_{dust}$\_err[M$_{\odot}$]} &
  \multicolumn{1}{c|}{M$_{star}$[M$_{\odot}$]} &
  \multicolumn{1}{c|}{M$_{star}$\_err[M$_{\odot}$]} &
  \multicolumn{1}{c|}{SFR[M$_\odot$yr$^{-1}$]} &
  \multicolumn{1}{c|}{SFR\_err[M$_\odot$yr$^{-1}$]} \\
\hline
\hline
  ID27 & 0.44246197945927773 & 6.351829317007038E9 & 8.122766397356078E9 & 3.759102726943607E10 & 1.8795513634718037E9 & 2.3349276312028714E6 & 2.985922756490198E6 & 1.549133240007793E8 & 1.7248810191199914E8 & 9.704597178584024 & 2.2550423829474844\\
  ID31 & 0.1545754935798121 & 4.827284242872368E9 & 6.5446152951358185E9 & 6.132915101988611E10 & 3.066457550994305E9 & 1.7745060202064675E6 & 2.4057956102960915E6 & 3.3419733416171426E8 & 3.2772640810542804E8 & 12.367747914140788 & 2.391003215562761\\
  lbg10 & 0.27152885848462854 & 7.966864024821577E10 & 4.876386833005684E10 & 1.7330863901013653E10 & 8.665431950506828E8 & 2.9286131627915405E7 & 1.792556095034349E7 & 6.409241721852171E8 & 3.424954004156766E8 & 13.590063151043188 & 10.543956026365155\\
  HZ4 & 0.5451264778885693 & 1.9650652918728564E11 & 6.3285514292513214E10 & 2.170516150937736E11 & 1.0852580754688679E10 & 7.223565083568165E7 & 2.3263706973489758E7 & 1.1694616670586159E10 & 2.4684663056047926E9 & 36.93322420771505 & 7.462129855394144\\
  HZ9 & 3.2598827758439892 & 5.2549024398548566E11 & 6.259185149975224E10 & 1.0646818214704771E11 & 5.3234091073523855E9 & 1.931698144539247E8 & 2.3008717057873107E7 & 6.644688760054354E9 & 1.472997680211925E9 & 52.27060466515386 & 6.61723426423018\\
  lbg53 & 1.1339474081519647 & 4.524955929042857E10 & 3.716360858634123E10 & 3.5207855534605385E10 & 1.7603927767302692E9 & 1.6633703617331814E7 & 1.3661314281716911E7 & 1.144498431884614E9 & 4.9672775613937956E8 & 8.346989937923304 & 6.023102930530617\\
  MACSJD1149 & 0.469825840388361 & 2.963348933263524E11 & 9.010253718993333E10 & 9.870043049911108E10 & 4.935021524955555E9 & 1.0893248164975664E8 & 3.3121624216658164E7 & 1.0377552316980312E10 & 2.5954810226897683E9 & 30.810283016775045 & 7.834596101121508\\
  566428 & 0.3092354200123932 & 2.5572945282478802E11 & 1.855944433678143E11 & 1.656862855546533E11 & 1.28314684331402E10 & 9.400595257089572E7 & 6.822437637877508E7 & 1.4038120495622028E10 & 3.351492696526416E9 & 33.09698643073471 & 21.002243060588448\\
  HZ2 & 2.2824328789722004 & 9.419552184225235E9 & 1.189106736767888E10 & 1.603188364210125E11 & 8.015941821050627E9 & 3.4626202265252857E6 & 4.371147330257748E6 & 1.9001931570374615E10 & 3.5868869331322007E9 & 15.59617130865702 & 2.7414914788008558\\
  HZ10 & 3.6516867890006592 & 1.275412445964528E12 & 8.743377616015245E10 & 1.2266869555694429E11 & 1.019264731685942E10 & 4.688406461566944E8 & 3.2140589689671077E7 & 3.0178368486999996E10 & 4.2250531522804804E9 & 69.68153566073447 & 6.826036144129\\
  HZ8 & 1.7345709663718178 & 9.815330845565304E9 & 1.1337117600107538E10 & 1.7467052483995953E11 & 8.73352624199798E9 & 3.60810816174563E6 & 4.1675158165554567E6 & 1.3222637563227901E10 & 2.0545290844094853E9 & 12.00582027653201 & 2.347286952430732\\
  HZ7 & 0.5926646041430074 & 2.8309036876573647E10 & 3.0341354957576046E10 & 1.518822845392539E11 & 7.594114226962697E9 & 1.0406380448364869E7 & 1.1153459030911177E7 & 1.341626421906366E10 & 2.4096821487920866E9 & 10.312381279433847 & 4.663063695657805\\
  HZ1 & 0.48355787354353647 & 1.29844766845842E10 & 1.544020902612193E10 & 1.7388843832745206E11 & 8.694421916372604E9 & 4.77308376444705E6 & 5.675809107482053E6 & 2.42280751295355E10 & 4.156888100142693E9 & 11.800535034335962 & 2.9319525711581127\\
  HZ6 & 2.195613763698042 & 3.475382590671902E11 & 1.7376912953359512E10 & 3.146981539543427E11 & 1.5734907697717134E10 & 1.2775479999492322E8 & 6.3877399997461615E6 & 2.922281681237203E10 & 3.1313053398655353E9 & 29.039061511403688 & 7.74791518498846\\
  lbg48 & 0.21569035945446857 & 3.614096370668909E10 & 5.014601959873017E10 & 3.3464065935634487E10 & 1.673203296781725E9 & 1.3285390800899407E7 & 1.8433638706634358E7 & 5.095834644418989E9 & 1.0210102550797567E9 & 4.1063601117537765 & 5.0016009956433845\\
  WMH5 & 1.951086320784788 & 4.283965123301382E11 & 5.827654611700889E10 & 2.0556908902648346E11 & 1.0506697579173874E10 & 1.5747823246325946E8 & 2.142241407768004E7 & 2.6770850584154087E10 & 3.743799486349329E9 & 22.694811838699128 & 2.4314840320650584\\
  HZ3 & 0.8374044285798142 & 3.309549823102239E10 & 3.986211235878893E10 & 1.3080268756649564E11 & 6.540134378324782E9 & 1.2165880005801527E7 & 1.465328221831163E7 & 1.4335565174737495E10 & 2.214071813693791E9 & 7.3561187704204345 & 4.944359877896662\\
  CLM1 & 0.9896535929623044 & 7.776350917813998E9 & 4.915773139071222E9 & 2.6925581793425623E11 & 2.1089477829670525E10 & 2.858580689395687E6 & 1.80703447122075E6 & 2.45142482730187E10 & 6.058527173057016E9 & 9.653828347448217 & 1.5000917612767681\\
\hline\end{tabular}}
   \caption{Parameters derived from fitting the SEDs of the Hi-z LBG sample. Note that the parameters for which the uncertainty (e.g., param\_err) is $\ge$ parameter (i.e., param) should be considered as upper limits.}%
    \label{Table2_Hi-z}
\end{table*}



By definition, CIGALE checks that the observed SED is consistent with the assumed SFH and we have tried several SFHs before selecting the delayed one. A first result from the SED fitting, both for the Hiz-LBGs and the Low-zZ DGS objects is that both $\tau_{SFH}$ and  the range of ages are quite short (Tabs.~\ref{Table1_Hi-z} and~\ref{Table2_Hi-z} and Tab.~\ref{Table_Low-zZ}) with ages in the range 100 - 500 Myrs for the Hiz-LBGs and 35 - 1300 Myrs for the Low-zZ sample.  From this result, two important points should be noted: i) we know that some young and even very young stellar populations are present in some of these objects. The most extreme one is the galaxy at z = 9.1 (\citealt{Hashimoto2018}) and observational evidences ([OIII]88.3$\mu$m line for instance) suggest the presence of a very young stellar population (a few Myrs). However, this paper deals with dust characteristics and any dust produced/destroyed over so short time-scales do not impact on the dust budget. This is why we assume a delayed (no final bursts) star formation history in the SED fitting. ii) this age range is in agreement with the properties of this sample as they are dwarf, low-metallicity star-forming galaxies (\citealt{Maiolino2008}, \citealt{Madden2013}, \citealt{Cormier2019}) This is the reason why they are often cited as being high-redshift LBG analogues.

Fig.~\ref{Fig_DFRD_Age} is colour-coded by age of the stellar population, assuming a delayed SFH. Both panels of Fig.~\ref{Fig_DFRD_Age} exhibit an age evolutionary sequence with younger objects at higher sSFR. To the right-hand side of both panels, sM$_{dust}$ presents (or reaches) an apparent maximum at sSFR $\sim$ 10$^{-8}$ yr$^{-1}$ with a decline at lower sSFRs. Fig.~\ref{Fig_DFRD_IRX} is colour-coded by IRX. Galaxies have low IRX to the right-hand side, i.e. at large sSFR but also to the left-hand side, i.e., to low sSFRs. In between, objects have larger IRX values. 

Combining the information from Figs.~\ref{Fig_DFRD_Age} and \ref{Fig_DFRD_IRX}, a qualitative scenario seems to emerge. We first focus on the Hi-z LBG sample which is the core objective of this work: Hi-z LBGs with low IRX, low sM$_{dust}$ and low ages are found at high sSFR. We also find Hi-z LBGs with low IRX and even lower sM$_{dust}$ but large ages are found at low sSFR. In between, we see an apparent maximum observed in IRX and sM$_{dust}$ for the present Hi-z LBG sample at sSFR $\sim$10$^{-8}$yr$^{-1}$ that needs to be confirmed by future observational data. Both IRX and sM$_{dust}$ decline and age increases to a locus at sSFR$\sim$10$^{-10}$ - 10$^{-9}$ yr$^{-1}$ where we find the LBGs with upper limits as those discussed in \citet{Capak2015}, \citet{Ferrara2017} and \citealt{Faisst2017}. This seems to define an evolutionary sequence from the right of the DFRD where the first dust grains form to the left where they are removed/destroyed. In Fig.~\ref{Fig_sMdust_Age}, we confirm the trends observed in Figs.~\ref{Fig_DFRD_Age} and \ref{Fig_DFRD_IRX} by plotting sM$_{dust}$ vs. age$_{main}$. The decline of sM$_{dust}$ is interpreted as an efficient dust removal (destruction or outflow) with a time-scale of $\sim$500Myrs. A word of caution, though: if the dust grains are carried away into the wider ISM, they could cool down or warm up from the injection of kinetic energy and therefore could be more difficult to detect given our wavelength coverage (Fig.~\ref{Fig_Template_Tdust}). We can try to quantify the removal trends: the distribution of Bayesian values $t/\tau$ found by the CIGALE SED fitting are in the range $0.5 \le t/\tau \le 7.5$. The galaxies with the maximum sM$_{dust}$ are in the range 0.4 - 2.9 corresponding to an integrated SFR = 12\% and 80\%, respectively. The maximum SFR occurs at $t/\tau = 1.0$ where about 26\% of the stars have formed. This means (and it is not unexpected) that sM$_{dust}$ correlates with SFR and therefore to the dust production by stars and not in the ISM.
 
\begin{figure}
   \centering
   \includegraphics[width=\columnwidth]{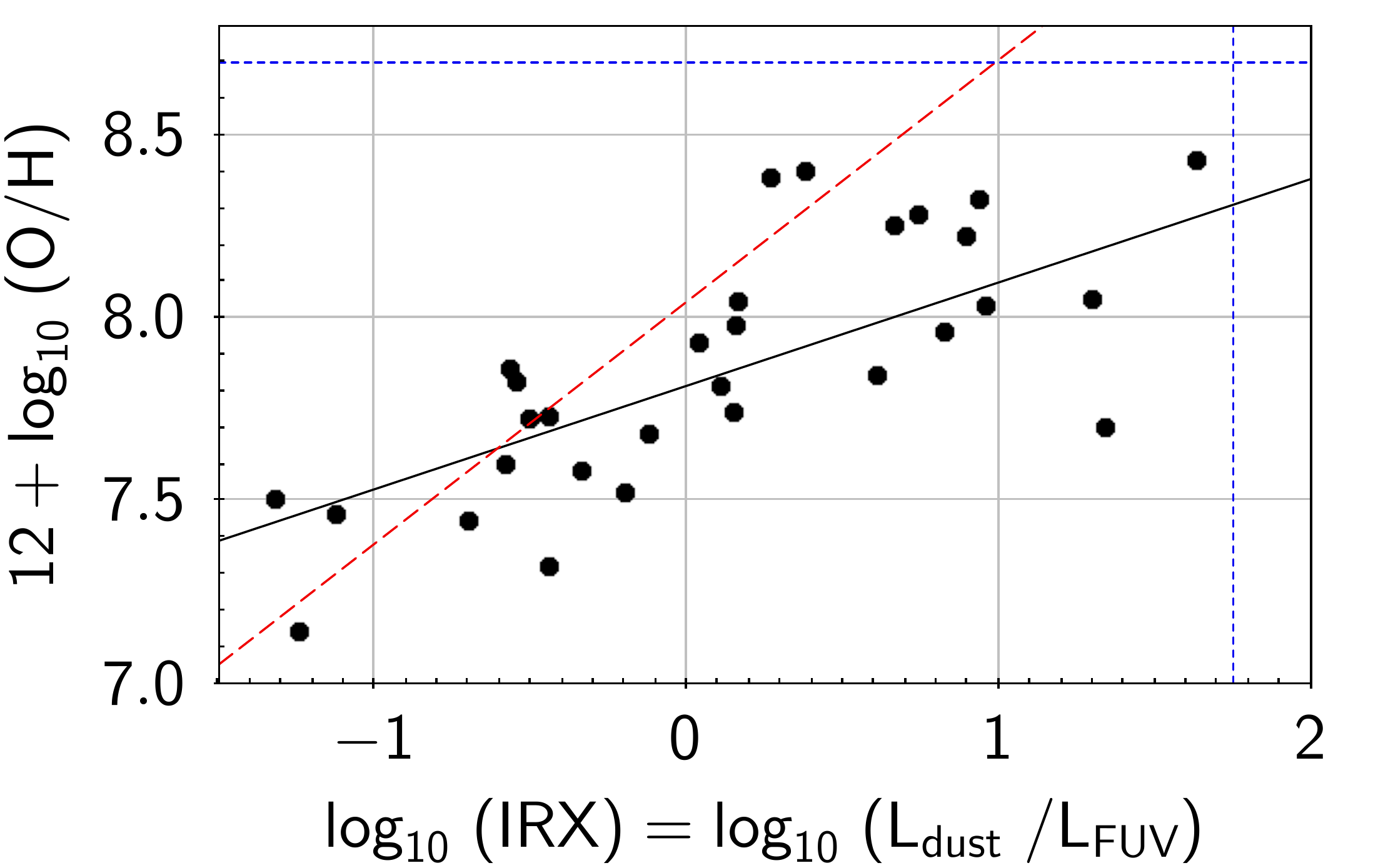}
   \includegraphics[width=0.95\columnwidth]{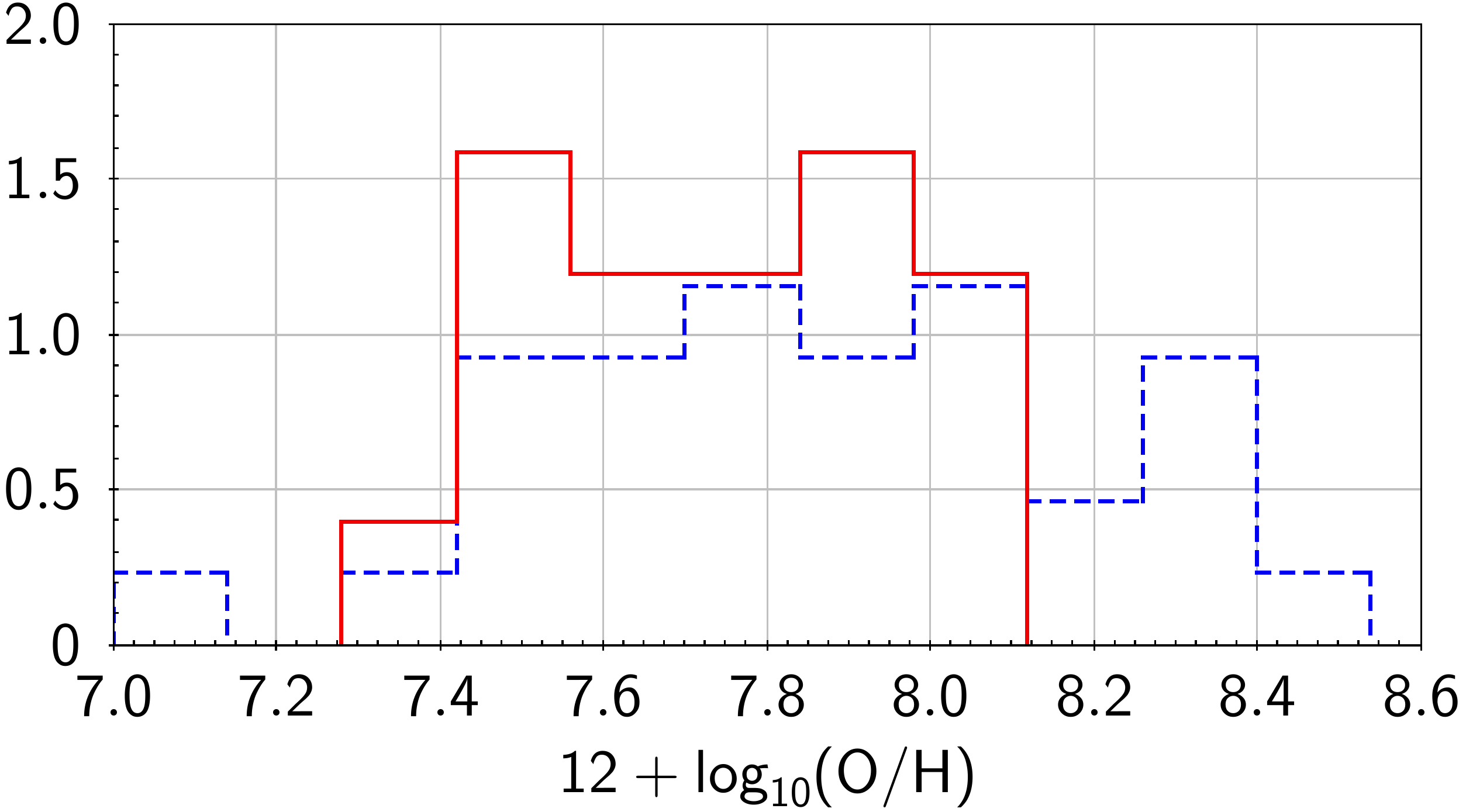}
   \caption{Top: We estimate the relation  (dashed red line) between the metallicity and IRX of our Low-zZ galaxies (black filled dots) 12+$\log_{10}$ (O/H) = 0.2829 $\times$ $\log_{10}$ (IRX) + 7.8155. We find that our fit resembles the starburst galaxy relation found by Heckman et al. (1998). However, our fit is less steep and closer to the average of the two data points corresponding to galaxies at z = 3.3 by Pannella et al. (2015). Bottom: from the derived relation between IRX and 12+$\log_{10}$(O/H) for the Low-zZ sample (dashed red line), we estimate the metallicity of the Hi-z LBG sample (blue continuous line). The distribution (normalized to the area) of the latter is lower than the Low-zZ one by about 0.5. This difference between the two samples could explain the upturn for Low-zZ objects not observed for Hi-z LBGs in Figs.~\ref{Fig_DFRD_Age} and~\ref{Fig_DFRD_IRX} because the dust growth in the ISM could contribute to the dust formation, especially in the highest metallicity of these Low-zZ galaxies.}%
    \label{Fig_12logOH}
    \end{figure}

\begin{figure*}
   \centering
   \includegraphics[width=2\columnwidth]{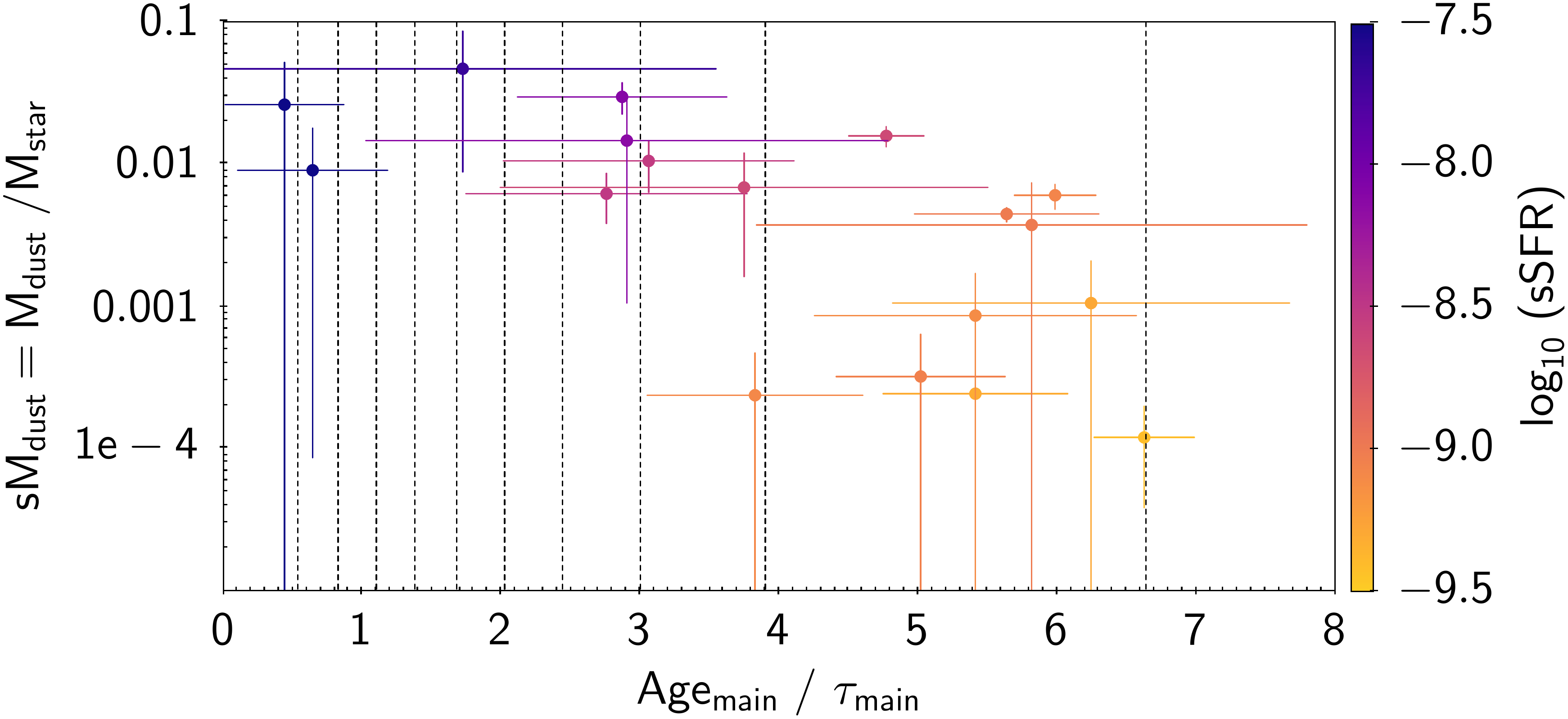}
   \caption{sM$_{dust}$ vs. Age$_{main}$ diagrams colour-coded in sSFR for the Hi-z LBGs. Low-sSFR objects are the oldest objects and low-sM$_{dust}$ objects are the youngest ones. The decline of sM$_{dust}$ is interpreted as an efficient dust removal (destruction or outflow) with a time-scale of $\sim$500Myrs. This diagram allows to obtain an estimate of the dust removal grain in these Hi-z galaxies. Symbols coded as in Fig. 1.  From left to right, the vertical dotted lines shows 10\%, 20\%, 30\%, 40\%, 50\%, 60\%, 70\%, 80\%, 90\% and 99\% of the integrated SFR assuming a delayed SFH.}%
    \label{Fig_sMdust_Age}
    \end{figure*}

If we now focus on the Low-zZ sample, our interpretation  is that separate parallel sM$_{dust}$ vs sSFR sequences could be stratified from bottom to top in addition to a right-to-left one.  Accounting for the age sequence for these low-zZ galaxies from Fig.~\ref{Fig_DFRD_Age}, it could there be possible that these local galaxies undergo an increase of sM$_{dust}$ when getting older. However, this interpretation is not valid for the Hi-z sample where we do not see low-IRX objects at intermediate sSFR and no high-IRX objects on the left-hand side of the plot, at low sSFR, while we still observe the evolutionary sequence from right to left. Under this hypothesis, we can suggest that the low-zZ and Hi-z LBG sample are not perfect analogues: it is possible that we observe for the Low-zZ galaxies the effect of another parameter, the metallicity (Fig.~\ref{Fig_DFRD_IRX}, lower panel) because these galaxies formed at much later cosmic time than the Hi-z LBGs. Indeed, we knew that the metallicity and IRX of Low-zZ galaxies  (black filled dots) are correlated (e.g. Heckman et al. 1998, Boissier et al. 2004). In Fig.~\ref{Fig_12logOH}, the linear fit we estimate for the Low-zZ sample is close to but above the starburst galaxy relation found by Heckman et al. (1998) and passes through the average of the two data points corresponding to galaxies at z = 3.3 by Pannella et al. (2015). Using the updated derived relation between IRX and 12+$\log_{10}$(O/H) for the Low-zZ sample ($\log_{10}$ (IRX) = 1.856 $\times$ (12+$\log_{10}$ (O/H)), we can estimate the metallicity of the Hi-z LBG sample. The normalized distribution of the latter shown in Fig.~\ref{Fig_12logOH} is offset to lower metallicities by about 0.3 as compared to the former. So, the origin of the difference between the two samples in Figs.~\ref{Fig_DFRD_Age} and~\ref{Fig_DFRD_IRX} might be due to the higher metallicity of the Low-zZ galaxies sample where dust growth in the ISM could contribute to the dust formation, especially in the highest metallicity of these Low-zZ galaxies as suggested by several papers for the DGS sample (e.g. \citealt{Asano2013}). \citep{Asano2013} define a metallicity as the critical metallicity Z$_{cr}$, which is the switching metallicity point at which the increasing rate of dust mass by the dust mass growth exceeds the dust production rate by stars. This Z$_{cr} \sim 0.1 - 0.5$ which approximately corresponds to 12+$\log_{10}$ (O/H)) = 8.0 - 8.6, in agreement with the Fig. 5 in \citep{Ma2016}. In other words, the Low-zZ sample might not be as clean and homogeneous sample than the Hi-z LBG one and are not complete analogues to galaxies in the early universe.

%
%


\begin{figure*}
   \includegraphics[width=1.9\columnwidth]{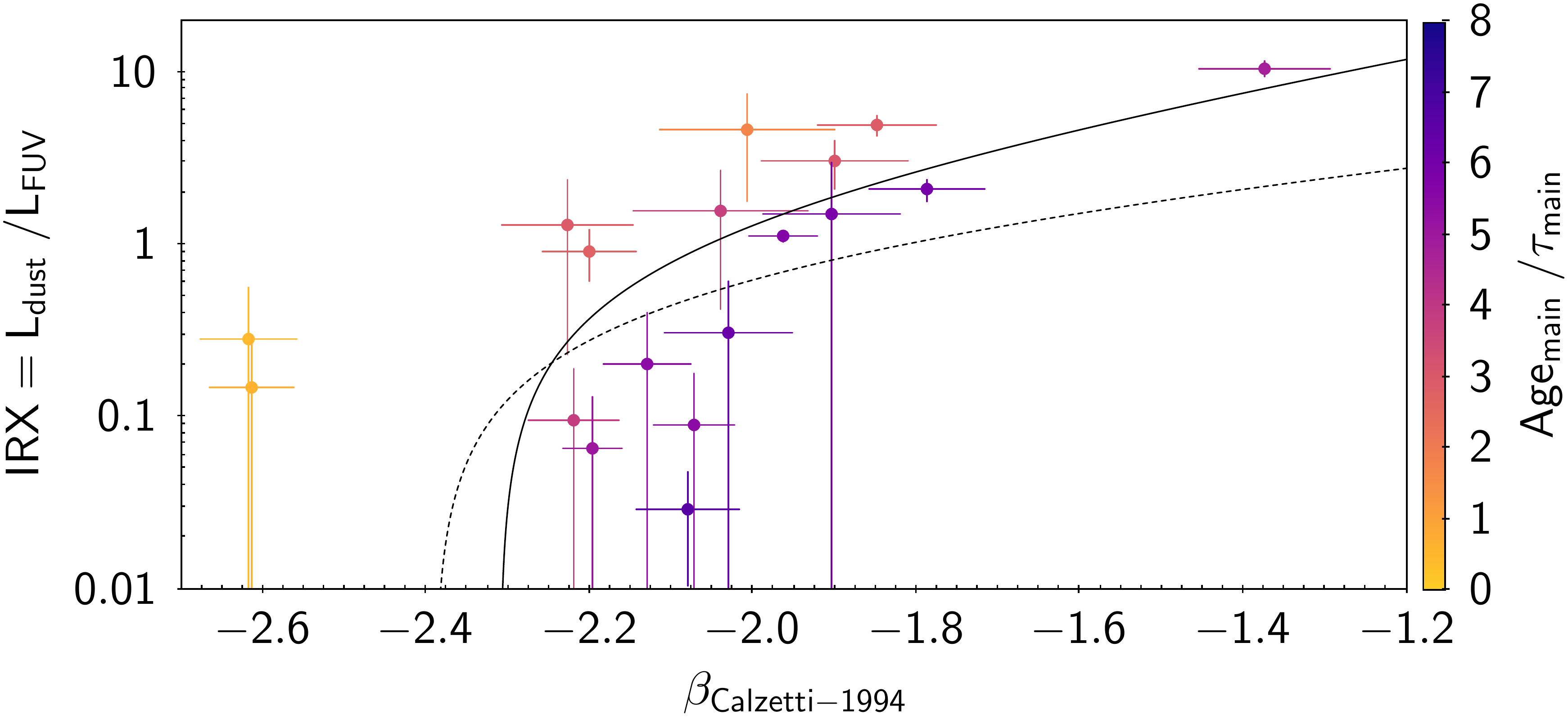}
   \includegraphics[width=2\columnwidth]{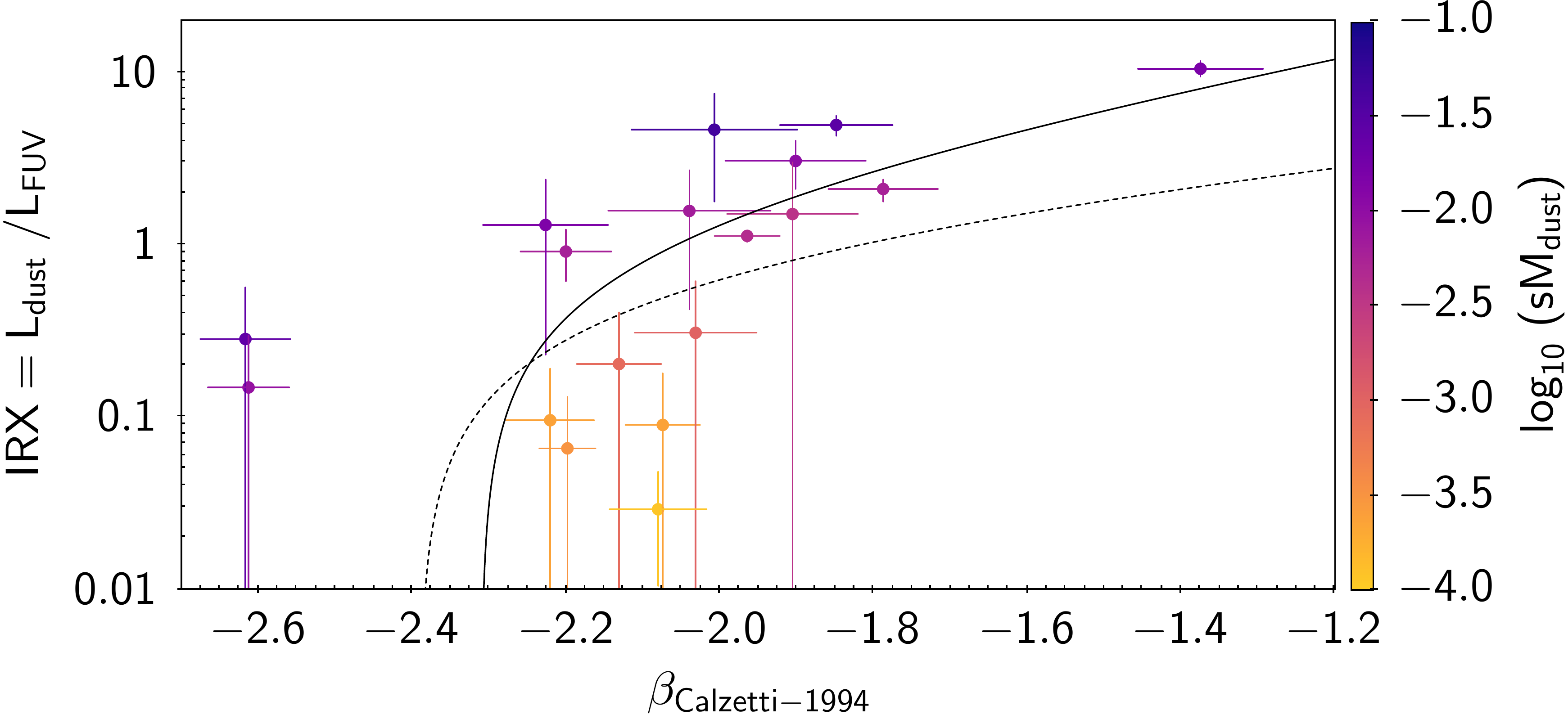}
   \caption{Top: the evolution of Hi-z LBGs in the IRX vs.$\beta$ diagram is an evolutionary sequence with younger galaxies being bluer and older galaxies redder. The continuous line is the Calzetti dust attenuation law and the dashed line is the SMC one. The trend observed here explains the location of galaxies not detected by \citet{Capak2015}. Bottom: The same objects are colour-coded with sM$_{dust}$. Hi-z LBGs with low dust masses are found below Calzetti and SMC dust attenuation laws.}%
    \label{Fig_IRX_beta}
    \end{figure*}

 In Fig.~\ref{Fig_IRX_beta}, we see that Hi-z LBGs with young stellar ages are found to the left and above Meurer's law (the local starburst law linking IRX and $\beta$, \citealt{Meurer1999}) while older and lower sM$_{dust}$ LBGs are found below this law, in the region previously identified from ALMA-undetected LBGs by \citet{Capak2015}. Therefore, the origin of the position of these Hi-z LBGs is consistent with having both low-IRX values and red UV slopes $\beta$. 

\section{Modelling the dust evolutionary sequence}

It is beyond the scope of this paper to describe the models in detail. They will be fully explained in an associated paper (Nanni et al. 2020, in prep.). In brief, we perform the calculations for the metal evolution using the OMEGA code (One-zone Model for the Evolution of GAlaxies, \citealt{Cote2017}). We assume the metal yields for type II SNe from \citet{Kobayashi2006} computed up to 40 M$_{\odot}$, and from the FRUITY database for low-mass stars with M > 1.3 M$_{\odot}$ evolving through the thermally pulsing AGB phase and developing stellar winds (\citealt{Cristallo2011}, \citealt{Piersanti2013}, \citealt{Cristallo2015}). The yields for population III stars are  from \citet{Heger2010} and are limited to the mass range 10 < M/M$_{\odot}$ < 30 in OMEGA. 

Dust removal from the galaxy through galactic outflow follows a rate proportional to the SFR through the "mass-loading factor": ML $\times$ SFR. This assumes that the galactic outflow is generated by the feedback of stars on the gas in the ISM (e.g., \citealt{Murray2005}). The SFH is a delayed one with $\tau$ assumed to be 83Myrs (in agreement with the average value found from the SED fitting). Two kind of IMFs are tested: top-heavy IMFs: m$^{\alpha}$ with $\alpha$ = -1.5, -1.35, -1.0 defined above in this paper and a Chabrier IMF. 

We assume that a fraction of silicates (olivine and pyroxene), iron and carbon grains ejected in the ISM are condensed. In different works (\citealt{Ventura2012},  \citealt{Nanni2013}, \citealt{Nanni2014}) it has been shown that for low-mass stars, the condensation fraction\footnote{the condensation fraction is the number of atoms of the least abundant elements forming a certain dust species divided by the total number of atoms initially available.} ($f_i$) of silicate and carbon dust can be up to 50-60\% or slightly more during the super wind phase, when most of the mass is lost by the stars through stellar winds. 


In a recent work in which dust dust condensation in SNe remnants is modelled \citealt{Marassi2019}), a fair estimate of the mass of dust over the mass of metals is found to be 0.3 $<$ $f_i$ $<$ 0.6. In this calculations, however the effect of the reverse shock on dust destruction is not taken into account. There is not yet a common agreement on the amount of dust produced by SNe, also on the observational viewpoint. \citet{Bocchio2016} argued that a only a small fraction (a few per cent) of the dust produced initially condensed in SNe remnants would survive the reverse shock, while \citet{Gall2014} found evidence of large dust grains ($>\  1\mu m$) in SNe remnants might survive the passage of the reverse shock. \citet{Matsuura2019} have also found that dust might be reformed after the passage of a shock.

In this work, we define the condensation fraction at each time-step as the ratio between the number of atoms locked into dust grains over the maximum that can condense according to its stoichiometric formula and to the chemical composition of the gas. On the basis of theoretical calculations we assume for TP-AGB stars: $f_{Si}$ = 0.6 for silicates, $f_{Fe\_iron}$ = 0.01 for solid iron and $f_{C\_car}$ = 0.5 for carbon dust grains. For SNe we consider two cases: a very high value of the condensation fraction: $f_{Si}$= 1.0 for silicates, $f_{Fe\_iron}$ = 1.0 and $f_{C\_car}$ = 0.5 for carbon to carbon dust grains, and an additional test case with half of the previously assumed condensation fraction and IMF with $\alpha$ = 1. These assumptions correspond to two very favourable scenarios for dust condensation in SNe remnants. 

For each dust species the destruction rate by SNe is computed as: SNe$_{destr}$ = M$_{dust,i}$ / $\tau_{destr}$ where M$_{dust, i}$ is the dust mass of each dust species, $\tau_{destr}$ is the destruction time-scale evaluated as: $\tau_{destr}$  = M$_{gas}$ / ($\tau_{destr}$ R$_{SNe}$ M$_{swept}$) where M$_{gas}$ is the gas mass in the galaxy that changes as a function of the outflow and star formation, $\epsilon_{destr}$ is the efficiency of the destruction. Here $\epsilon_{destr}$ = 0.1 (\citealt{Hirashita2019}) R$_{SNe}$ is the SNe rate and M$_{swept}$ is the mass of gas swept by the SNe blast wave. We explore values for M$_{swept}$ = 1000 and 6800 M$_{\odot}$. The latter case represents the typical value assumed in the literature (e.g. \citealt{Dwek2007}), while the former assumption considers that much lower values for M$_{swept}$ are also possible (\citealt{Nozawa2006}). 

In the dust removal budget, outflows are dominant ($\gtrsim$ 70 – 80\%) over SN destruction ($\lesssim$ 20 – 30\%).

\begin{figure*}
   \centering
   \includegraphics[width=2\columnwidth]{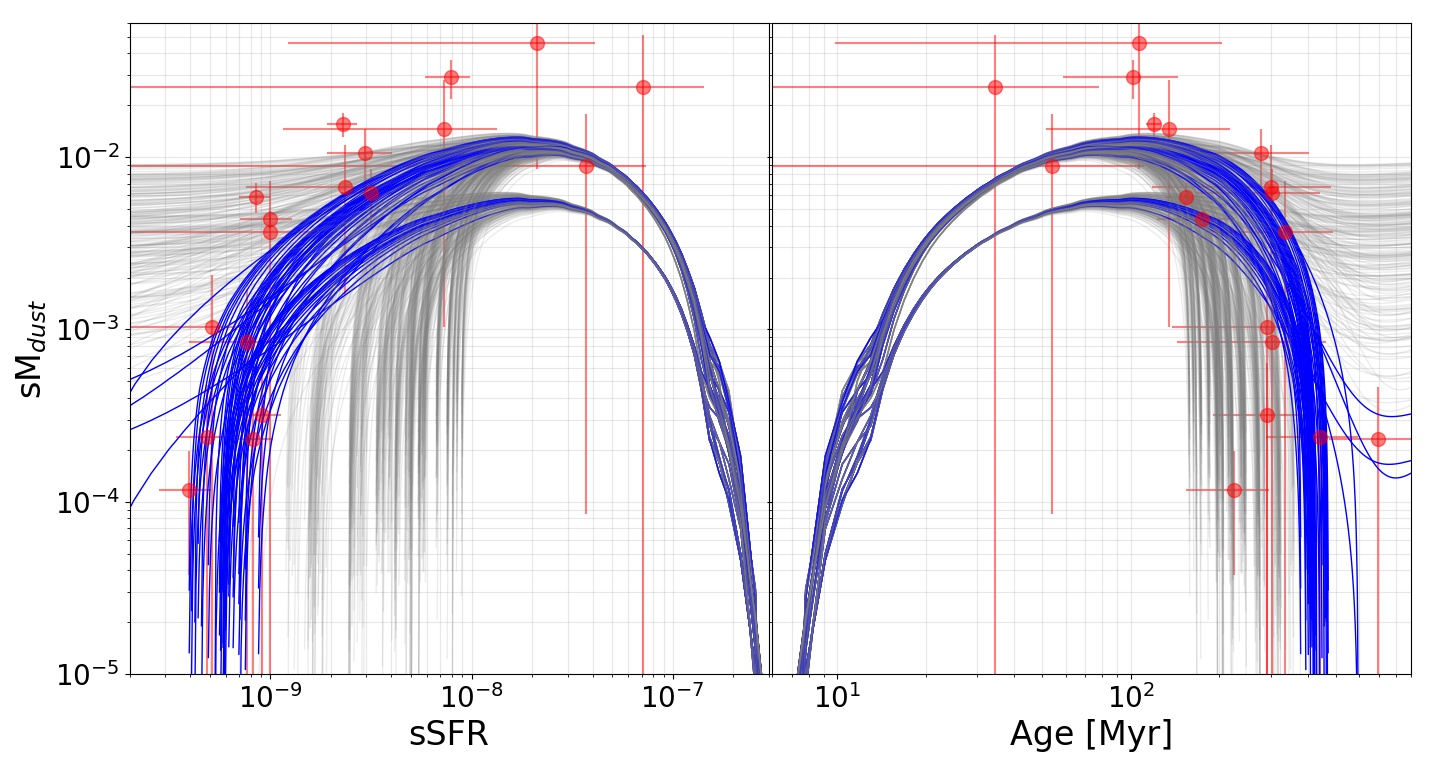}
   \caption{Left: Dust formation rate diagram: sM$_{dust}$ vs. sSFR in the Hi-z LBGs (red data points). Only models with a delayed star formation history are plotted. Consistently, a delayed star formation history only was used in the SED fitting with CIGALE to estimate sM$_{dust}$ and sSFR. The models shown do not have any grain growth in the ISM. We select the best models (shown in blue) by computing the $\chi^2$ of the models with respect to sM$_{dust}$, sSFR and age and finally use the combined total $\chi^2$ as quality of the fit to select the best models. We also show the other models in light grey to show which parameter space the models cover. Right: Specific dust mass vs. age of the dominant stellar population. Models with larger condensation fraction provide better fits, in average. The Hi-z LBGs are young galaxies. The steep decline of sM$_{dust}$ for Hi-z LBGs with respect to Low-zZ galaxies suggest a fast removal (destruction + outflow) of grains in about 0.5 Gyr for Hi-z LBGs that can explain the various detectability of these Hi-z LBGs. Colours have the same meaning that in the left panel.}
     \label{Fig_Models}
   \end{figure*}

The best models selected by computed the $\chi^2$ with respect to sM$_{dust}$, sSFR and age  are presented in Fig.~\ref{Fig_Models}. The blue models show that we can explain the structure of the DFRD diagram. However, even though the models are consistent with the data if we account for the uncertainties, we cannot reach the top data points. This might or might not mean that we are still missing some important physics about the dust cycle in these early galaxies. The fact that the models are low in the DFRD could be relevant given the uncertainties in the theoretical metal yields presently available in the literature and quoted earlier in this paper. Further observational constraints and deeper theoretical studies need to be devoted to the understanding of the dust cycle in the first galaxies.

The average $\tau_{SFH}\ 83Myrs$ found by fitting individual LBGs is used for the SFH in our modelling. The calculations have been performed by normalizing the total stellar mass formed in the galaxy after 13 Gyr to 1 M$_\odot$. We adopted different initial total baryonic mass of the galaxy between M$_{gas}$ = 20 M$_\odot$ and 100 M$_\odot$. The larger the mass of the gas the lower the dust destruction efficiency of SNe. Different values of the mass loading factor (ML) have also been tested, ranging from 0.5 to 0.98 $\times$ M$_{gas}$. 

For the best models shown in blue in Fig.~\ref{Fig_Models}, we have computed the mean and standard deviation values: M$_{gas}$ = 57.9 $\pm$ 26.7, $\alpha$ = -1.14 $\pm$ 0.18, ML = 38.3 $\pm$ 18.2 and M$_{swept}$ = 3951 $\pm$ 2900.


\section{Conclusions}

The major result from this work is a consistent model that explains the (sub-)mm detections (or non-detections) of Hi-z LBGs in the EoR. The non-detections are explained by dust destruction by shocks produced by SNe plus removal in the circum- and intergalactic media by outflows (Fig.~\ref{Fig_Models}). Such large-scale outflows of interstellar material are a generic feature of LBGs (\citealt{Shapley2003}, \citealt{Pettini2002}). The models here presented do not include grain growth in the ISM of these galaxies since the efficiency of such a process can be highly inhibited at the very low temperature characterizing the ISM (\citealt{Ferrara2016}, \citealt{Ceccarelli2018}). Our modelling suggests that a Chabrier IMF cannot produce enough dust mass to reach the upper part of the sequence close. For these LBGs, a top-heavy IMF is preferred. 

The possible rise, suggested from data, and decline of sM$_{dust}$ with sSFR would be the formation and removal (dust destruction and outflow) of the first dust grains in the universe. If a coeval and massive burst of star formation occurred in the early universe (pop.III), we could predict an {\it earlier bump} at higher redshift in the evolution of the cosmic average dust attenuation \citep{Burgarella2013}. However, this bump is likely to be blurred by a non-coeval galaxy formation.

There are still large uncertainties in our modelling because of a poor knowledge of the physical conditions in the early universe. It is crucial to further constrain the models by securing new observational data and invest in works on dust modelling to fully explain the dust cycle in Hi-z LBGs  (and in other objects).

\begin{acknowledgements}
      Part of this work was supported by the Centre National d'Etudes Spatiale (CNES) through a post-doctoral fellowship for AN.
\end{acknowledgements}

%
%

\bibliography{FirstDust_arXiv2}{}
\bibliographystyle{aa} 
 
\begin{appendices}

\section{Individual fits}

\begin{figure*}
   \centering
   \includegraphics[width=1.0\columnwidth]{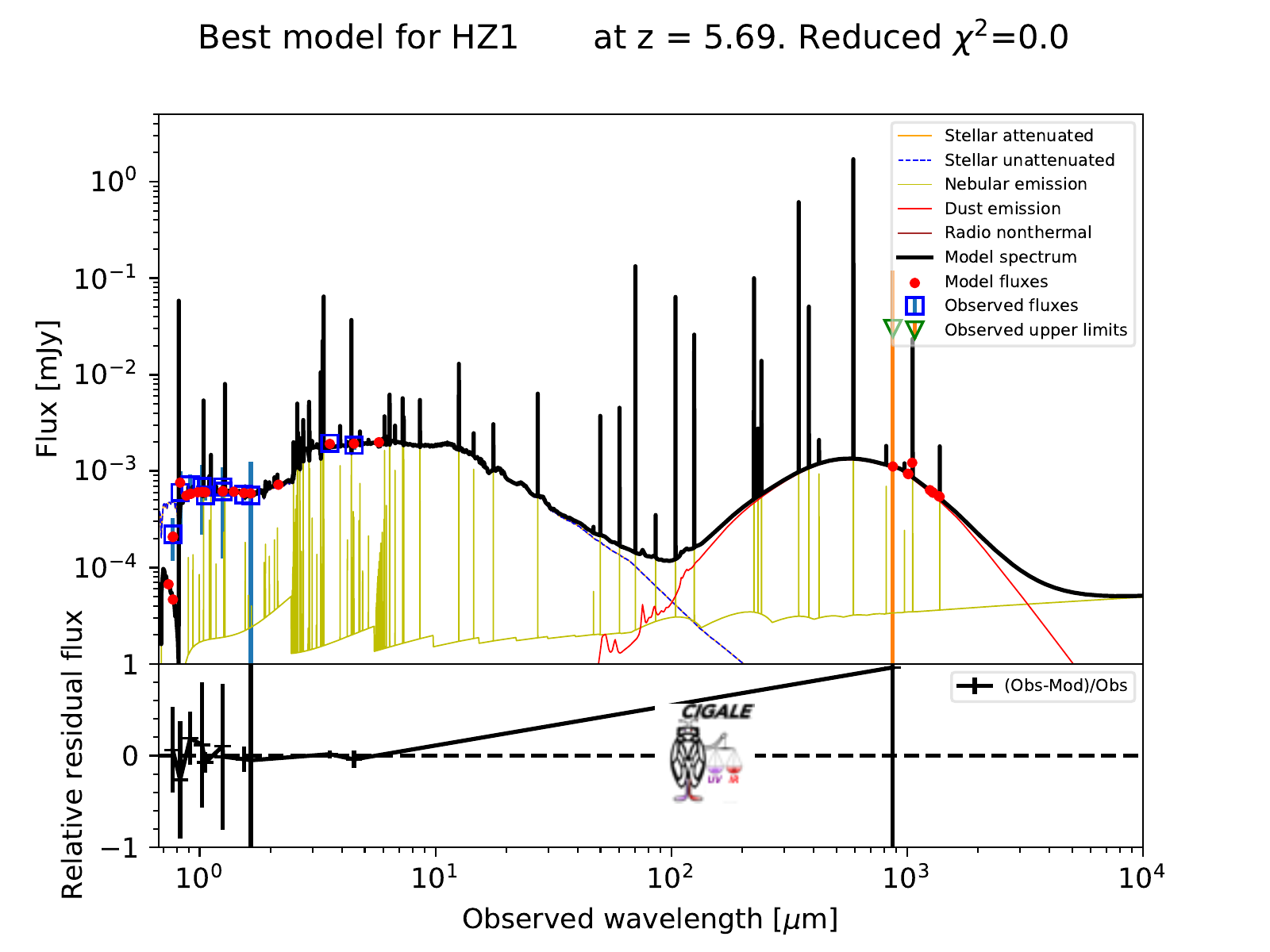}
   \includegraphics[width=1.0\columnwidth]{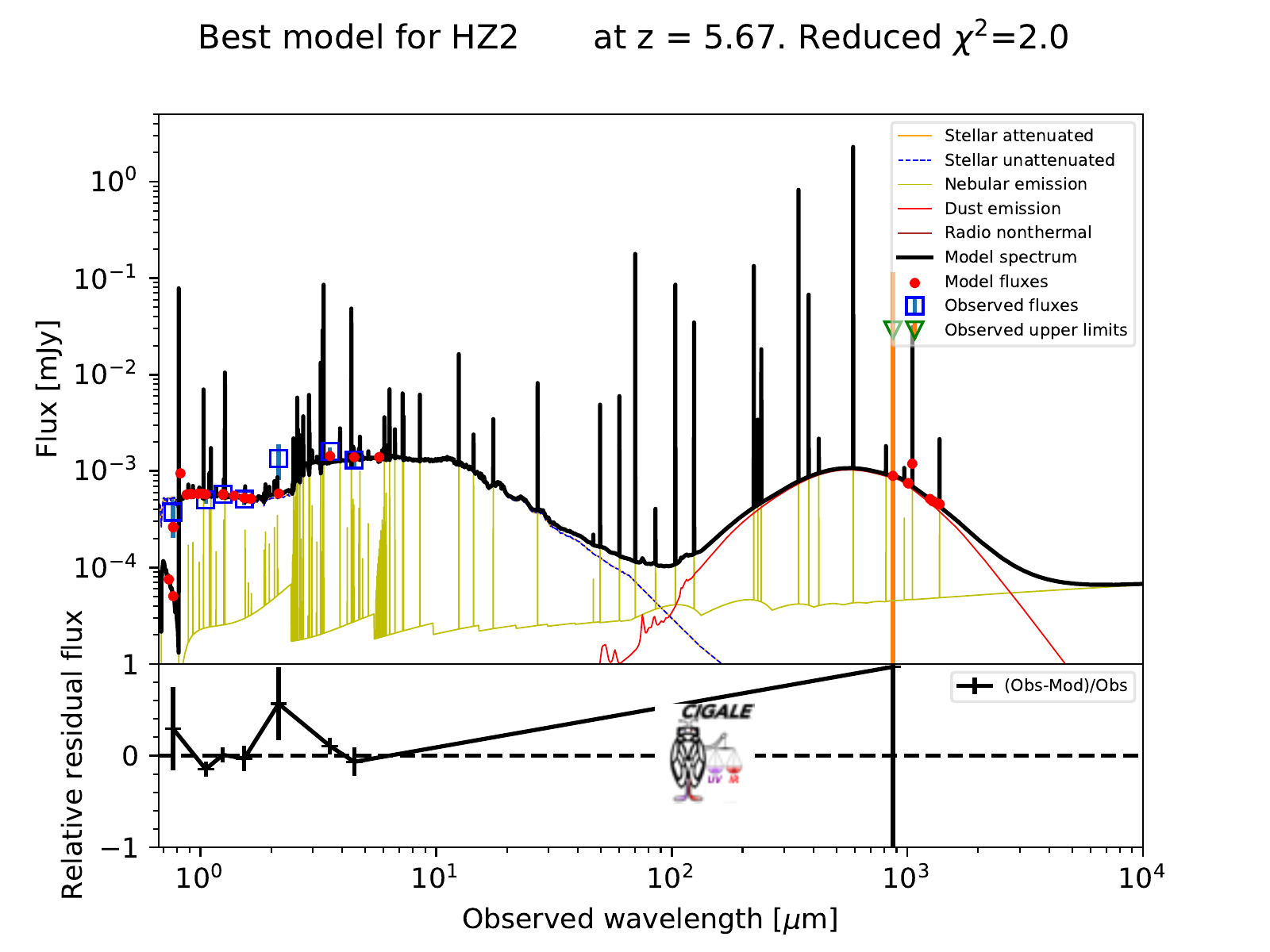}
   \includegraphics[width=1.0\columnwidth]{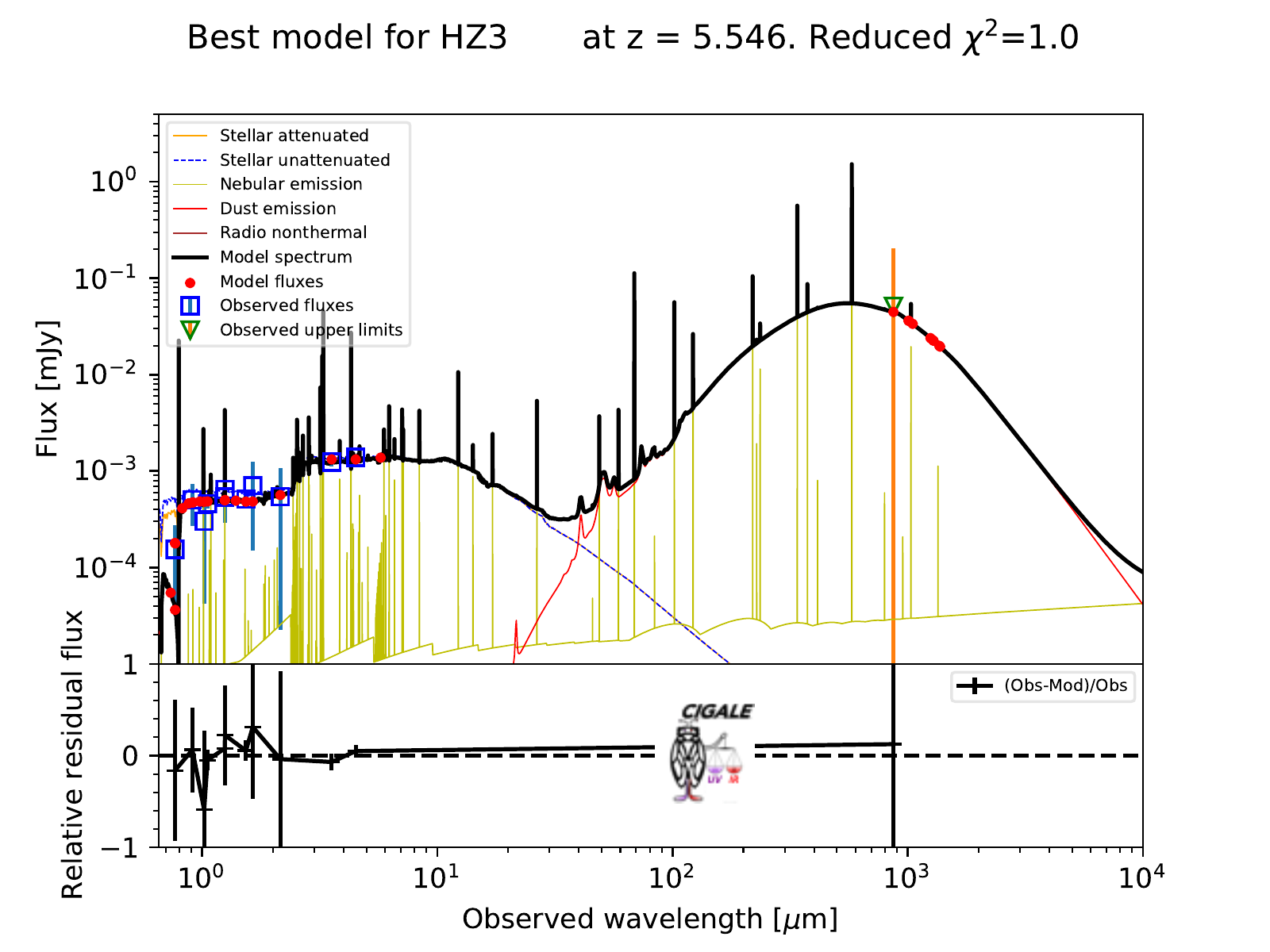}
   \includegraphics[width=1.0\columnwidth]{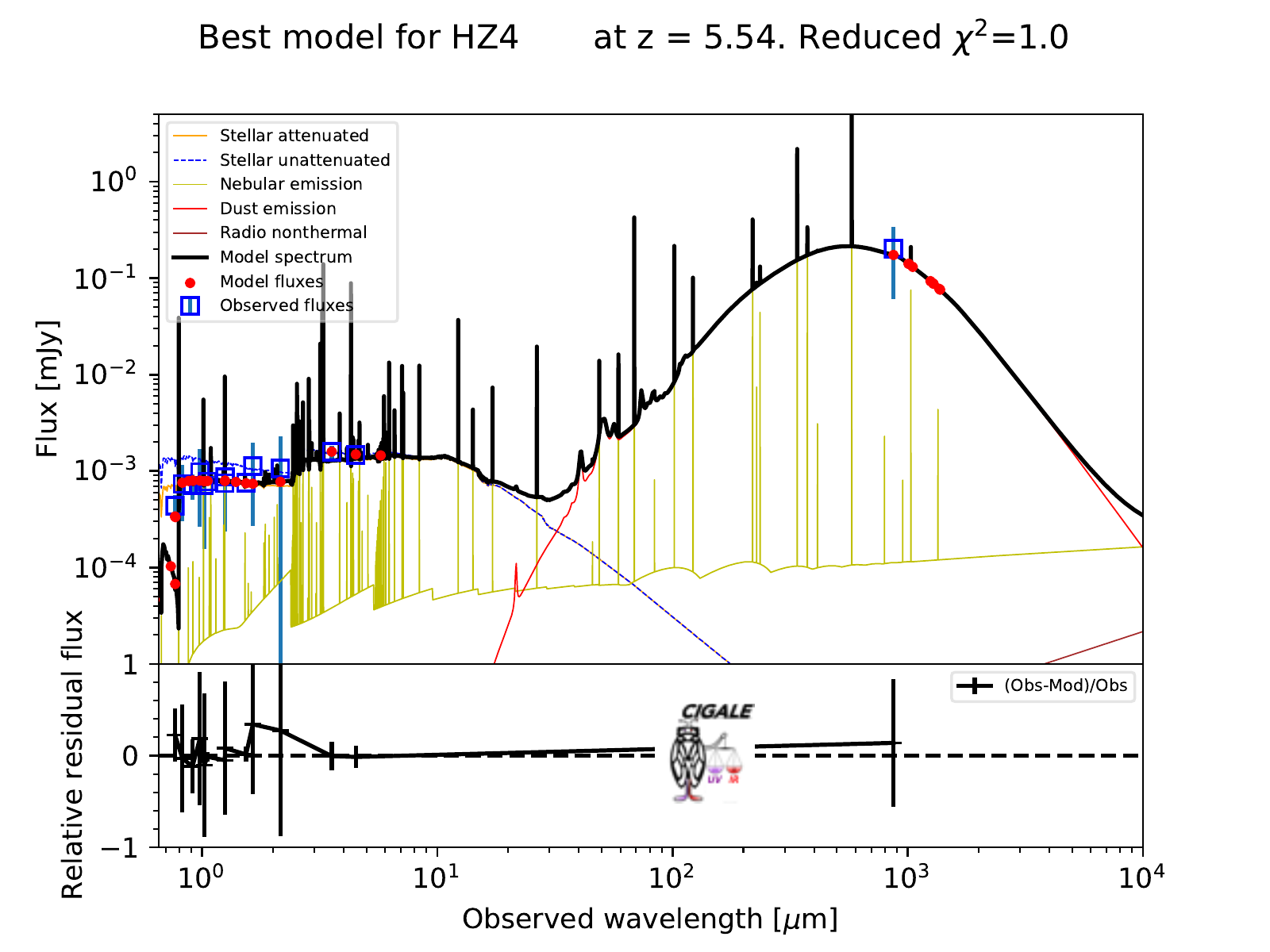}
   \includegraphics[width=1.0\columnwidth]{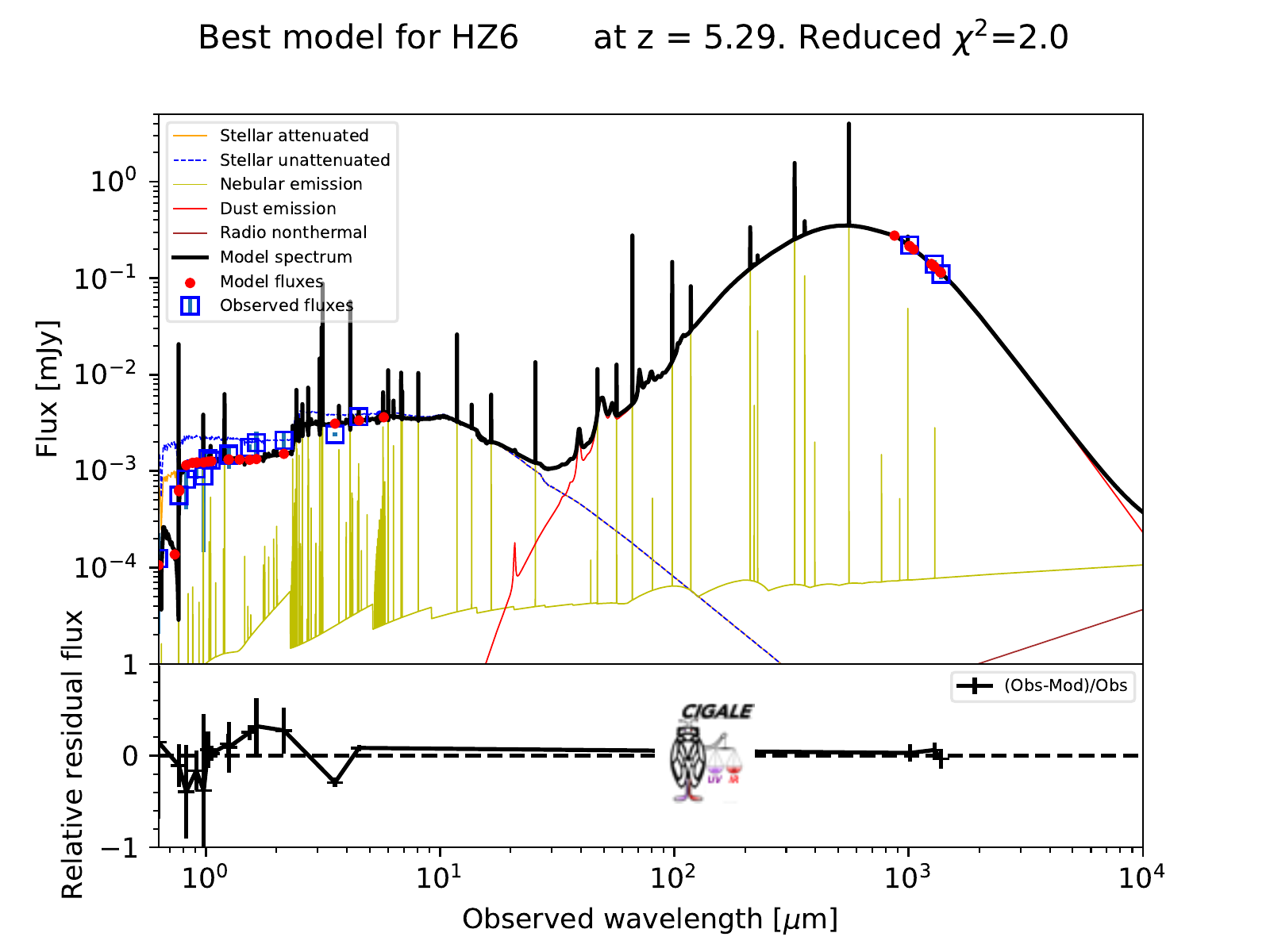}
   \includegraphics[width=1.0\columnwidth]{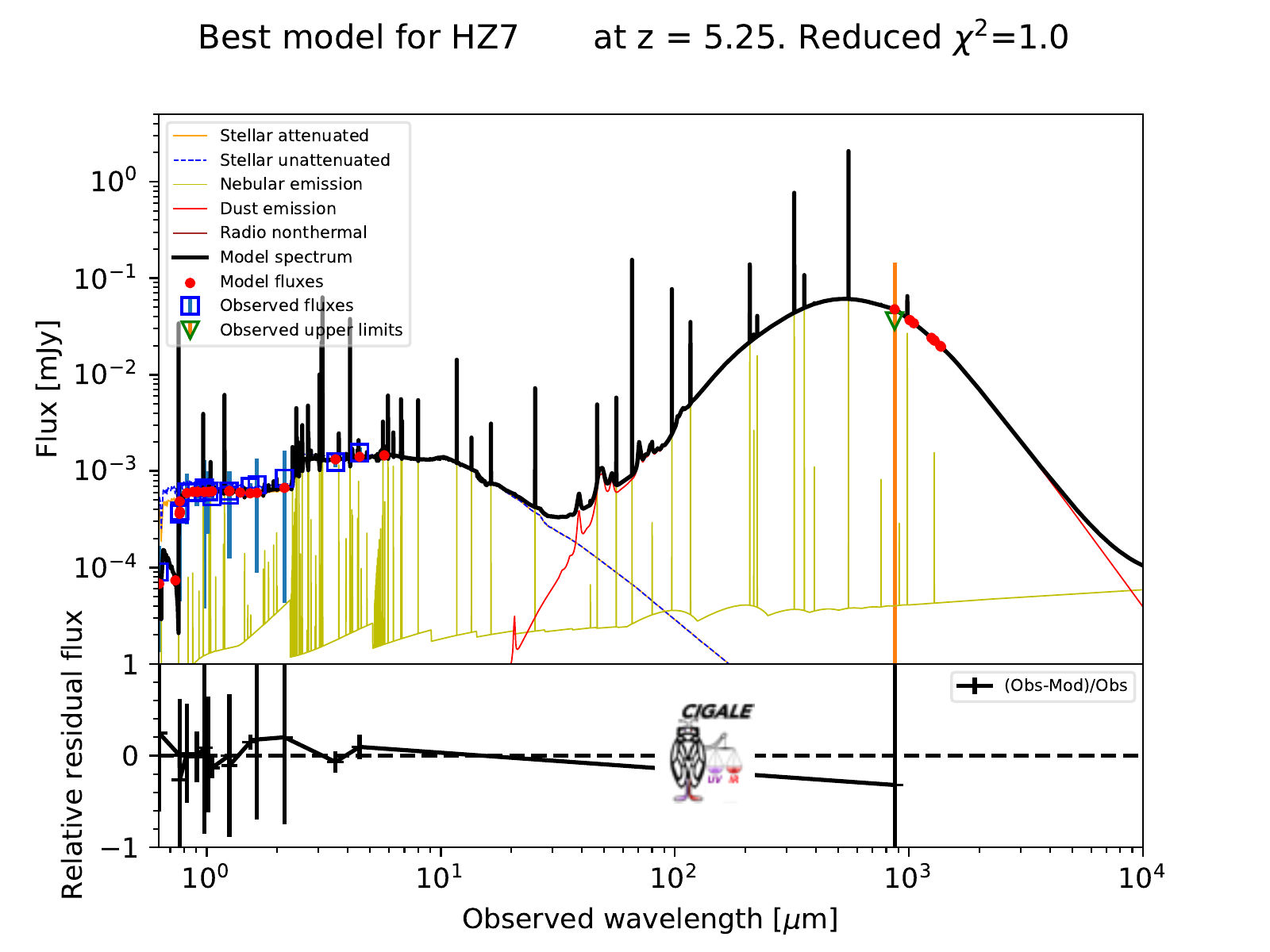}
    \caption{Final fits for all the Hi-z SEDs. Only fits with reduced $\chi^2 \le 5.0$ are kept.}%
   \label{Fig1_individual_fits}
  \end{figure*}

\begin{figure*}
   \centering
   \includegraphics[width=1.0\columnwidth]{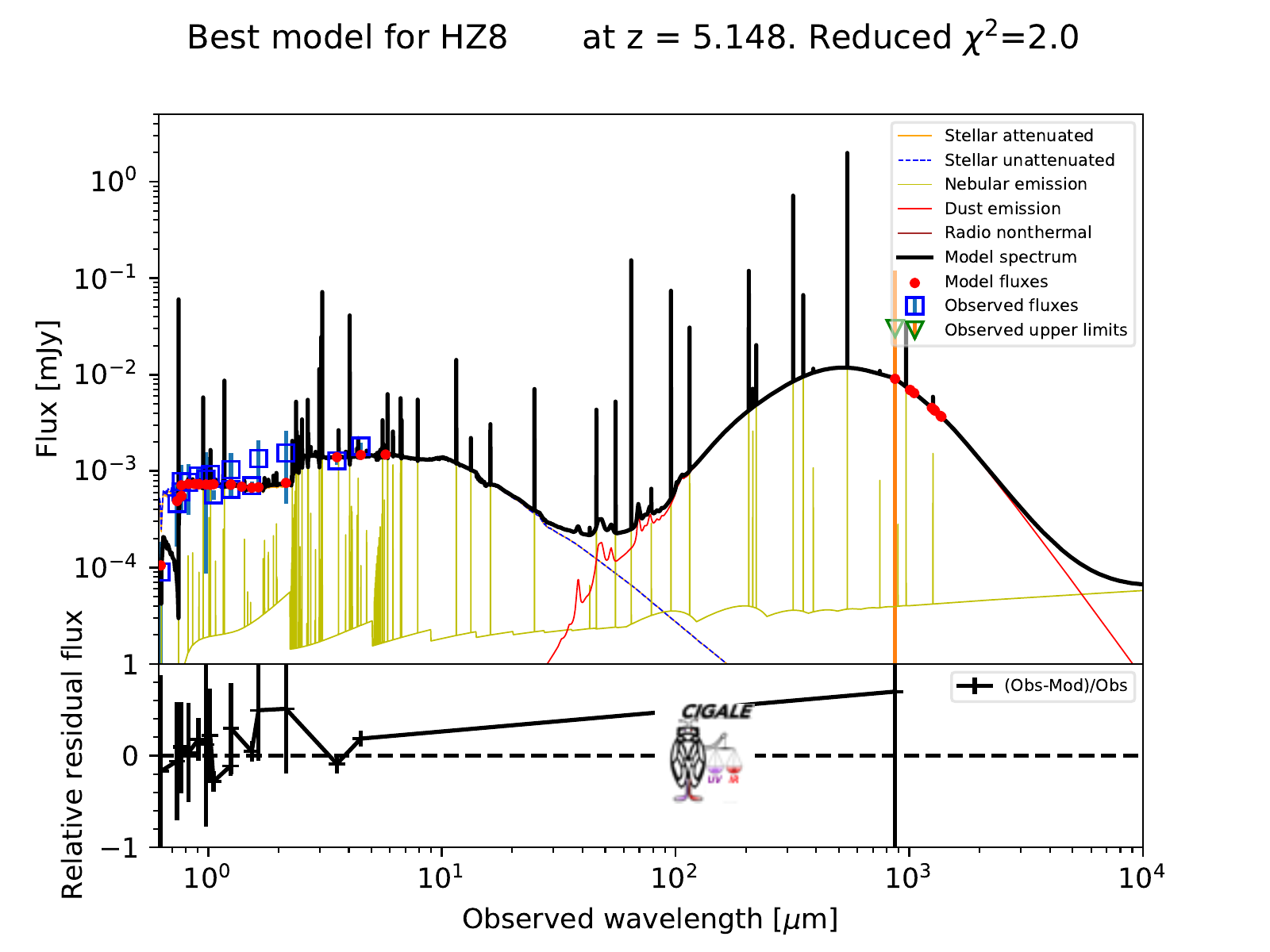}
   \includegraphics[width=1.0\columnwidth]{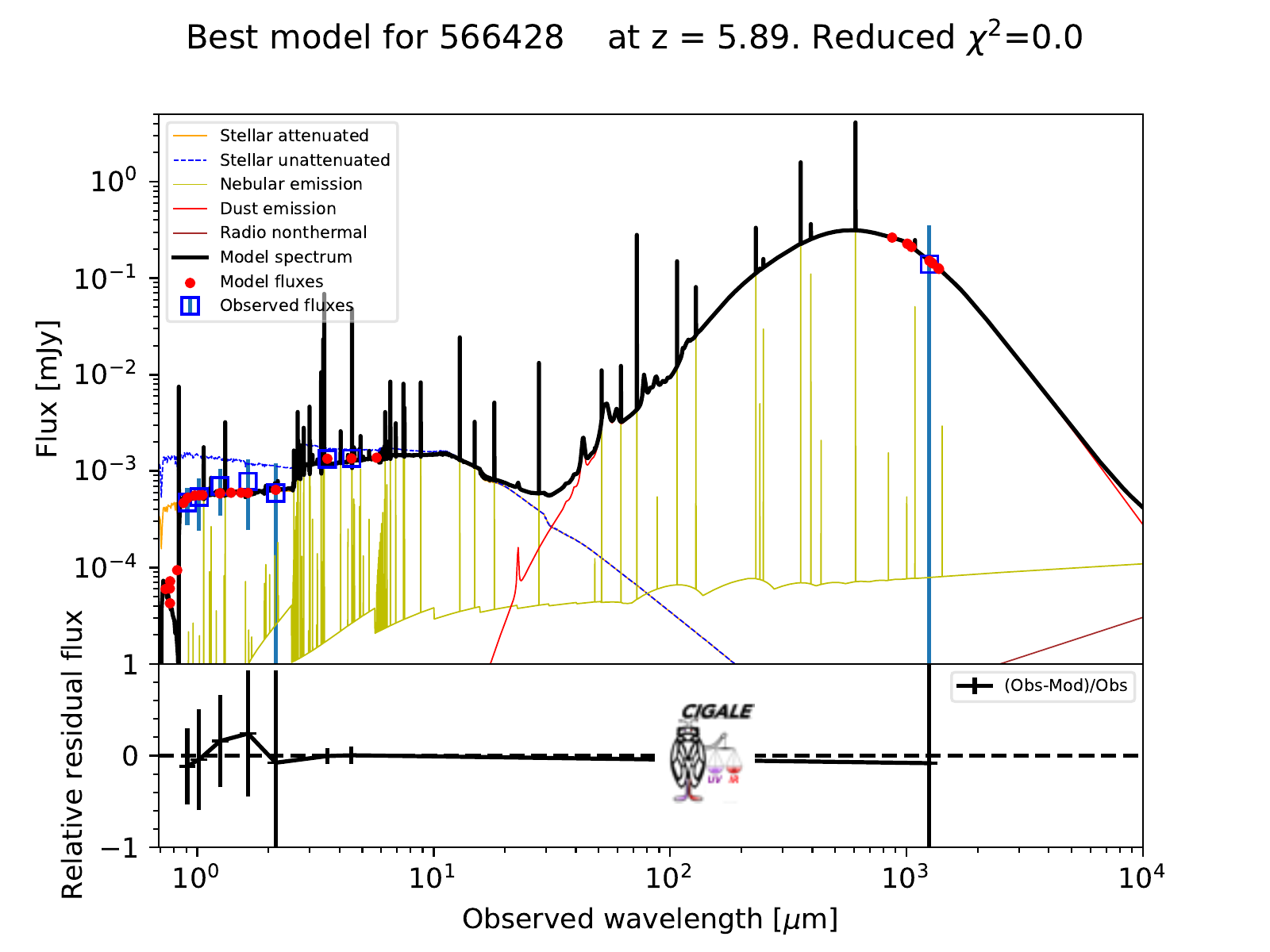}
   \includegraphics[width=1.0\columnwidth]{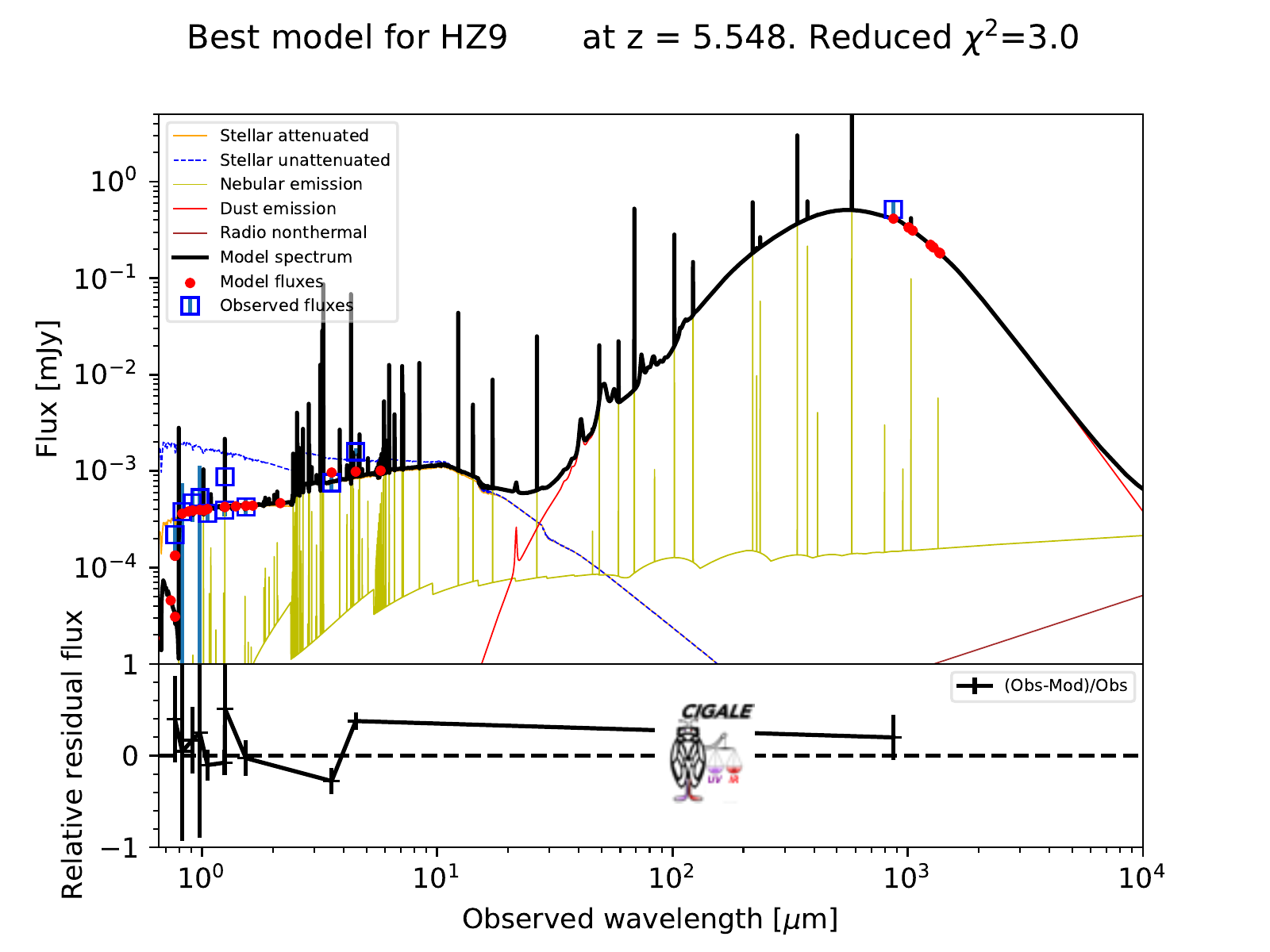}
   \includegraphics[width=1.0\columnwidth]{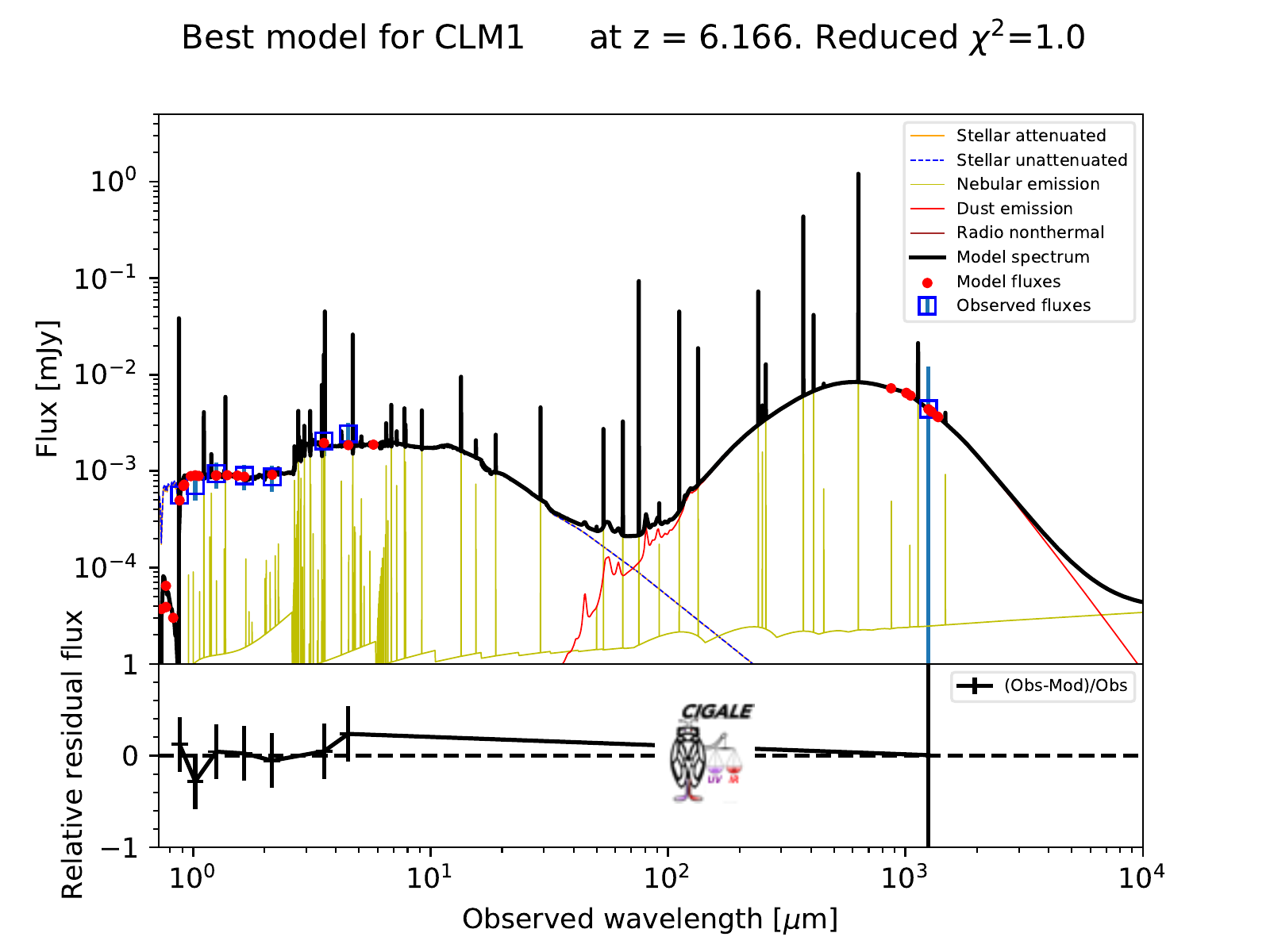}
   \includegraphics[width=1.0\columnwidth]{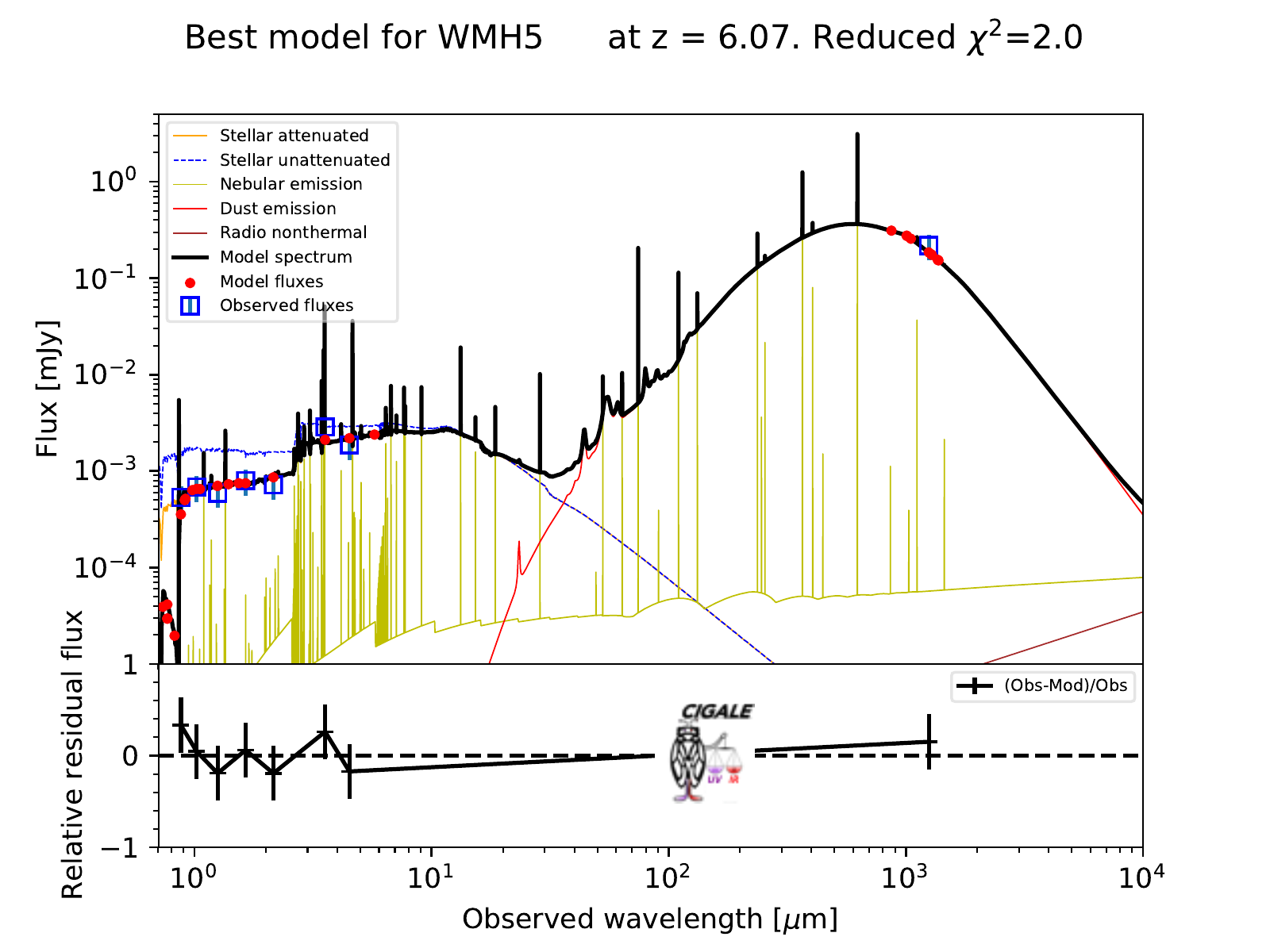}
   \includegraphics[width=1.0\columnwidth]{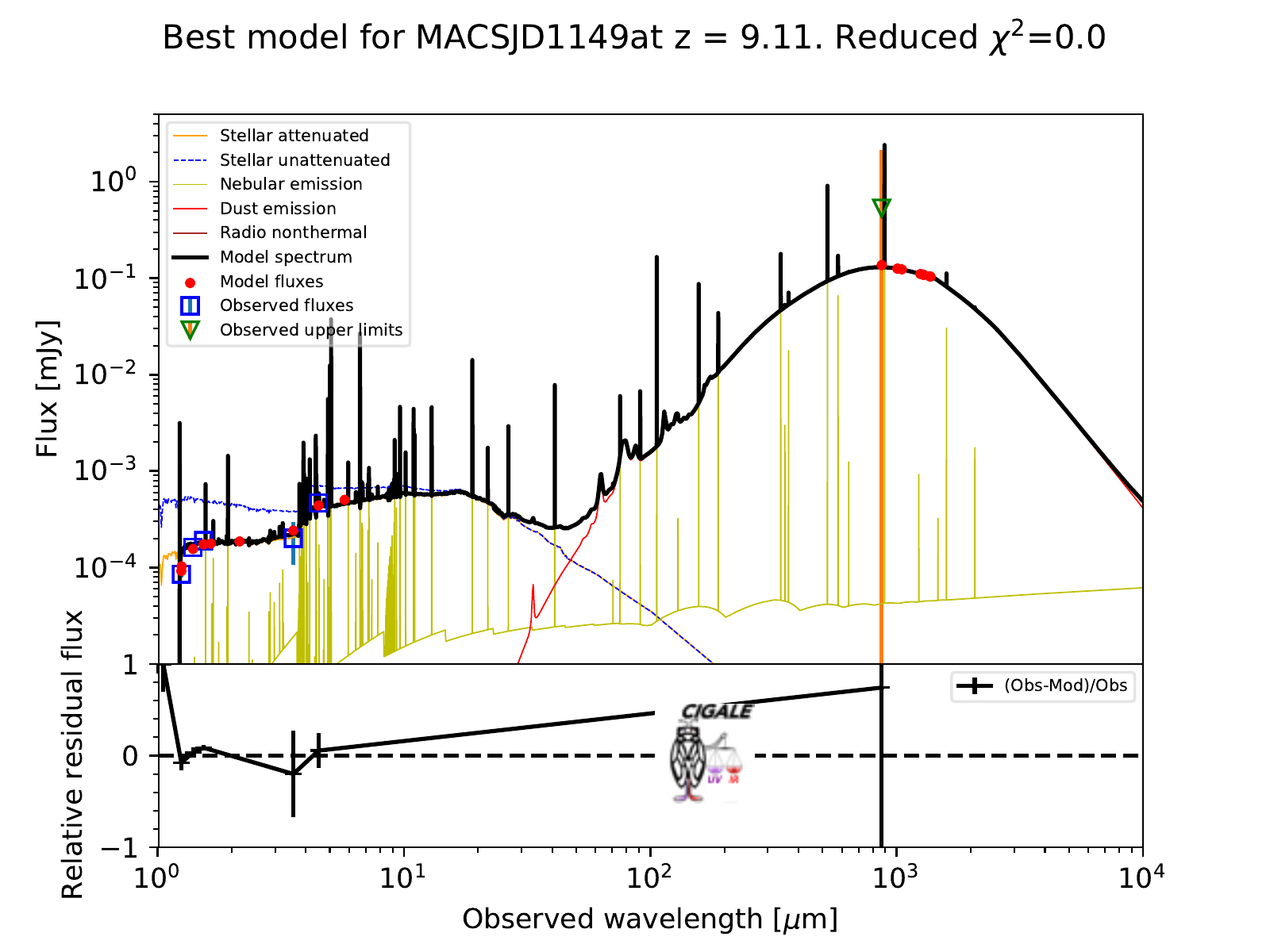}
    \caption{Final fits for all the Hi-z SEDs (continued). Only fits with reduced $\chi^2 \le 5.0$ are kept.}%
   \label{Fig2_individual_fits}
  \end{figure*}

\begin{figure*}
   \centering
   \includegraphics[width=1.0\columnwidth]{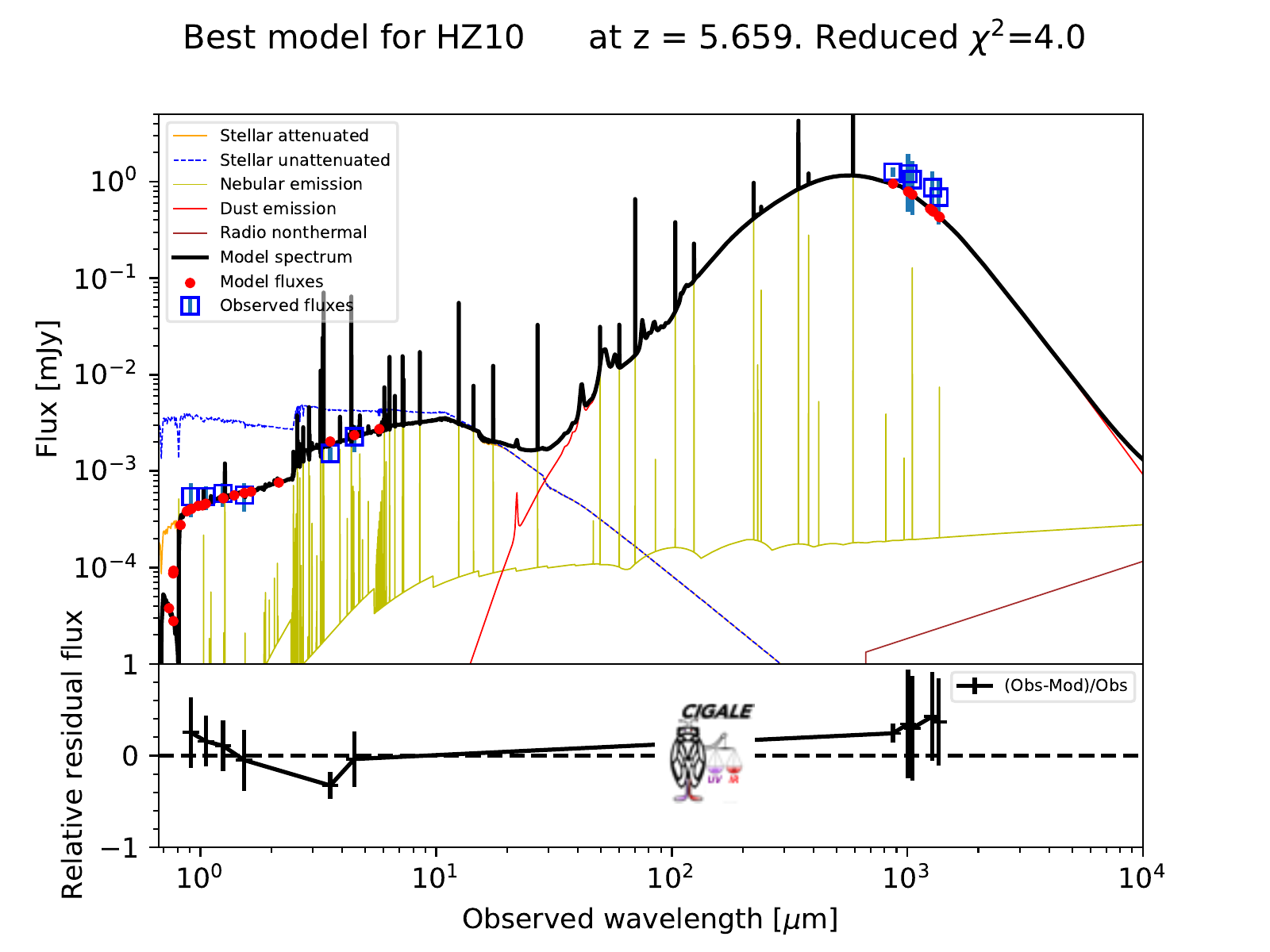}
   \includegraphics[width=1.0\columnwidth]{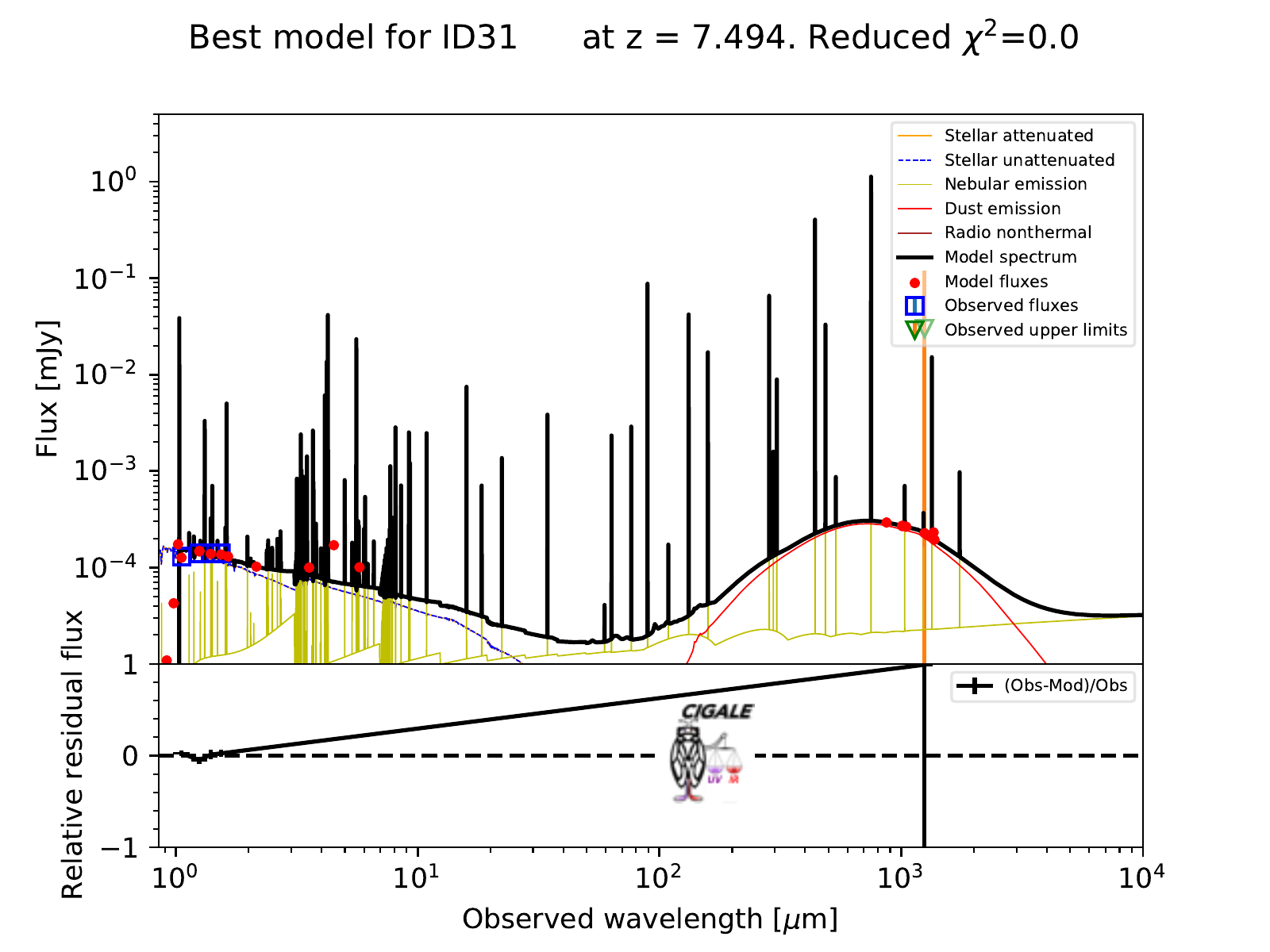}
   \includegraphics[width=1.0\columnwidth]{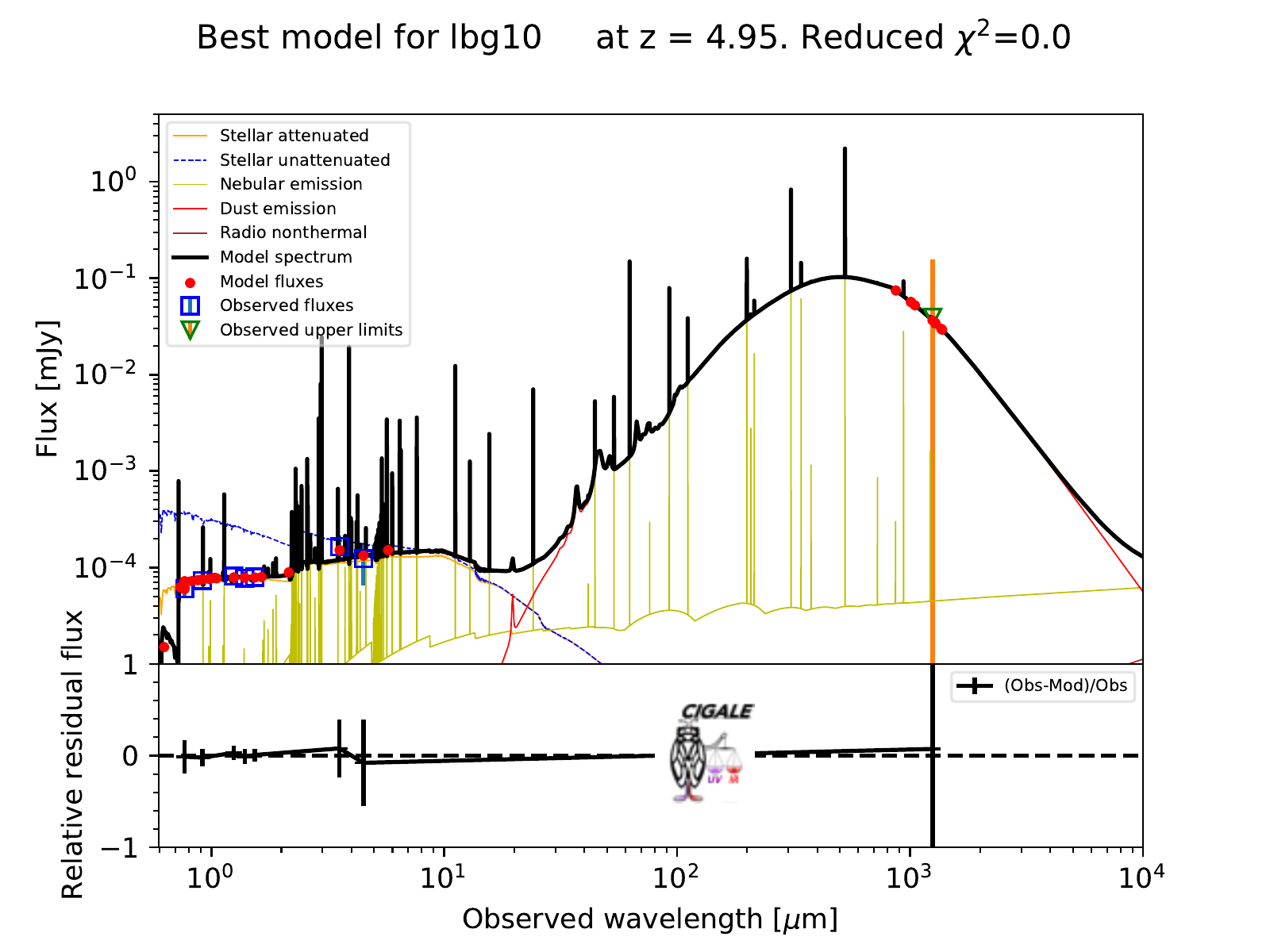}
   \includegraphics[width=1.0\columnwidth]{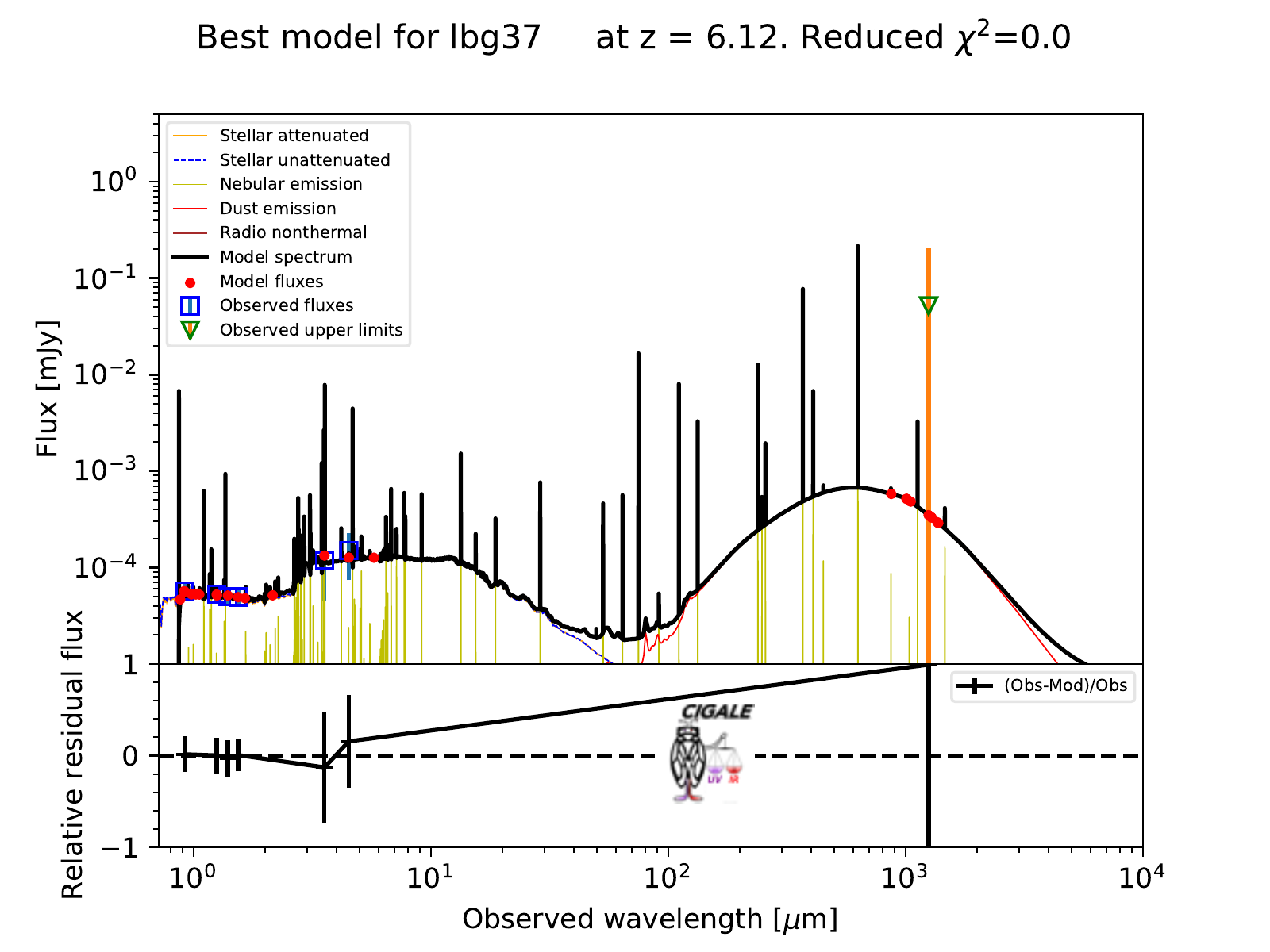}
   \includegraphics[width=1.0\columnwidth]{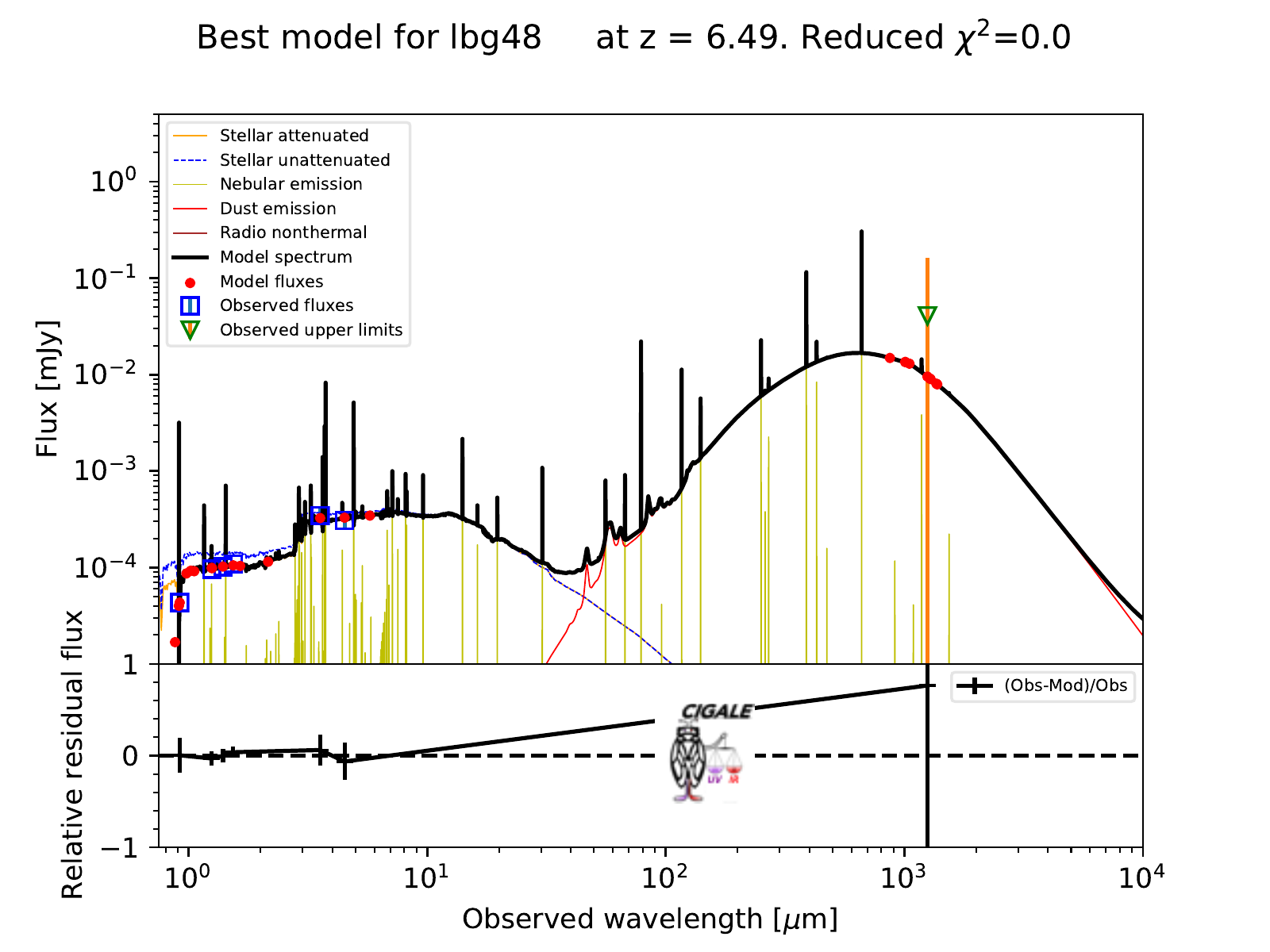}
   \includegraphics[width=1.0\columnwidth]{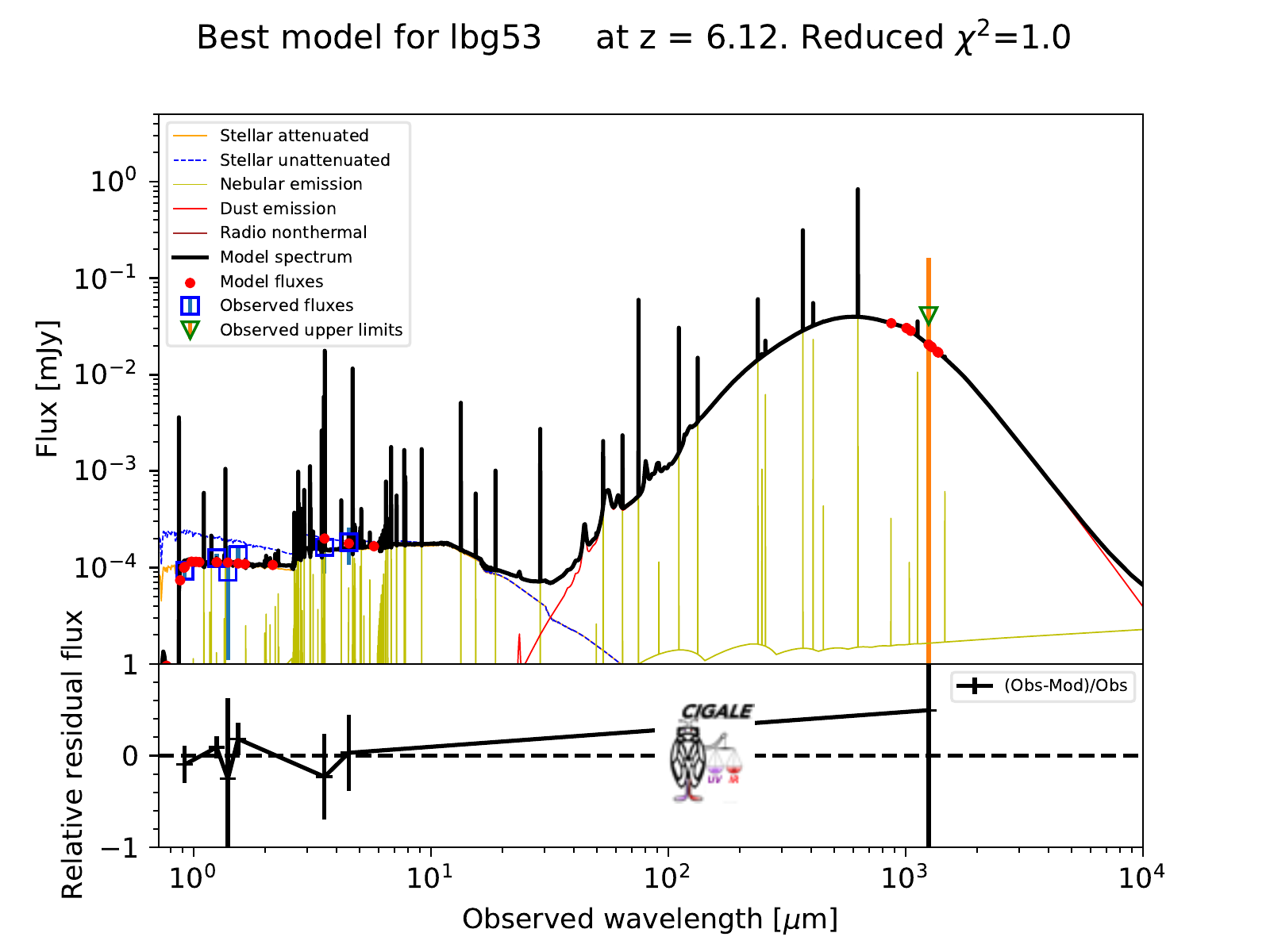}
    \caption{Final fits for all the Hi-z SEDs (continued). Only fits with reduced $\chi^2 \le 5.0$ are kept.}%
   \label{Fig3_individual_fits}
  \end{figure*}

Individual fits are shown in Fig.s~\ref{Fig1_individual_fits}, \ref{Fig2_individual_fits} and \ref{Fig3_individual_fits}. 

\section{Quality of the fits}

  \begin{figure*}
   \centering
   \includegraphics[width=1.\columnwidth]{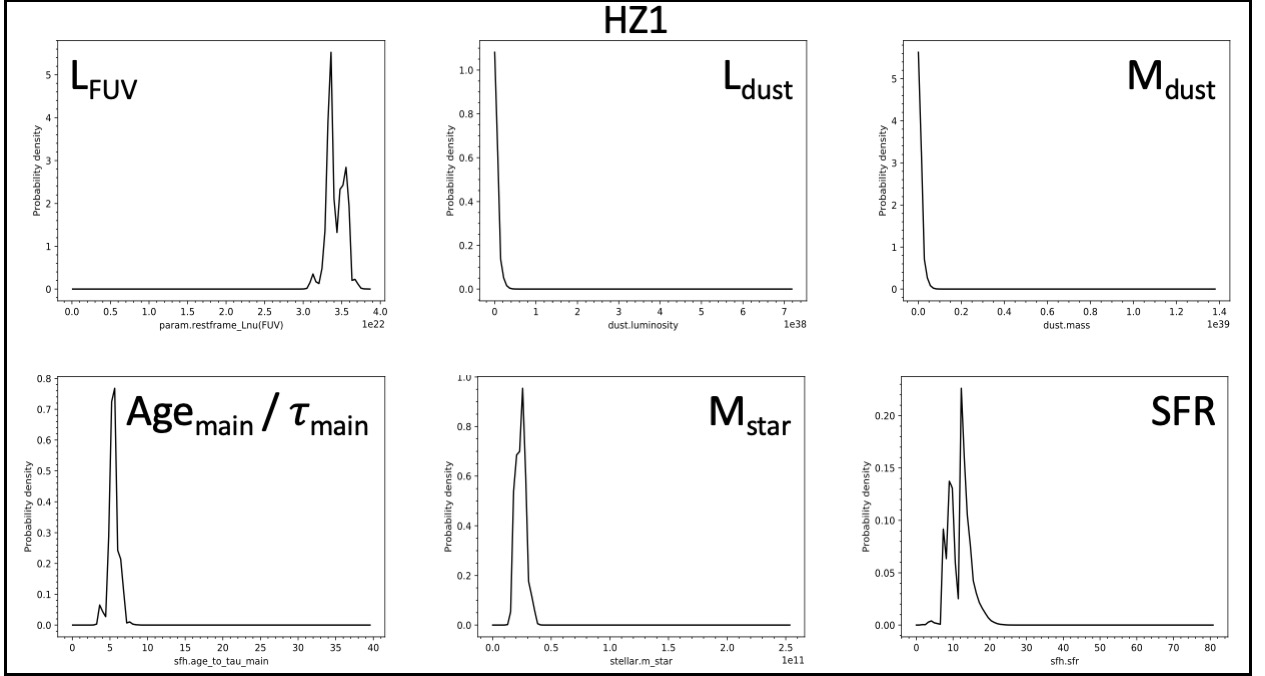}
   \includegraphics[width=1.\columnwidth]{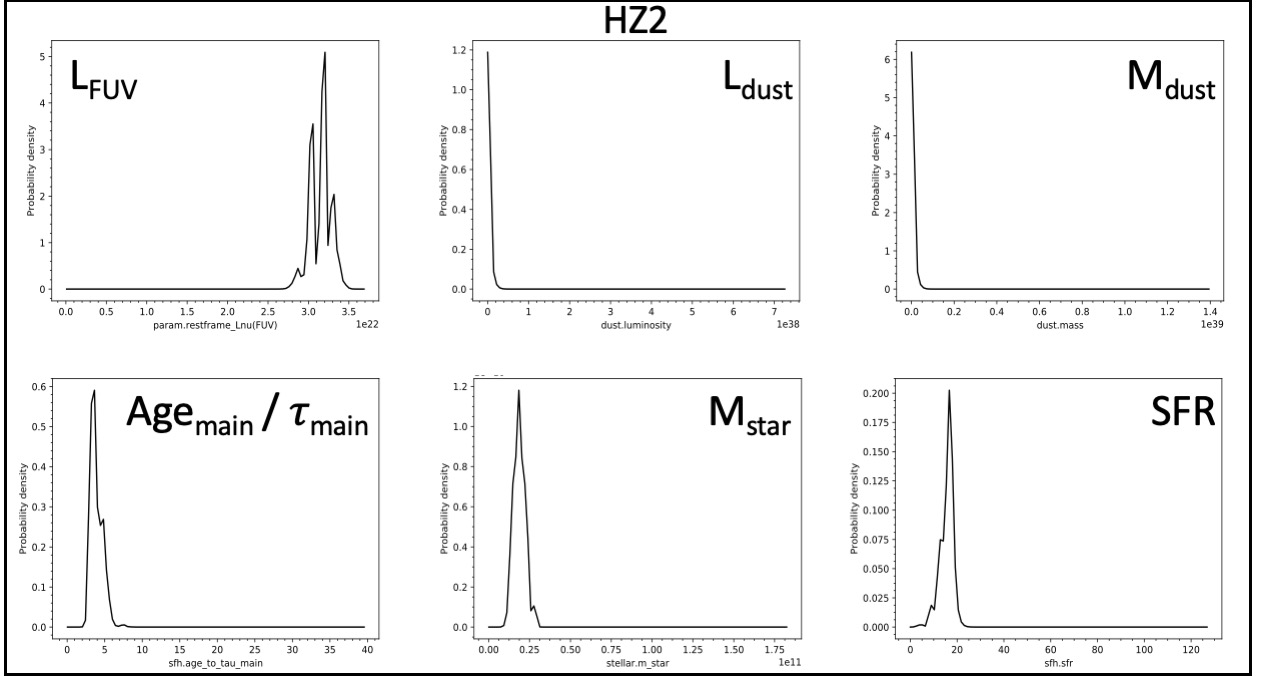}
   \includegraphics[width=1.\columnwidth]{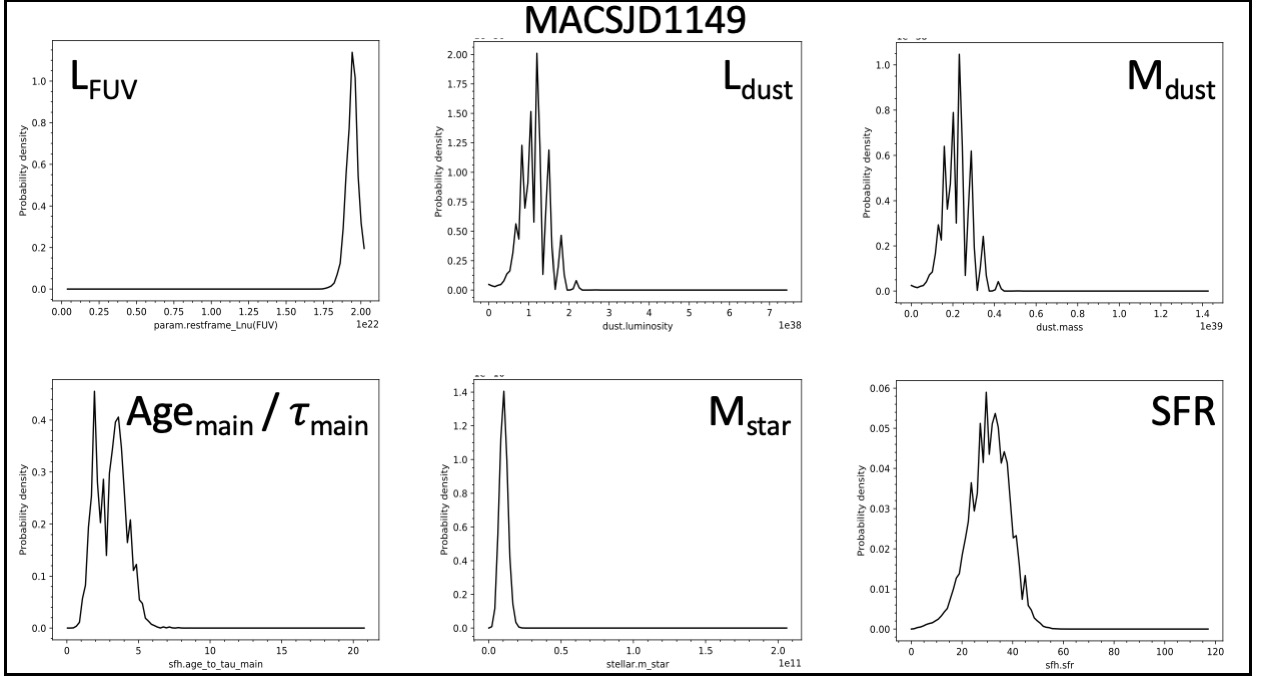}
    \caption{Probability Distribution function (PDF) derived from the fit for each of the important parameters for this paper: L$_{FUV}$, L$_{dust}$, M$_{dust}$, Age$_{main}$, M$_{star}$ and SFR for each representative Hi-z LBGs.}%
   \label{Quality_fits}
  \end{figure*}

  \begin{figure*}
   \centering
   \includegraphics[width=1.\columnwidth]{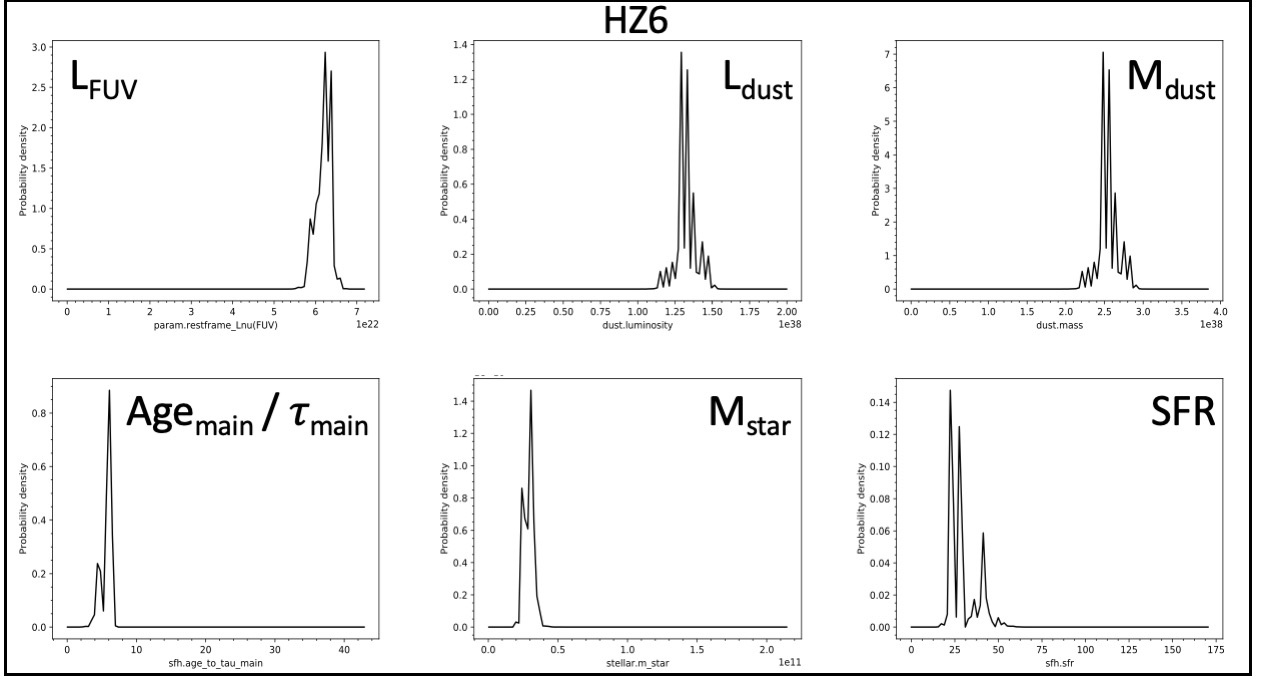}
   \includegraphics[width=1.\columnwidth]{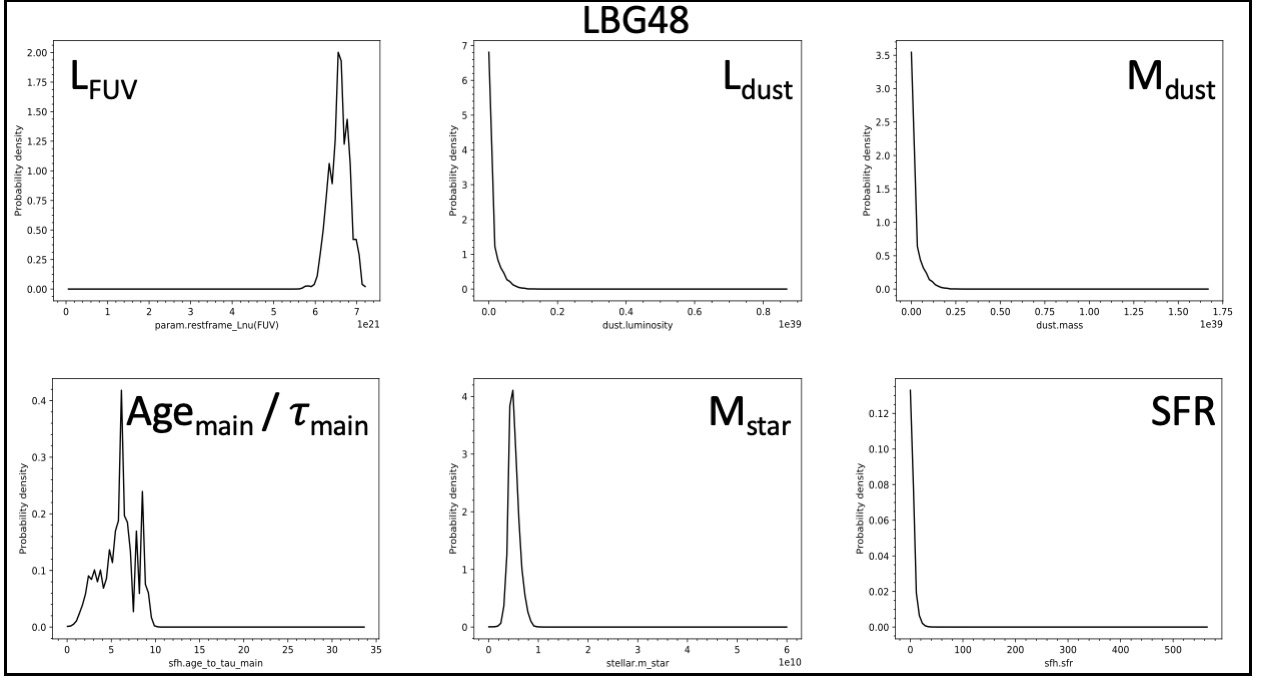}
   \includegraphics[width=1.\columnwidth]{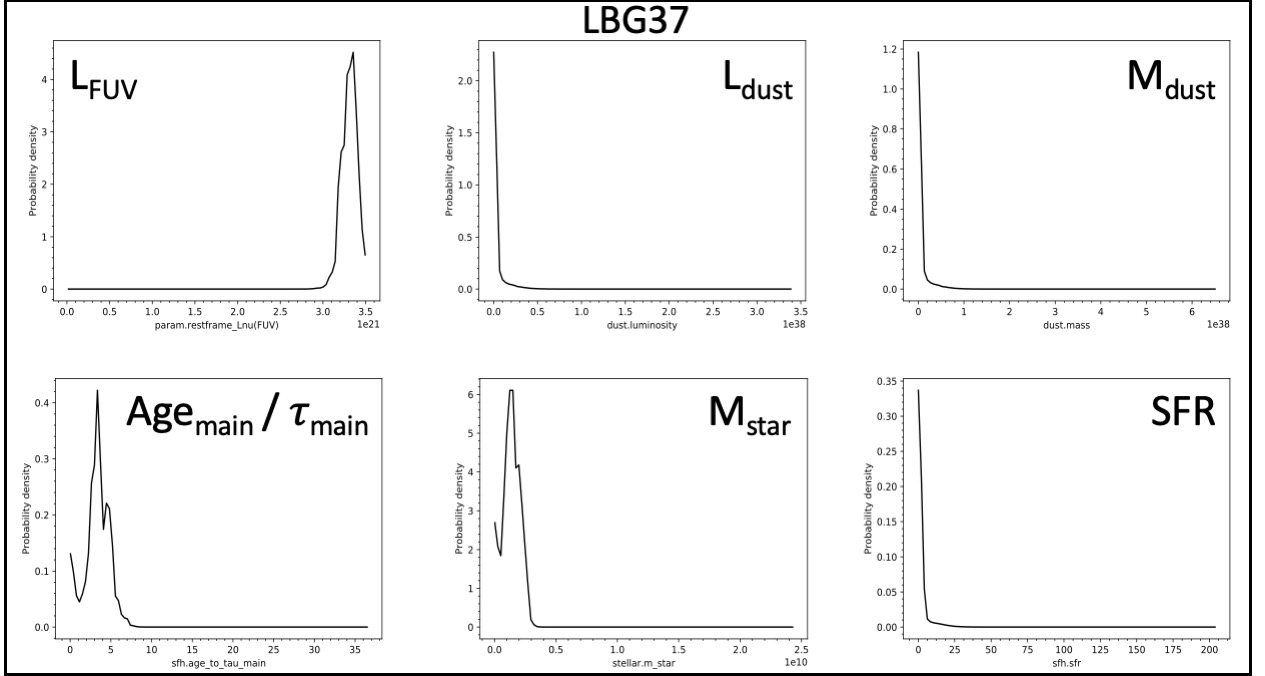}
    \caption{Fig.~\ref{Quality_fits} continued.}%
  \end{figure*}
  
  \begin{figure*}
   \centering
   \includegraphics[width=1.0\columnwidth]{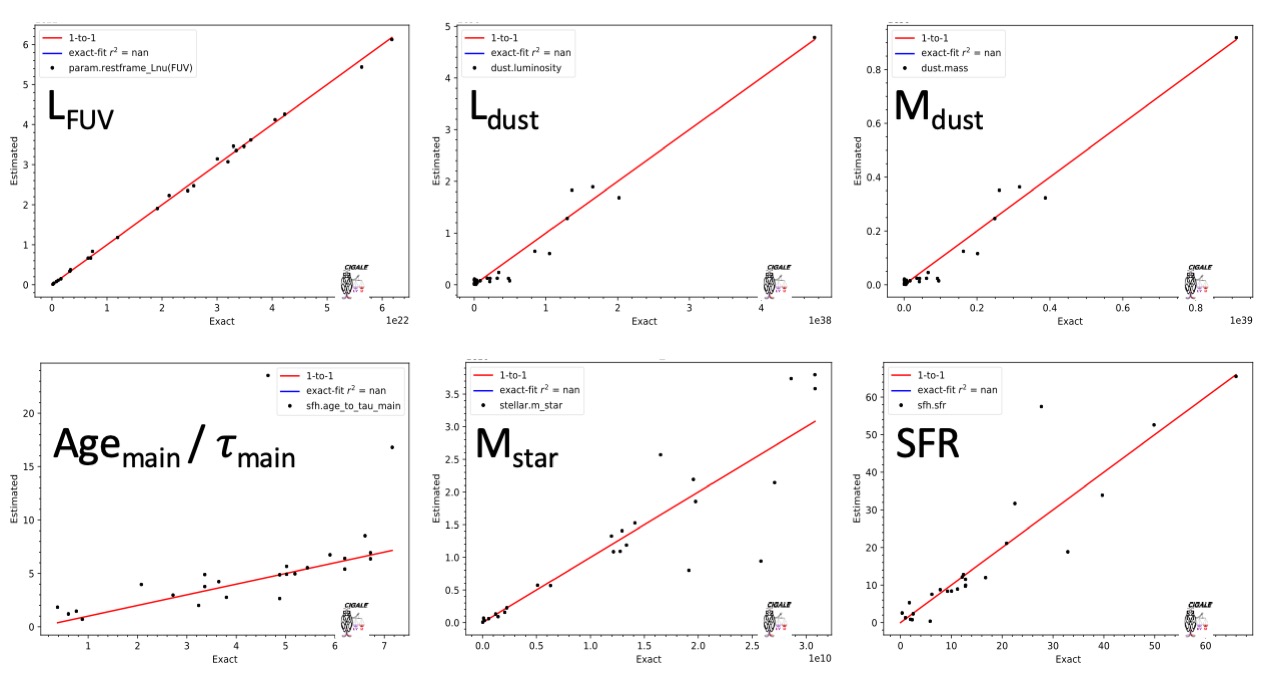}
    \caption{Mock analysis derived from the fit for each of the important parameters for this paper: L$_{FUV}$, L$_{dust}$, M$_{dust}$, Age$_{main}$, M$_{star}$ and SFR for each representative Hi-z LBGs. To build these plots, we use the best-fit modelled SED for each of the Hi-z LBGs. Then, we add the observational noise to the observed data points and we refit the noised best models. Finally, we compare the input and the output parameters. These plots in this figure shows that CIGALE is able to re-derive the input parameters quite nicely and also provide a estimate of the degeneracy between these parameters.}%
   \label{Mock_fits}
  \end{figure*}

\section{Physical parameters derived for the final individual fits of the Low-zZ sample}

\begin{sidewaystable}

\sisetup{table-column-width=12ex,    
         round-mode=places,
         round-precision=3,
         tight-spacing,
         table-format = 2.3e-2,
         table-number-alignment = center
         }
  \resizebox{\linewidth}{0.6\height}{%

\begin{tabular}{l|S|S|S|S|S|S|S|S|S|S|S|S|S|S|S|S|S|S|S|S|S|S|S|}

\hline
  \multicolumn{1}{|c|}{id} &
  \multicolumn{1}{c|}{$\chi_\nu^2$} &
  \multicolumn{1}{c|}{ssfr[yr$^{-1}$]} &
  \multicolumn{1}{c|}{ssfr\_err[yr$^{-1}$]} &
  \multicolumn{1}{c|}{sm$_{dust}$} &
  \multicolumn{1}{c|}{sm$_{dust}$\_err} &
  \multicolumn{1}{c|}{IRX} &
  \multicolumn{1}{c|}{IRX\_err} &
  \multicolumn{1}{c|}{A$_{FUV}$} &
  \multicolumn{1}{c|}{A$_{FUV}$\_err} &
  \multicolumn{1}{c|}{$\beta_{calz94}$} &
  \multicolumn{1}{c|}{$\beta_{calz94}$\_err} &
  \multicolumn{1}{c|}{Age/$\tau$} &
  \multicolumn{1}{c|}{Age/$\tau$\_err[Myr]} &
  \multicolumn{1}{c|}{L$_{dust}$[L$_{\odot}$]} &
  \multicolumn{1}{c|}{L$_{dust}$\_err[L$_{\odot}$]} &
  \multicolumn{1}{c|}{L$_{FUV}$[L$_{\odot}$]} &
  \multicolumn{1}{c|}{L$_{FUV}$\_err[L$_{\odot}$]} &
  \multicolumn{1}{c|}{m$_{dust}$[L$_{\odot}$]} &
  \multicolumn{1}{c|}{m$_{dust}$\_err[L$_{\odot}$]} &
  \multicolumn{1}{c|}{m$_{star}$[M$_\odot$]} &
  \multicolumn{1}{c|}{m$_{star}$\_err[M$_\odot$]} &
  \multicolumn{1}{c|}{sfr[M$_\odot$yr$^{-1}$]} &
  \multicolumn{1}{c|}{sfr\_err[M$_\odot$yr$^{-1}$]} \\
\hline
\hline
  Haro3 & 0.8299179267933174 & 6.440527677418304E-10 & 2.26161128468502E-10 & 0.003483516180040833 & 5.541179377332281E-4 & 5.5624721723502475 & 111.2503498589364 & 1.6587225605834333 & 0.08271924284047054 & -1.4752424890402607 & 0.1133140985425108 & 4.4720664478067285 & 1.3399940686949652 & 3.524911866839427E9 & 1.7624559334197134E8 & 6.336951912066981E8 & 5.1161135270187005E7 & 1.2957549242391195E6 & 6.4787746211955986E4 & 1.8598376715791958E-22 & 2.8084663488775046E-23 & 0.2395667199862205 & 0.07594895744829178\\
  He2-10 & 1.6163215099262476 & 2.2358543565059964E-10 & 1.4503719087547024E-10 & 0.0027427304065287157 & 4.8033635092333655E-4 & 43.23638997314557 & 864.8572632002451 & 3.4518133080971123 & 0.31650453351449614 & -0.3858003492434866 & 0.28622219271234534 & 6.615775939390816 & 2.1737301476441626 & 6.698749082587574E9 & 3.3493745412937874E8 & 1.5493312662662667E8 & 5.362161533734574E7 & 2.4624550734619936E6 & 1.231227536730997E5 & 4.489057815526522E-22 & 7.534501969889094E-23 & 0.20073758946904532 & 0.12578182159072343\\
  HS0052+2536 & 1.882866352721776 & 1.05917399759042E-8 & 1.0455572045037899E-8 & 0.017213480562399443 & 0.010322104941253627 & 1.4739267005685863 & 29.513792645394133 & 1.5612870597632553 & 1.2479271635335845 & -1.877063204553993 & 0.6457749054201105 & 1.1706510799774081 & 1.150104680626877 & 1.8535187381174507E10 & 9.267593690587255E8 & 1.2575379341472218E10 & 1.230483267352845E10 & 6.813520799425344E6 & 3.406760399712672E5 & 1.9791235057680818E-22 & 1.1826534014175572E-22 & 4.192472310659092 & 3.294147387111761\\
  HS0822+3542 & 0.1230782526686612 & 1.7606492712448656E-9 & 4.1108892949945343E-10 & 0.0017537547173579727 & 3.7708367198283294E-4 & 0.36283032296095946 & 2.2272221610108978 & 0.24536450588441147 & 0.043481152311112846 & -2.2665664657706523 & 0.05576308124021396 & 2.2973641120769566 & 1.2847309839091503 & 3.0186707243231284E6 & 4.9178984697019914E5 & 8.319786228693838E6 & 5.344210888974528E5 & 1.1096610648609856E3 & 1.807815741145925E2 & 3.1636723593042913E-25 & 4.439280861966439E-26 & 0.001114023486773325 & 2.0789717873083874E-4\\
  HS1236+3937 & 0.06351813067950332 & 5.624688990498985E-11 & 5.6246E-11 & 2.3016697750666274E-4 & 2.301E-4 & 0.3161995774597723 & 0.21407054678700135 & 0.14680465351236224 & 0.20239265512782212 & -1.2209443992977627 & 0.3052339843326822 & 12.790605937387781 & 5.575002757585253 & 5.577527196618466E7 & 8.405742991243863E7 & 1.763926201744546E8 & 2.370591353336644E7 & 2.050294760015119E4 & 3.0899447374151507E4 & 6.826680573487184E-23 & 1.9065613095015898E-23 & 0.004250484340226587 & 0.007587289478123867\\
  HS1304+3529 & 1.858350265243585 & 1.2358134182758631E-8 & 1.2358E-8 & 0.016107164467410222 & 0.010912729696388224 & 1.0931666441829946 & 21.883400010448604 & 1.2335039887943093 & 1.1249298360534368 & -2.0193801312261597 & 0.6072434767887269 & 1.2269055211110058 & 1.3141841227435618 & 1.154348636939129E9 & 5.7717431846956454E7 & 1.0559676725243118E9 & 9.050647714773438E8 & 4.243376819357876E5 & 2.1216884096789378E4 & 1.3172327221043594E-23 & 8.900018457797852E-24 & 0.3141170796569589 & 0.2468846365043803\\
  HS1319+3224 & 0.3350021535830532 & 7.601774990132374E-10 & 3.46805570357753E-10 & 0.002309937916943497 & 5.762979284722418E-4 & 1.284051560631936 & 12.224948485951112 & 0.6881883597056268 & 0.0925384976824176 & -1.9160364126689877 & 0.1389565140746108 & 4.298788847382507 & 1.7953867497906009 & 1.9054418947034004E8 & 2.0014954390161227E7 & 1.4839294255175015E8 & 1.4687337577269107E7 & 7.004389928555115E4 & 7.357482027692978E3 & 1.516142463651859E-23 & 3.430967541345728E-24 & 0.023050747723332772 & 0.009131231394319553\\
  HS1330+3651 & 2.2215310977712783 & 5.744143331144236E-10 & 5.744E-10 & 0.0014161199253683626 & 4.735451349736672E-4 & 1.4397127442981863 & 28.884823050750523 & 1.9193339810399583 & 1.3851695172772458 & -0.5316734949242482 & 1.5069686338460988 & 11.268221096498856 & 8.382909587724946 & 1.0345363109280722E9 & 5.172681554640362E7 & 7.18571336556707E8 & 1.1407513167197964E9 & 3.802947619201532E5 & 1.9014738096007663E4 & 1.3427350152609093E-22 & 4.439578752771015E-23 & 0.08593567135635768 & 0.15161784449298318\\
  HS1442+4250 & 0.14214737519180135 & 5.532531780069078E-10 & 5.147937955391706E-10 & 6.628378552293109E-4 & 1.7259978278433784E-4 & 0.26462993235550386 & 2.111092441759449 & 0.21647507848262304 & 0.09973729964871904 & -1.9484474362505997 & 0.3112022381020671 & 6.1316411037586285 & 3.2379568074084464 & 2.1029951623400677E7 & 2.6405032816581363E6 & 7.946928541382477E7 & 3.6392160936758354E7 & 7.7305942394993845E3 & 9.706470002456786E2 & 5.83143688800405E-24 & 1.3302905197205315E-24 & 0.0064525219812699075 & 0.005820739329846258\\
  HS2352+2733 & 0.5291293301210153 & 1.974720152519649E-9 & 1.897849018531267E-9 & 0.005732697785140216 & 0.004722084432713481 & 2.413070470993118 & 28.8822113473741 & 1.0763388848259374 & 0.12473772937252146 & -1.9336030085575242 & 0.23903603676240054 & 2.769462077524101 & 2.2897230082736506 & 7.354568313785857E8 & 6.144895645866037E7 & 3.0478050277408713E8 & 3.299204324586509E7 & 2.7035337245994003E5 & 2.258858970906905E4 & 2.3579942863962385E-23 & 1.932287253370622E-23 & 0.09312757673745682 & 0.046762823289165166\\
  IZw18 & 0.023977342456789223 & 1.5204112264311973E-8 & 3.421960078989992E-9 & 0.0021661794997737097 & 4.815440133859023E-4 & 0.058033257672305764 & 0.5081955401028743 & 0.04260299296821588 & 0.006953054088945587 & -2.558036843794125 & 0.023145869592524616 & 0.7796650124236215 & 0.7108030607598648 & 6.524641062034224E6 & 7.45097371911473E5 & 1.1242934351327772E8 & 6.782536960632237E6 & 2.3984530973831384E3 & 2.7389722783554015E2 & 5.536136542778872E-25 & 1.0558895258806325E-25 & 0.016834408301393988 & 0.0020116185182913107\\
  Mrk1450 & 1.074300332409487 & 2.3332089853585824E-9 & 1.1258896867546344E-9 & 0.008053988183626345 & 0.002341195494417348 & 4.085551094076617 & 81.80218252722905 & 1.8888490109321534 & 0.8552277700787447 & -1.6072702268748087 & 0.4475503861036963 & 1.9797211972174629 & 1.2373132401997995 & 1.3500255731928724E8 & 6.7501278659643615E6 & 3.30439037991763E7 & 3.122632058505393E7 & 4.962683750393727E4 & 2.481341875196864E3 & 3.080885914684355E-24 & 8.822279837925465E-25 & 0.014376701398012462 & 0.005583919368570188\\
  Mrk153 & 0.09543314749570261 & 1.957048494727529E-9 & 5.710680604504942E-10 & 0.001633737359573327 & 2.685487114571756E-4 & 0.271946426320087 & 5.438960906358257 & 0.1918182132803094 & 0.020182584694921606 & -2.2900053220707433 & 0.060093771736182786 & 2.6680700349677475 & 1.5702971907611285 & 6.488901397261491E8 & 3.2444506986307457E7 & 2.3860954839773884E9 & 1.646701622381257E8 & 2.3853152237654795E5 & 1.1926576118827397E4 & 7.300179584521582E-23 & 1.1431198056333031E-23 & 0.28573610934257204 & 0.07035601692130228\\
  Mrk209 & 0.39857559497635364 & 1.8651652434494181E-10 & 1.8651E-10 & 9.930064514476974E-4 & 1.950566794545398E-4 & 1.4226221262658731 & 28.462599995417317 & 0.7350021338367122 & 0.21055229032489323 & -1.4105097357556964 & 0.377697077450592 & 8.824307605728855 & 3.951586051115838 & 1.7309601636874318E7 & 8.654800818437157E5 & 1.216739239274234E7 & 6.503007630834449E6 & 6.3629964109452E3 & 3.1814982054726E2 & 3.2039048697360577E-24 & 6.086146994999039E-25 & 9.968301487613906E-4 & 0.0011800663788545558\\
  Mrk930 & 2.2748604893443307 & 2.228856948975916E-9 & 1.1097211418061527E-9 & 0.00986623090663942 & 0.002653969301257565 & 9.100949481949403 & 182.49785737639235 & 2.8441278944746493 & 1.0368786831764412 & -1.1711561030923876 & 0.5286075161662069 & 2.547579227229824 & 1.58585546474545 & 2.2252663762534428E10 & 1.1126331881267214E9 & 2.4450925484939575E9 & 3.5495654641684656E9 & 8.180062292903535E6 & 4.0900311464517674E5 & 4.14548492241288E-22 & 1.0956826996676032E-22 & 1.8479385752389668 & 0.779720637818464\\
  NGC1140 & 0.41605911220775393 & 4.605513184784708E-10 & 1.646986959318506E-10 & 0.0018145117706947263 & 2.832179699844259E-4 & 1.876148588240084 & 37.52321900402666 & 0.8617455301702796 & 0.05923187021961713 & -1.7436542926691545 & 0.10778552427079452 & 4.89433616593002 & 1.3060305879610343 & 4.763574129662977E9 & 2.3817870648314884E8 & 2.5390175168009667E9 & 1.843408049109572E8 & 1.7510862309937174E6 & 8.755431154968588E4 & 4.8252269819205275E-22 & 7.134569269499923E-23 & 0.4444529296962783 & 0.14471967527248975\\
  NGC1569 & 1.9929107610434476 & 2.7851481090677065E-11 & 2.7851E-11 & 0.001640289281910433 & 1.8379315163946026E-4 & 120.52338650497414 & 2410.488886723137 & 3.9989323470344784 & 0.04776933112997698 & 0.49457224235655456 & 0.13449641195976844 & 11.658386475204914 & 4.429954272987841 & 1.7539584477690402E7 & 8.769792238845201E5 & 1.455284736540864E5 & 1.219456857451914E4 & 6.447537928502387E3 & 3.223768964251194E2 & 1.965366109383153E-24 & 1.9707663978118964E-25 & 1.0339476355716489E-4 & 1.0898466797473173E-4\\
  NGC5253 & 0.723526128866812 & 2.9890570044403053E-10 & 1.2488254182276024E-10 & 0.0019416292314291066 & 3.5803898431790955E-4 & 4.6513093586783985 & 93.02701981914906 & 1.4353249834980686 & 0.0801751492348405 & -1.3814595720995035 & 0.12308633167200003 & 5.605474544419001 & 1.4098949488381327 & 1.5035530259243493E9 & 7.517765129621747E7 & 3.232537141652435E8 & 2.7353756032841884E7 & 5.527049500227558E5 & 2.763524750113779E4 & 1.4233019906069958E-22 & 2.5262646128278623E-23 & 0.0850866156891534 & 0.03218167351779979\\
  NGC625 & 0.6303796535889011 & 1.3451786347187161E-11 & 1.3451E-11 & 5.80238529023639E-4 & 7.298558297661988E-5 & 7.938347180412756 & 158.83897248783083 & 1.3097469674252353 & 0.17989205262591693 & 0.17854768945759336 & 0.7588944748411238 & 18.620868973867463 & 8.639340434177514 & 4.4210735360393435E8 & 2.2105367680196717E7 & 5.5692620082779855E7 & 3.355561343710519E7 & 1.6251832729885375E5 & 8.125916364942687E3 & 1.4004441205612591E-22 & 1.6164054077929298E-23 & 0.0014576430940722418 & 0.0037639367989981045\\
  NGC6822 & 0.6317589972502442 & 1.4497720842191996E-11 & 1.4497E-11 & 7.632180991556319E-4 & 6.663186971389948E-5 & 6.670688983429836 & 130.0090894094408 & 1.3332388688695596 & 0.08282871839507926 & -0.45587818467500496 & 0.12513441361425176 & 15.06266971184447 & 5.651983371278256 & 4.8756438146683335E7 & 2.501692107310612E6 & 7.309055821339533E6 & 6.774580132687407E5 & 1.792282962058038E4 & 9.196201180157142E2 & 1.1741617265372032E-23 & 8.293643520231944E-25 & 2.1656285841571997E-4 & 3.4010955599816136E-4\\
  Pox186 & 1.7885404858233342 & 8.540105323633891E-10 & 4.851204378640763E-10 & 0.0057913641560429765 & 0.0016919451481035173 & 21.937535765937312 & 438.9832353590333 & 3.1360315653934627 & 0.5936347843437859 & -0.8563223799109269 & 0.39844219932197494 & 4.376822485147006 & 2.1340548096373464 & 3.963829027008088E7 & 1.9819145135040442E6 & 1.8068706847023244E6 & 1.1766596835838652E6 & 1.457100538854879E4 & 7.285502694274395E2 & 1.257994230370122E-24 & 3.6210011132907055E-25 & 0.00214868064477692 & 0.0010522592702887125\\
  SBS1159+545 & 0.06231689681398642 & 1.6770865963482512E-8 & 1.677E-8 & 0.005859529771894387 & 0.005070861228950745 & 0.20006693871737943 & 1.8469529013970214 & 0.2935874818555122 & 0.6171377301034321 & -2.4027011549231125 & 0.5285591308443431 & 1.346529298733249 & 2.579590342928294 & 1.4632171333234474E8 & 1.5865576850398662E7 & 7.313637838935657E8 & 2.994561936756596E8 & 5.378775065473924E4 & 5.832173996517893E3 & 4.589766819919207E-24 & 3.940702696788988E-24 & 0.1492837188309498 & 0.0852758631433329\\
  SBS1211+540 & 0.1655687002741206 & 1.96543103560938E-9 & 6.895692357683207E-10 & 0.0027094679458938946 & 5.094530935186647E-4 & 0.4644300346947036 & 6.258348970826542 & 0.31415523652273936 & 0.03762290581188288 & -2.2296341170821776 & 0.07597517100595835 & 3.13429383937225 & 1.761637006665983 & 1.0717058436854925E7 & 7.953202258402037E5 & 2.3075722145962976E7 & 1.6217592046425238E6 & 3.9395825392265783E3 & 2.923591107835087E2 & 7.270029795327307E-25 & 1.2559893824358627E-25 & 0.0028577484379082394 & 8.726573632771174E-4\\
  SBS1249+493 & 1.8347597152906037 & 1.0514525695996389E-8 & 1.2137798542265944E-8 & 0.011295067841184366 & 0.009440056809044922 & 0.7594347464385826 & 7.7718430559541405 & 0.982889510564771 & 1.0424593261900346 & -2.107554860815968 & 0.5792715234517312 & 1.4974938248324703 & 1.8315828407106602 & 7.317654024330935E8 & 7.173155831601718E7 & 9.635658703591768E8 & 7.825671239896232E8 & 2.6899640598409757E5 & 2.6368466339745475E4 & 1.1907693241260402E-23 & 9.883380098331887E-24 & 0.25040749313055005 & 0.2009046621321163\\
  SBS1533+574 & 1.2940599695266044 & 2.7198337804812697E-10 & 1.6775127645099588E-10 & 0.0027436849753318744 & 5.06278939322368E-4 & 20.10037930340061 & 402.1722610505226 & 2.8428356901166785 & 0.506337817813171 & -0.7076688824064195 & 0.3723361217874379 & 6.242250518063985 & 2.1108938150700642 & 1.8397341227722435E9 & 9.198670613861218E7 & 9.15273336389724E7 & 5.240068492015006E7 & 6.762848658143388E5 & 3.381424329071694E4 & 1.2324389860620492E-22 & 2.1890815219999432E-23 & 0.06704058373347292 & 0.03959687792149364\\
  Tol1214-277 & 0.19674874802739073 & 2.5159523306077408E-9 & 7.493931069691348E-10 & 0.003621726729519737 & 9.562252041877468E-4 & 0.633082405919707 & 8.178110417351757 & 0.40216873421030447 & 0.044870551106345996 & -2.2479808803756134 & 0.05602747457672733 & 1.872291029010497 & 1.221401966579015 & 5.94362418781903E8 & 4.601127308761224E7 & 9.38838946121787E8 & 6.17765394754928E7 & 2.1848717358424887E5 & 1.6913709030473274E4 & 3.0163398552880545E-23 & 7.613870434842796E-24 & 0.15177934577633995 & 0.023999524084468503\\
  UGC4483 & 0.029229819097546288 & 6.943615443429681E-8 & 1.9975065199810555E-8 & 0.008301289869716628 & 0.0017499519561172058 & 0.07524551954026262 & 0.6165835732009624 & 0.05308461547286938 & 0.014374719269725198 & -2.634861277221576 & 0.020314822819660315 & 0.3139379901173674 & 0.3432443661512078 & 2.786506548630406E6 & 3.400893798061703E5 & 3.7032192290723614E7 & 4.334394012263486E6 & 1.0243176902603902E3 & 1.2501659763777228E2 & 6.169629698133624E-26 & 1.0604371235804834E-26 & 0.008567907210440606 & 0.0019764634977275336\\
  UGCA20 & 0.002249358836924399 & 3.742399203071043E-9 & 3.7423E-9 & 3.7638531309660856E-4 & 3.76E-4 & 0.048127124700497736 & 0.06421758313421301 & 0.13106341602726365 & 0.3877264931501059 & -1.6476763388979063 & 1.3644152575967048 & 7.822259731369842 & 10.411417437430329 & 9.696306064603437E5 & 1.0047860409391595E6 & 2.014727895120474E7 & 1.8566073703925572E7 & 3.5643547426917513E2 & 3.6935858527465257E2 & 5.801492607850201E-25 & 3.792719695949259E-25 & 0.003032377536127929 & 0.0038633674143098624\\
  UM133 & 0.13654721475029621 & 1.1433267462738665E-8 & 8.061317792491239E-9 & 0.006809297076460686 & 0.0037670935697224238 & 0.28527688882967667 & 3.564756352881579 & 0.21506509004828817 & 0.08542894506358002 & -2.4588065255583613 & 0.12044553244016534 & 1.3104615897446648 & 1.3489169728495505 & 1.3412472722967036E8 & 1.073607082758298E7 & 4.70156302460761E8 & 1.2592054155068168E8 & 4.9304147829915375E4 & 3.9465714796137668E3 & 3.620355175893025E-24 & 1.981806073745359E-24 & 0.08278497807219047 & 0.036788369530716354\\
  UM448 & 1.052408880151527 & 9.475939091059099E-10 & 2.726955634853973E-10 & 0.004789167545899492 & 7.894411518151542E-4 & 8.618377764209326 & 172.36865768602613 & 2.0591112115500083 & 0.07891032707283742 & -1.3991759147364147 & 0.09071181449091763 & 3.5087359264582694 & 1.2874432728619714 & 6.762915308343776E10 & 3.381457654171888E9 & 7.847086184164527E9 & 5.613016437104697E8 & 2.4860425293004252E7 & 1.2430212646502126E6 & 2.5954850247711493E-21 & 4.076801110993057E-22 & 4.918931601297485 & 1.1861015325706854\\
  UM461 & 0.12782763056497426 & 6.5430039735121646E-9 & 4.369625885121012E-9 & 0.005376988639846539 & 0.0023735921477550342 & 0.36609360889956555 & 7.323788633836668 & 0.33844619881865107 & 0.29941074792748884 & -2.343393279978197 & 0.25322432660313304 & 1.6767436410415757 & 1.5668716402184402 & 6.973141230013943E7 & 3.486570615006971E6 & 1.9047426834285328E8 & 8.716618959429131E7 & 2.563321418389663E4 & 1.2816607091948315E3 & 2.3836031560435108E-24 & 1.0454350047974094E-24 & 0.031191849842537655 & 0.015708873779557295\\
\hline\end{tabular}}
   \caption{Parameters derived from fitting the SEDs of the DGS Low-zZ sample. Following the IAU recommendations (https://www.iau.org/publications/proceedings\_rules/units/), this table gives values in the International System. To convert to solar luminosity and mass, one can use the following values L$_{\odot} = 3.846 \times 10^{26}$ W and M$_{\odot} = 1.988 \times 10^{30}$ kg, respectively.}%
    \label{Table_Low-zZ}
\end{sidewaystable}

\end{appendices}

\end{document}